\documentclass[11pt]{article}
\usepackage{amsmath}
\usepackage[margin=1in]{geometry}  
\usepackage[T1]{fontenc}          
\usepackage{lmodern}              
\usepackage{amsmath,amssymb}      
\usepackage{graphicx}             
\usepackage{hyperref}             
\usepackage{booktabs}  
\usepackage{siunitx}
\usepackage{pdfpages}
\usepackage{tikz}
\usepackage{soul}
\usepackage{xcolor,colortbl} 
\usepackage{changepage,threeparttable} 
\usepackage{booktabs}
\usepackage{subcaption}
\usepackage{threeparttable}
\usepackage{lscape}
\usepackage{amsfonts}
\usepackage[numbers,sort&compress]{natbib}
\usepackage{authblk}
\usepackage{float}
\usepackage{placeins}
\usepackage{chngcntr}
\definecolor{ModernTeal}{HTML}{4E8A84} 
\definecolor{LightTeal}{HTML}{A2C9C5}
\definecolor{VeryLightTeal}{HTML}{D9E9E7}
\definecolor{ModernOrange}{HTML}{D98244} 
\definecolor{LightOrange}{HTML}{E8B38A}
\definecolor{NeutralFill}{HTML}{F5F5F5}  
\definecolor{NeutralBorder}{HTML}{A9A9A9} 
\definecolor{DarkText}{HTML}{4F4F4F}    
\usetikzlibrary{calc} 
\usetikzlibrary{decorations.pathreplacing,calc,matrix,positioning,arrows.meta}
\definecolor{ModernTeal}{RGB}{0,128,128}
\definecolor{MidGray}{RGB}{128,128,128}
\definecolor{DarkText}{RGB}{40,40,40}
\usetikzlibrary{
    shapes.geometric, 
    arrows.meta, 
    positioning,     
    shadows,
    calc}
\usetikzlibrary{matrix, positioning}
\usetikzlibrary{positioning, arrows.meta, shadows, calc, fit}

\newcommand{\norm}[1]{\left\lVert#1\right\rVert}

\title{
  \begin{center}
    \includegraphics[height=5em]{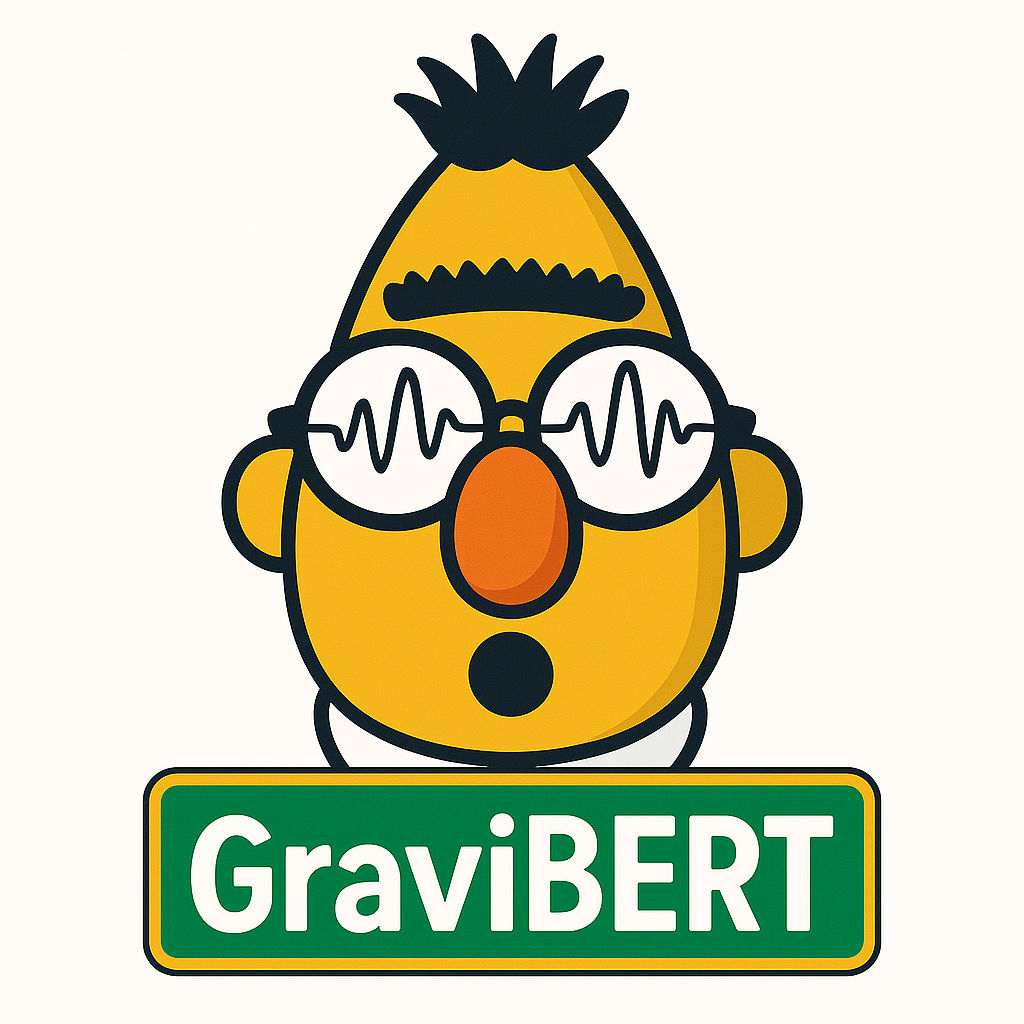}\\[0.5em]
    \texttt{GraviBERT}: Transformer-based inference for gravitational-wave time series
  \end{center} }
  
\author[1]{Martin Benedikt}
\author[2]{Ippocratis D. Saltas\thanks{Corresponding author: saltas@fzu.cz}}
\affil[1]{Faculty of Information Technology, Czech Technical University, Prague, Czechia}
\affil[2]{CEICO, Institute of Physics, Czech Academy of Sciences, Prague, Czechia}

\begin{document}
\date{}
\maketitle

\begin{abstract}
We introduce \texttt{GraviBERT}, a novel deep learning framework for gravitational wave inference, built on a multi-scale feature extractor with a transformer encoder and a suitable regression head. A key novelty of \texttt{GraviBERT} is its staged training: a BERT-style self-supervised pretraining phase to learn transferable representations, followed by supervised fine-tuning on labeled data. \texttt{GraviBERT} demonstrates consistent transfer learning across detector configurations and waveform models. On in-domain data, pretraining reduces the MAE by up to 31\% and accelerates convergence by $\sim 6.6 \times$, with mean relative precision for point estimates reaching the few-percent level and MAE in effective spin of $\sim 10^{-3}$ at SNR = 10. For domain adaptation to new detector noise profiles, the pretrained model converges up to $15\times$ faster on small target datasets and reduces estimation errors by up to $\sim 47$\%, demonstrating detector-agnostic learning. Cross-waveform approximant transfer achieves up to $44\%$ MAE reductions and up to $15\times$ training speedups, with $R^2$ scores consistently exceeding $0.9$ for mass parameters at SNR = 10 compared to $0.74$ - $0.87$ when training from scratch. \texttt{GraviBERT} works directly with noisy waveforms, and in its current form quantifies predictive uncertainty through MC dropouts. After pretraining, the regression head could be adapted to multiple downstream inference tasks in gravitational-wave astronomy.
\end{abstract}

\tableofcontents
\section{Introduction}

Time series analysis lies at the heart of numerous scientific and applied domains, from fundamental physics to finance and climate modelling.  Traditionally, auto-regressive models and recurrent neural networks (RNNs) have been widely employed for this task, with RNNs and their variants such as Long Short-Term Memory (LSTM) networks \cite{Hochreiter1997} the dominant choice. 

More recently, transformer architectures, originally developed for natural language processing, have redefined the landscape of time series modelling by offering significantly enhanced scalability and parallelism, as well as the ability to capture subtle global dependencies through the so--called self-attention mechanism \cite{vaswani2017attention}. Yet, despite their growing success, the application of transformers to the computationally demanding domain of gravitational-wave astronomy, remains mostly unexplored. In this work, we introduce a transformer-based architecture tailored for analysis of gravitational waveforms. Our approach demonstrates that an architecture based on a transformer encoder combined with self-supervised pretraining can improve both the efficiency and accuracy for gravitational-wave parameter estimation. This approach suggests a potential pathway towards foundation-style models for gravitational-wave science that could be extended to multi-modal tasks.

The discovery of gravitational waves (GWs) marked a key moment in physics, opening an unprecedented observational window into the Universe \cite{Abbott2016, TheLIGOScientificCollaboration2021}. This makes them invaluable for probing the nature of gravity over astrophysical and cosmological distances, the physics of stellar populations, testing theories related to dark matter or environmental effects, the physics of dark energy, or exotic fields \cite{Berti2015, Saltas:2018fwy, LISA:2022kgy, Barausse:2020rsu}. Reconstructing this wealth of information from noisy observational data constitutes a challenging inverse problem. Aside the challenges arising from the non-linear nature of Einstein’s equations, the low signal-to-noise ratios (SNR) and various instrumental and modelling systematics present hurdles with the waveform reconstruction. 
Consequently, the development of fast, accurate, and robust methods for GW signal analysis is essential to fully exploit the scientific potential of current and next-generation detectors. 

Machine learning (ML) has emerged as a promising tool to tackle the above challenge. GW data analysis relies on parameter estimation using Bayesian inference \cite{skilling2006nested, veitch2015parameter, ashton2019bilby} for confirmed detections. ML-based approaches can substantially accelerate GW data analysis\footnote{We should note here that it is impossible to provide here an exhaustive list of previous works, and for reviews we refer to \cite{cuoco2021enhancing, stergioulas2024machine}.}. GW parameter estimation conventionally relies on matched filtering \cite{Abbott2016, allen2005chi2} and Bayesian inference \cite{skilling2006nested, veitch2010bayesian, vandersluys2008parameter, cornish2015bayeswave, veitch2015parameter, ashton2019bilby, GWsequential} (see also \cite{thrane2019introduction} for reviews), requiring up to days per event. Recent neural network approaches achieve dramatic speedups: normalising flows accelerate posterior sampling \cite{williams2021nested, williams2023importance, green2020gravitational}, neural posterior estimation enables real-time inference through frameworks like DINGO, which solves the full 15-parameter BBH problem \cite{dax2023neural, Dax:2024mcn}, and related NPE methods \cite{dax2023neural,dax2021real}, conditional variational autoencoders reduce inference time from weeks to seconds \cite{gabbard2022bayesian}, and neural network emulators enable rapid likelihood evaluation \cite{chua2019reducing, williams2021nested}. Recent transformer-based approaches for GWs include work on overlapping signal parameter estimation in Einstein Telescope (ET) scenarios \cite{papalini2025transformer} and pretrained audio transformers \cite{chatterjee2024transformer} (see also \cite{Wang:2025aqk,Chan:2025pdf,Inguglia:2025cig} for similar applications with alternative architectures). Our work differs in its focus on self-supervised masked pretraining directly on noisy GW data for learning generalisable representations and demonstrating transfer learning across detector configurations, rather than addressing a specific astrophysical application. Very recently, Ref.~\cite{kofler2025} introduced Dingo-T1, a transformer-encoder–based model that achieves flexibility via tokenisation of frequency-segmented data, enabling adaptation at inference time to different detector configurations. In contrast, GraviBERT combines a multi-scale feature extractor with a transformer encoder and advances a distinct transfer-learning paradigm: BERT-style masked pretraining on unlabeled GW data to learn transferable representations of the underlying physics. These representations can, in principle, support multiple downstream tasks and accelerate adaptation to new detectors and waveform approximants, demonstrating for the first time that self-supervised pretraining enables efficient transfer learning in GW analysis.

ML methods come with limitations -- they usually depend on assumptions about the underlying noise, waveform or detector modelling. Training must be typically repeated whenever detector configurations, noise characteristics, or theoretical models evolve or change. In this regard, most methods lack a fair amount of generalisability across different signal types or parameter ranges, creating brittleness in operational deployment. Often, the black-box nature of these approaches obscures uncertainty quantification and makes systematic error detection challenging. 

Transformers revolutionised natural language processing through their self-attention mechanism, which enables dynamic weighting of temporal components while processing data in parallel \cite{vaswani2017attention}. This architecture produces rich embeddings that serve as versatile foundations for diverse downstream tasks, forming the basis of foundation models like BERT \cite{Devlin2018} and the GPT series \cite{Radford2018,Radford2019}, which achieve broad performance through pretraining and fine-tuning paradigms.

Building on this paradigm, we develop \texttt{GraviBERT}, which adapts transformer architectures to gravitational wave analysis. Our multi-scale convolutional feature extractor downsamples the time-domain waveform while progressively upsampling feature dimensions, hence learning local temporal features. The compressed representations are passed to the transformer encoder, which encodes transferable GW features through self-attention. {\bf Our primary contributions are}: {\bf (i)} demonstrating that this architecture captures transferable representations of GW signals, {\bf (ii)} showing that self-supervised pretraining improves training efficiency and accuracy compared to training from scratch, and {\bf (iii)} establishing promising transfer learning capabilities for domain and waveform adaptation. The name GraviBERT reflects our model's conceptual inspiration from BERT's masked pretraining approach, adapted for continuous-valued time series.

Our focus is on demonstrating for the first time the effectiveness of transformer-based architectures with self-supervised pretraining for GW parameter inference. Therefore, we work in the simplified setting of point estimates backed by with MC Dropout uncertainty quantification \cite{gal2016dropoutbayesianapproximationrepresenting} as a computationally efficient framework to evaluate our architectural and training novelties. Extension to full Bayesian inference represents a natural, but separate direction which deserves its own study.

{\it We structure the paper as follows\footnote{Our code and pretrained models are publicly available at https://zenodo.org/records/18671949.}}: In Section \ref{sec:data} we discuss our choice of waveform modelling and data generation. In Section \ref{sec:model} we introduce our model architecture, and proceed discussing its training in Section \ref{sec:training}. In Section \ref{sec:results} we present our results and discuss our conclusions in Section \ref{sec:conclusions}. 

\section{Modelling and data generation} \label{sec:data}

The GW signal from a binary merger, such as that of two black holes or neutron stars, exhibits three distinct evolutionary phases: inspiral, merger, and ringdown \cite{Maggiore2000,Blanchet2014}. 
At leading order in the post-Newtonian expansion, the GW strain can be approximated as \cite{Blanchet2014},
\begin{equation}
h(t) \simeq \frac{\mathcal{M}_c^{5/3}}{d} \, \omega(t)^{2/3} \cos\left[2\Phi(t)\right],
\end{equation}
where $\mathcal{M}_c \equiv (m_1 m_2)^{3/5}/(m_1 + m_2)^{1/5}$ is the chirp mass, $\omega(t)$ represents the centre-of-mass orbital frequency, $\Phi(t)$ denotes the orbital phase, and $d$ is the luminosity distance to the source. The orbital phase evolution is governed by \cite{Blanchet2014,Buonanno99}:
\begin{equation}
\frac{d\Phi}{dt} \propto  \left( \frac{G \mathcal{M}_c}{c^3} \right)^{-5/8} \cdot (t_c - t)^{-3/8} \cdot \left[ 1 + \cdots + \chi v^3 + \mathcal{O}(v^n) \right],
\end{equation} 
where $t_c$ denotes the coalescence time, $v \equiv \left( G (m_1 + m_2) \omega \right)^{1/3}$ is the orbital velocity, and $\chi_{\rm{eff}} \equiv (m_1 s_1 + m_2 s_2)/(m_1 + m_2)$ represents the effective spin parameter. Notice that higher-order corrections introduce mass-spin coupling terms that create parameter degeneracies during inference \cite{Cutler1994,Vallisneri2008}.

We utilise the Effective One Body (EOB) formalism through the \texttt{SEOBNRv4} approximant \cite{Bohe2017}, which delivers accurate modelling of both inspiral and merger phases through a  combination of analytical post-Newtonian theory and numerical relativity calibration. 
We will focus on binary black hole (BBH) mergers with mass ratios $q = m_2 / m_1$ ranging between $q \sim 0.1 - 1$. These systems fall within the expected validity of our approximant and lie within the detectable range of current and next-generation GW observatories, such as Advanced LIGO and the ET. 

GWs manifest as time-series strain measurements in interferometric detectors. Given the limited number of confirmed detections from current observatories \cite{Abbott2021}, training robust neural networks necessitates extensive simulated datasets. For waveform generation, we employ the open-source \texttt{PyCBC} library, which provides access to state-of-the-art waveform approximants. Our training dataset is constructed using a grid-based approach over source masses, spins and luminosity distance. We exclude source orientation or other parameters from our grid, as we are here interested in a first proof-of-concept analysis. We employ a fixed sampling rate of 4096 Hz with a lower frequency cutoff of $f_{\rm low} = 20$ Hz and simulate ET detector noise using as our reference the ET-B power spectral density (PSD) \cite{Hild2011}. While ET typically uses $f_{\rm low} = 5$ Hz to maximise sensitivity, we use 20 Hz for this proof-of-concept to reduce computational costs and waveform duration, noting that the lower cutoff primarily affects low-mass systems and does not fundamentally limit the applicability of our method.

The duration of each waveform depends on the source parameters, with more massive systems producing shorter signals. We write,
\begin{equation}
h = h\left[p_i; t \right], \quad p_i = \{m_1, m_2, \chi_\text{eff}, d\},
\end{equation}
with the effective spin $\chi_{\rm eff}$ defined as $\chi_{\rm eff} \equiv (m_1 s_1+m_2 s_2)/(m_1+m_2)$, where $m_1$ and $m_2$ are the component masses (with $m_1 \geq m_2$ by convention), and $s_1$ and $s_2$ are the dimensionless spin parameters aligned with the orbital angular momentum, and the parameter ranges
\begin{equation}
m_1, m_2 \in [5, 150]~M_\odot, \quad s_1, s_2 \in [-0.4, 0.4].
\end{equation}
The distance $d$ denotes an effective luminosity distance used to rescale the signal amplitude and enforce a fixed target SNR. When detector noise is included, $d$ is adjusted for each waveform realization using the scaling $h(t) \propto 1/d$. We focus on two fixed SNR values (10 and 30) to enable controlled comparison of model performance under different (conservative) noise conditions. This assumption must be compared with the continuous SNR distributions expected in realistic astrophysical observations. We fix the length of our input data to $4096$ using appropriate padding as needed.

\begin{figure}[htbp] 
  \centering
  \includegraphics[width=0.7\linewidth]{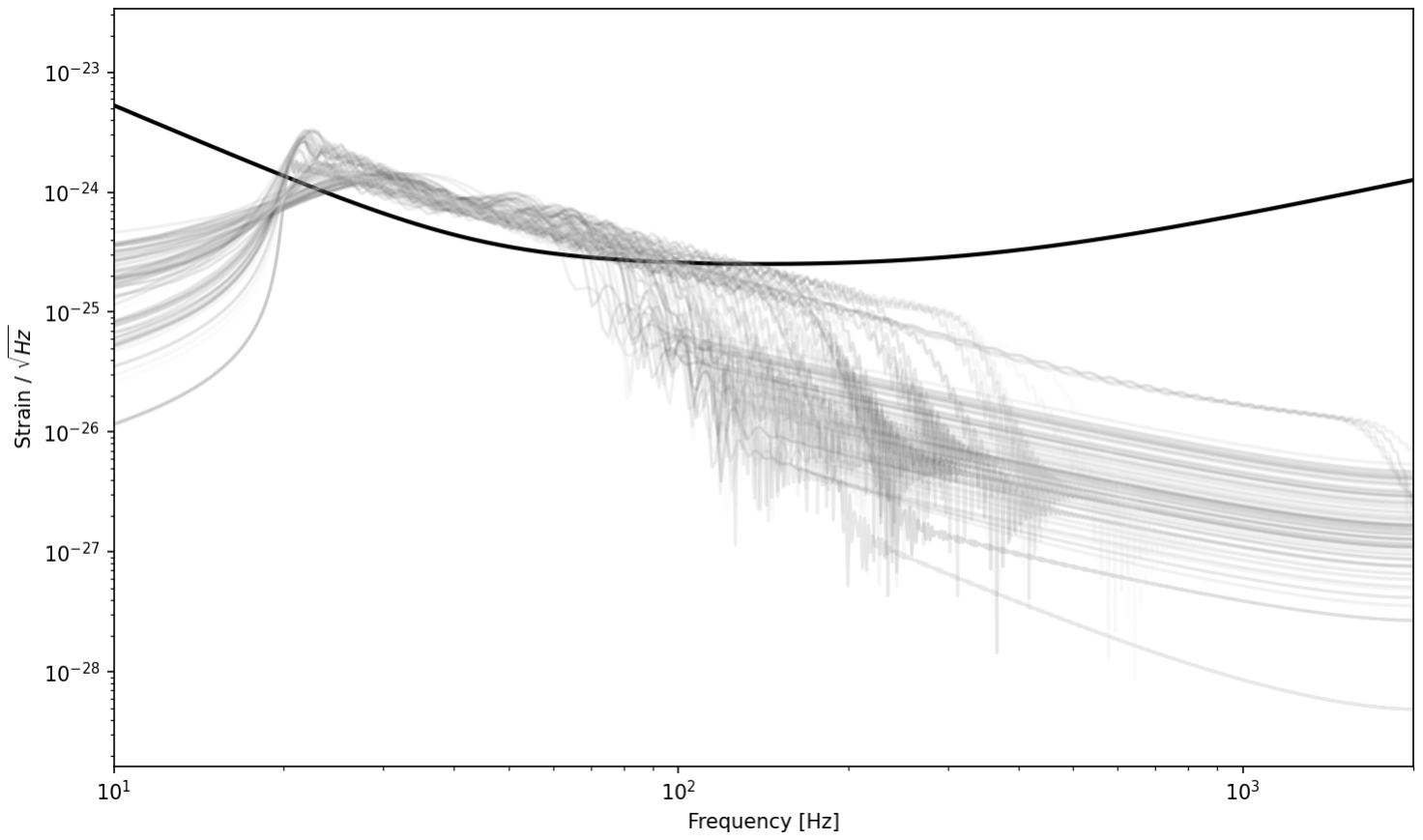}
  \caption{The ET's power spectral density (black, ET-B) overlaid with the frequency-domain representations of selected waveforms from our sample at fixed SNR = 30. Different waveforms intersect the PSD at different stages of their evolution. Our analysis uses waveforms with a low-frequency cutoff at $f_{\rm low}$ = 20 Hz, as explained in Section \ref{sec:data}.}
 \label{fig:PSD_waveforms}
\end{figure}

We model single-channel strain data compatible with either the L-shaped configuration or a single arm of the triangular configuration. The full triangular ET design provides three independent interferometric data streams, which would require multi-channel architectures for optimal analysis. For this proof-of-concept, we focus on single-channel analysis to isolate the benefits of our pretraining strategy and an extension to multi-channel processing for the complete triangular configuration is straightforward and remains for future work. We generate the noise realization $n(t)$ by colouring Gaussian noise based on the given PSD and construct the observed signal (s) as
\begin{equation}
s(t) = h(t) + n(t).
\end{equation}
A sample of our waveforms in frequency space is shown in Figure \ref{fig:PSD_waveforms}. 

We emphasise that we focus on the four intrinsic parameters and do not model extrinsic parameters such as sky location ($\alpha, \delta$), inclination angle ($\iota$), polarization angle ($\psi$), or coalescence phase ($\phi_c$). Consequently, we do not apply the detector antenna pattern function $F^+(\alpha,\delta,\psi,t)$ and $F^\times(\alpha,\delta,\psi,t)$ that would project the GW polarizations onto the detector frame. Our strain data represents the waveform $h(t)$ in the source frame at optimal orientation, rather than the detector-frame strain $s(t) = F^+ h_+(t) + F^\times h_\times(t)$ that would be observed in a real interferometer. This simplification significantly reduces the complexity of the parameter space but limits direct applicability to observational data. However, it serves our primary objective: demonstrating that self-supervised pretraining and transfer learning are effective strategies for GW analysis. The objectives of this work are orthogonal to whether the full parameter space is modelled, that is, the same pretraining approach can be applied to architectures handling all 15 parameters.

We note that the $(m_1, m_2)$ parameterisation used here exhibits stronger correlations than alternative parameterisations such as chirp mass and mass ratio. Future implementations could benefit from using $({\cal M}_c, q)$ or similar parameterisations that align more naturally with the physical variations encoded in the waveform, potentially improving training efficiency and reducing parameter degeneracies.

We generate training data on a regular grid over the parameter space rather than drawing randomly from astrophysically motivated priors. While production systems typically use prior-based sampling, grid-based generation offers advantages for our proof-of-concept study: (i) it ensures uniform coverage of the parameter space, (ii) it enables systematic evaluation of model performance across parameter ranges, and (iii) it provides a controlled setting for isolating the effects of our pretraining and transfer learning strategies. The network performs regression on continuous values, i.e the model outputs continuous predictions, and is not constrained to return grid values. However, since our test set shares the same grid structure, we do not explicitly evaluate off-grid generalisation.

\section{Model architecture} \label{sec:model}

\subsection{Overview}
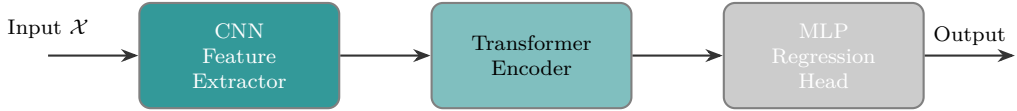
\begin{figure}
    \centering
    \begin{tikzpicture}[
    node distance=0.5cm and 1.2cm, 
    >=Stealth, 
    block/.style={ 
        rectangle, 
        draw=gray,      
        thick, 
        text centered, 
        minimum height=1.4cm, 
        text width=2.4cm,   
        font=\scriptsize,        
        text=black,            
        rounded corners=1.5mm,
        align=center
    },
    cnn_style/.style={
        block, 
        fill=teal!80,
        text=white
    },
    transformer_style/.style={
        block, 
        fill=teal!50
    },
    mlp_style/.style={
        block, 
        fill=gray!40,
        text=white
    },
    arrow/.style={
        ->, 
        thick, 
        rounded corners=1mm, 
        color=black!80
    },
    coord/.style={coordinate}
]
\node (start) [coord] {};
\node (input_label) [above=0.1cm of start, font=\scriptsize, text=black] {Input $\mathcal{X}$};
\node[cnn_style, right=of start] (CNN) {CNN\\Feature\\Extractor};
\node[transformer_style, right=of CNN] (Transformer) {Transformer\\Encoder};
\node[mlp_style, right=of Transformer] (MLP) {MLP\\Regression\\Head};
\node (end) [coord, right=of MLP] {};
\draw[arrow] (MLP) -- node[above, font=\scriptsize, text=black] {Output} (end);
\draw[arrow] (start) -- (CNN);
\draw[arrow] (CNN) -- (Transformer);
\draw[arrow] (Transformer) -- (MLP);
\end{tikzpicture}
    \caption{Overview of our model architecture.}
    \label{fig:Model_Overview}
\end{figure}

Our model architecture, schematically shown in Figure \ref{fig:Model_Overview}, is structured as a sequential combination of three primary components: {\bf i)} a convolutional neural network (CNN) feature extractor, {\bf ii)} a transformer encoder, and {\bf iii)} a regression multi-layer perceptron (MLP) head. This hybrid architecture is the core strength of our model. The model combines a CNN feature extractor ($\mathcal{F}$) with a Transformer encoder ($\mathcal{T}$), leveraging the complementary strengths of both architectures for analysing time-series data. Whereas CNNs are highly effective at efficiently identifying local patterns and hierarchical structures within the input signal, the Transformer's self-attention mechanism excels at modelling global context and capturing subtle long-range temporal dependencies.

The processing of information in our model can be viewed as a composite function operating on the input data. Let $\mathcal{X}$ represent the input, which could be potentially padded or pre-processed. The data flows through the components as follows:

\begin{enumerate}

    \item \textbf{CNN Feature Extractor ($\mathcal{F}$):} $\mathcal{X}$ is first processed by the CNN feature extractor ($\mathcal{F}$) which maps the time series to a sequence of latent feature vectors ($L$). The feature extractor reduces the temporal dimension while increasing the feature dimensionality, capturing local patterns in the signal and potentially making the training faster due to working with shorter sequences.

    \item \textbf{Transformer Encoder (T):} The latent sequence of feature vectors ($L$) is first linearly projected to match the transformer's input dimensionality, to be followed by the addition of positional encodings to inject information about the position of each element in the sequence. To facilitate sequence-level aggregation, a special token, often denoted as \texttt{[CLS]}, is also prepended to the sequence before the addition of positional encodings. This approach, popularised by models like BERT \cite{Devlin2018}, allows the model to learn an aggregate representation of the entire input sequence into the output state corresponding to this \texttt{[CLS]} token. The encoder then processes this augmented sequence ($L'$) using stacked layers of self-attention and feed-forward networks to capture global dependencies and contextual relationships within the data, outputting an encoded sequence representation ($E$). We write,
    \begin{equation}
    E = \mathcal{T}(L').
    \end{equation}

    \item \textbf{Aggregation (Agg):} The aggregation step involves extracting the final hidden state vector from the transformer encoder's output sequence $E$ that corresponds to the position of the prepended \texttt{[CLS]} token. This vector ($E_{agg}$) carries the aggregated information from the entire sequence and is suitable for downstream tasks such as classification or source parameter inference. Formally, we write,
    \begin{equation}
    E_{agg} = \text{Agg}(E),
    \end{equation}
    where the function $\text{Agg}$ specifically selects the output vector at the \texttt{[CLS]} token's position.

    \item \textbf{Regression MLP Head ($\mathcal{H}$):} Finally, the aggregated encoded representation $E_{agg}$ is passed through the regression MLP head ($\mathcal{H}$) which maps this fixed-size vector to e.g the predicted GW source parameters ($\mathbf{y}$) using linear layers. We write,   
    \begin{equation}
    \mathbf{y} = \mathcal{H}(E_{agg}).
    \end{equation}
\end{enumerate}
In summary, the entire model can be described as a composite function where the input $\mathcal{X}$ is sequentially transformed according to
\begin{equation}
\mathcal{X} 
\xrightarrow{\text{(1)}} \; 
\mathcal{F}(\mathcal{X}) \;
\xrightarrow{\text{(2)}}  \;
\text{concat}(\texttt{[CLS]}, L)  \;
\xrightarrow{\text{(2)}}  \;
\mathcal{T}(L')  \;
\xrightarrow{\text{(3)}}  \;
\text{Agg}(E)  \;
\xrightarrow{\text{(4)}}  \;
\mathcal{H}(E_{\text{agg}}).
\end{equation}

\subsection{Feature extractor}
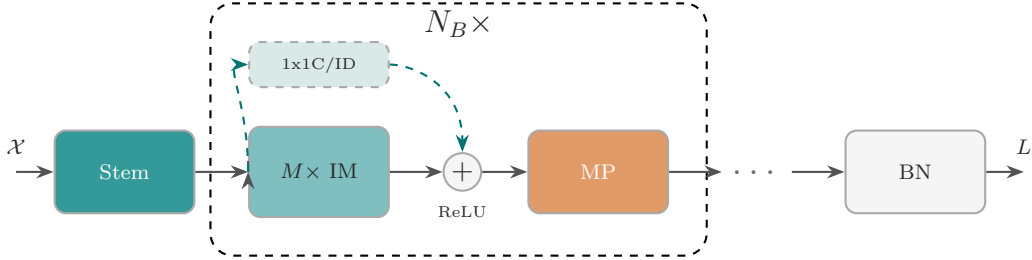
\begin{figure}[h]
    \centering

    \begin{tikzpicture}[
        node distance=0.4cm and 0.7cm, 
        >=Stealth, 
        block/.style={
            rectangle, 
            draw=NeutralBorder,         
            thick, 
            text centered, 
            minimum height=1.1cm, 
            minimum width=1.8cm,   
            text width=1.6cm,    
            font=\scriptsize,    
            text=DarkText,            
            rounded corners=1.5mm     
        },
        stem_style/.style={block, fill=ModernTeal!80, text=white, node contents={Stem}}, 
        it_module_stack/.style={block, fill=ModernTeal!50, text=DarkText, minimum height=1.2cm, node contents={$M \times$ IM}}, 
        res_proj_style/.style={block, fill=VeryLightTeal, dashed, text=DarkText, minimum width=1.6cm, minimum height=0.6cm, font=\tiny, node contents={1x1C/ID}}, 
        add_relu_style/.style={ 
            circle, 
            draw=NeutralBorder, 
            thick, 
            fill=NeutralFill,         
            minimum size=5mm,
            inner sep=0pt, 
            node contents={+},
            text=DarkText,
            label={[font=\tiny, yshift=-0.05cm, text=DarkText]below:ReLU}
        },
        downsampler_style/.style={block, fill=ModernOrange!80, text=white, node contents={MP}}, 
        final_bn_style/.style={block, fill=NeutralFill, text=DarkText, node contents={BN}}, 
        coord/.style={coordinate}, 
        arrow/.style={->, thick, rounded corners=1mm, color=DarkText!80}, 
        res_arrow/.style={->, thick, dashed, rounded corners=1mm, color=ModernTeal!90!black}, 
        ellipsis_style/.style={font=\Large, text=DarkText!70}, 
        block_label_style/.style={font=\scriptsize\bfseries, text=ModernTeal!90!black} 
    ]
     
    \node (start_coord) [coord] {};
    \node (input_label) [above=0.1cm of start_coord, font=\scriptsize, text=DarkText] {$\mathcal{X}$}; 

    \node (stem) [stem_style, right=0.5cm of start_coord] {}; 

    \coordinate (block1_input_anchor) at ($(stem.east) + (0.7cm, 0)$);

    \node (it_stack1) [it_module_stack, right=0cm of block1_input_anchor] {};
    \node (res_proj1) [res_proj_style, above=0.5cm of it_stack1] {}; 
    \node (add_relu1) [add_relu_style, right=0.7cm of it_stack1] {};
    \node (ds1)       [downsampler_style, right=0.6cm of add_relu1] {};

    \node (repeating_block_box) [
        draw=black, 
        thick, 
        rounded corners=3mm, 
        inner sep=0.5cm, 
        dashed, 
        fit=(it_stack1) (res_proj1) (add_relu1) (ds1)
    ] {};
    
    \node[font=\large, text=DarkText] at (repeating_block_box.north) [yshift=-8pt] {$N_B \times$};

    \node (ellipsis) [ellipsis_style, right=0.7cm of ds1] {\dots}; 
    \node (final_bn) [final_bn_style, right=0.7cm of ellipsis] {}; 
    \node (end_coord)   [coord, right=0.5cm of final_bn] {}; 
    \node (out_label) [above=0.1cm of end_coord, font=\scriptsize, text=DarkText] {$L$}; 

    \draw [arrow] (start_coord) -- (stem.west);
    \draw [arrow] (stem.east) -- (block1_input_anchor);
    \draw [arrow] (block1_input_anchor) -- (it_stack1.west);
    \draw [arrow] (it_stack1.east) -- (add_relu1.west);
    \draw [arrow] (add_relu1.east) -- (ds1.west);

    \draw [res_arrow] (block1_input_anchor) .. controls +(0,0.9cm) and +(-0.3cm,0cm) .. (res_proj1.west); 
    \draw [res_arrow] (res_proj1.east) .. controls +(0.3cm,0cm) and +(0,0.9cm) .. (add_relu1.north);     

    \draw [arrow] (ds1.east) -- (ellipsis.west);
    \draw [arrow] (ellipsis.east) -- (final_bn.west);
    \draw [arrow] (final_bn.east) -- (end_coord);

    \end{tikzpicture}
    \caption{Architecture of our feature extractor ($\mathcal{F}$). The highlighted dashed box represents the main processing block, which is repeated $N_B$ times. IM stands for \texttt{InceptionModule}, and MP denotes MaxPool layer. $M$ is the number of modules stacked within each main block.}
    \label{fig:inception_time_network}
\end{figure}

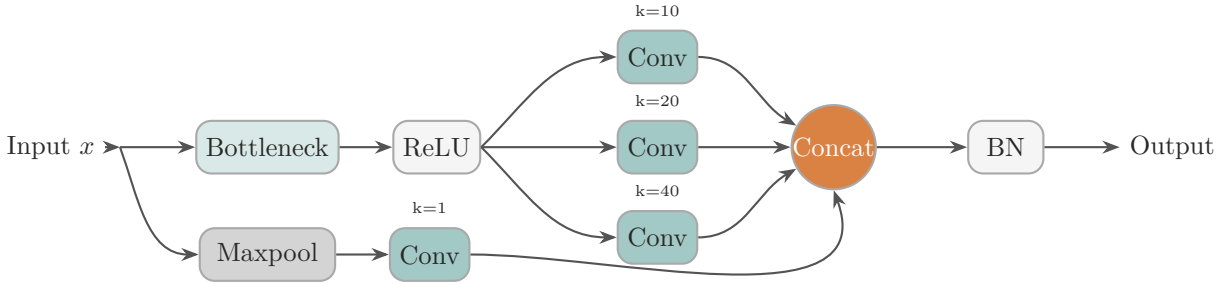
\begin{figure}[h!] 
    \centering
    \begin{tikzpicture}[
        node distance=0.5cm and 0.8cm, 
        >=Stealth, 
        op/.style={
            rectangle, 
            draw=NeutralBorder, 
            thick, 
            minimum height=7mm, 
            text centered, 
            font=\small, 
            fill=NeutralFill,         
            rounded corners=2mm,
            text=DarkText           
        },
        conv/.style={op, fill=LightTeal, node contents={Conv}},
        bottleneck/.style={op, fill=VeryLightTeal, minimum width=18mm, node contents={Bottleneck}}, 
        relu/.style={op, fill=NeutralFill, node contents={ReLU}},
        pool/.style={op, fill=NeutralBorder!50, minimum width=18mm, node contents={Maxpool}}, 
        bn/.style={op, fill=NeutralFill, minimum width=10mm, node contents={BN}},
        concat/.style={ 
            circle, 
            draw=NeutralBorder, 
            thick, 
            fill=ModernOrange,      
            minimum size=8mm, 
            inner sep=0pt, 
            font=\small,
            text=white,             
            node contents={Concat}
        },
        io_label/.style={font=\small, text=DarkText}, 
        arrow/.style={->, thick, rounded corners=2mm, color=DarkText!80}
    ]
    \node (input_label) [io_label] {Input $x$};
    \coordinate (input_start_coord) at ($(input_label.east)+(0.2cm,0)$);

    \node (bottle_node) [bottleneck, right=1cm of input_start_coord] {};
    \node (bottle_relu) [relu, right=0.7cm of bottle_node] {};
    
    \node (cb1_conv) [conv, right=1.8cm of bottle_relu, yshift=1.2cm] {}; 
    \node (cb2_conv) [conv, right=1.8cm of bottle_relu] {};
    \node (cb3_conv) [conv, right=1.8cm of bottle_relu, yshift=-1.2cm] {};
    \node[above=0.05cm of cb1_conv, font=\tiny, text=DarkText]{k=10};
    \node[above=0.05cm of cb2_conv, font=\tiny, text=DarkText]{k=20};
    \node[above=0.05cm of cb3_conv, font=\tiny, text=DarkText]{k=40};

    \node (pool_node)  [pool, below=0.7cm of bottle_node] {};
    \node (pool_conv1) [conv, right=0.7cm of pool_node] {}; 
    \node[above=0.05cm of pool_conv1, font=\tiny, text=DarkText]{k=1};

    \path let 
        \p1 = (cb1_conv.east),
        \p2 = (pool_conv1.east),
        \p3 = (bottle_relu),
        \n1 = {max(\x1,\x2)}
    in
        node (concat_node) [concat, at={(\n1+1.8cm, \y3)}] {};

    \node (final_bn_node) [bn, right=1.2cm of concat_node] {};
    \node (output_label) [io_label, right=1cm of final_bn_node] {Output};

    \draw [arrow] (input_label.east) -- (input_start_coord);
    \draw [arrow] (input_start_coord) -- (bottle_node.west);
    \draw [arrow] (input_start_coord) to[out=-70, in=180] (pool_node.west);
    \draw [arrow] (bottle_node.east) -- (bottle_relu.west);
    \draw [arrow] (bottle_relu.east) to[out=45, in=180] (cb1_conv.west);
    \draw [arrow] (bottle_relu.east) to[out=0,  in=180] (cb2_conv.west);
    \draw [arrow] (bottle_relu.east) to[out=-45, in=180] (cb3_conv.west);
    \draw [arrow] (pool_node.east) -- (pool_conv1.west);
    \draw [arrow] (cb1_conv.east) to[out=0, in=150] (concat_node.150); 
    \draw [arrow] (cb2_conv.east) to[out=0, in=180] (concat_node.west);
    \draw [arrow] (cb3_conv.east) to[out=0, in=210] (concat_node.210);
    \draw [arrow] (pool_conv1.east) to[out=0, in=-70] (concat_node.south); 
    \draw [arrow] (concat_node.east) -- (final_bn_node.west);
    \draw [arrow] (final_bn_node.east) -- (output_label.west);

    \end{tikzpicture}
    \caption{Architecture of our \texttt{InceptionModule}. The bottleneck layer first reduces the channel dimensionality of the input $x$. This output is then processed by multiple parallel convolutions to extract multi-scale features. Concurrently, the original input is processed by a max-pooling pathway. The outputs of all pathways are concatenated and normalised.}
    \label{fig:inception_time_module}
\end{figure}

The CNN feature extractor ($\mathcal{F}$), illustrated in Figures \ref{fig:inception_time_network} and \ref{fig:inception_time_module}, is based on the InceptionTime architecture \cite{inceptiontime}, adapted for our hybrid pipeline. A stem layer (convolution with $k=7$, followed by BN and ReLU) projects the single-channel input to 64 feature channels. The core consists of $N_B=3$ main blocks, each containing a stack of $M=3$ \texttt{InceptionModule} sub-blocks with a residual connection, followed by a MaxPool layer (kernel size and stride of 2) for temporal downsampling. This progressive downsampling is a departure from the original InceptionTime design, which uses Global Average Pooling for classification. Here, halving the sequence length at each block reduces the quadratic cost of the subsequent Transformer's self-attention calculation by a factor of four per block, making the hybrid architecture more computationally feasible for potentially long GW time series. While we use the standard self-attention, we note that significant research has been dedicated to mitigating this quadratic bottleneck, leading to more efficient Transformer variants with linear-time attention mechanisms \cite{tay2022efficienttransformerssurvey}. Each \texttt{InceptionModule} uses a bottleneck ($k=1$, $d_{\text{bottle}}=32$) followed by three parallel convolutions with kernels $k \in \{10,20,40\}$  alongside a MaxPool pathway ($k=3$, stride 1, preserving sequence length) for local shift invariance. We make two modifications relative to the original InceptionTime formulation \cite{inceptiontime}: (i) we apply a single Batch Normalization layer after concatenating all branch outputs, rather than after each internal convolution, reducing computational cost while providing a unified normalization of the multi-scale features; and (ii) we progressively increase the number of filters per branch across blocks, $d_{\text{branch}} \in \{ 32, 48, 64 \}$, to compensate for the temporal compression introduced by MaxPool downsampling.

A convolutional kernel of size $k$ at sampling rate $f_s$ spans a time window $\Delta t = k/f_s$ and can resolve features with frequencies approximately $f \gtrsim f_s/k$. For GW signals, whose frequency increases from inspiral to merger, the branches with $k \in \{10,20,40\}$ act as complementary band-pass filters: larger kernels have the capacity to capture the slowly varying inspiral chirp (chirp-mass dominated), while smaller kernels resolve the rapid merger–ringdown oscillations encoding mass ratio and spin. This multi-scale design matches the  chirping structure of compact binaries. The kernel choices were empirically motivated. The total downsampling $2^{N_B}=8$ yields a latent resolution of $\sim$2\,ms per token, adequate for our mass range. Future extensions could employ adaptive kernels and learnable downsampling.
The fixed sampling rate of 4096\,Hz and input length of $4096$ samples limit generality. Future improvements include (i) multi-rate training with augmentation to promote rate-invariant representations, and (ii) an input resampling layer that maps arbitrary sampling to a standardised representation with adjusted positional encodings.

\subsection{Transformer encoder and regression head}
The feature extractor produces a latent sequence $L \in \mathbb{R}^{B \times N \times d_{\mathcal{F}}}$, which is linearly projected to the transformer's model dimension $d_{\text{model}}=256$. A learnable \texttt{[CLS]} token \cite{Devlin2018} is prepended and standard sinusoidal positional encodings \cite{vaswani2017attention} are added, forming the input sequence $L'$. Our encoder follows the standard Transformer architecture \cite{vaswani2017attention} with 4 layers, 4 attention heads, feed-forward dimension $d_{ff}=512$, GELU activations, and a dropout rate of 0.10. The output vector at the \texttt{[CLS]} position serves as the sequence-level representation $E_{\text{agg}}$, which is passed to the regression MLP head $\mathcal{H}$: two hidden layers (dimensions 256 and 64) with GELU and dropout, followed by a linear output layer mapping to the four inferred parameters, $\mathbf{y} = \mathcal{H}(E_{\text{agg}})$.

\section{Training} \label{sec:training}
\subsection{Data pre-processing} 
\label{dataproc}
 For numerical stability, the original input is uniformly scaled by a factor of $10^{21}$ as, $\mathbf{x}'_i = \mathbf{x}_i \times 10^{21}$, which transforms the typical strain amplitude of a GW closer to $\mathcal{O}(1)$. We ensure that all waveforms consist of 4096 samples, through appropriate padding. 
For the target parameters, $y_i = [m_1, m_2, \chi_{\rm{eff}}, d]_i$, each component is independently normalised to [0,1] using Min-Max scaling based on the training set range. Model predictions are inverse-transformed to physical units for evaluation. We remind that distances are not sampled from an astrophysical distribution, but are instead rescaled to enforce a fixed network SNR.

The entire dataset was initially shuffled randomly to ensure an unbiased distribution of samples, and then partitioned into a training set ($D_{train}$), a validation set ($D_{val}$), and a test set ($D_{test}$) under a split ratio of $80$\%, $10$\% and $10$\% respectively. $D_{train}$  was used for learning the model parameters $\theta$, $D_{val}$ was used to monitor the model's performance on data not seen during training and to guide the selection of the best-performing model checkpoint. $D_{test}$ was held out during all training and model selection stages and was used at the end to provide an unbiased estimate of the final selected model's generalisation performance, including the test error and other relevant evaluation metrics.

\subsection{Pretraining and training}

We adopt a self-supervised pretraining strategy inspired by pretraining task proposed in \cite{zerveas2020transformerbasedframeworkmultivariatetime}. This self-supervised objective is conceptually similar to the Masked Language Model (MLM) pretraining task popularised by BERT \cite{Devlin2018}, in that the model learns contextual representations by reconstructing a corrupted input. However, our implementation is adapted for the continuous time-series domain and differs in several key aspects. First, while BERT operates on a sequence of discrete text tokens from a finite vocabulary, our model processes a continuous-valued signal. Second, our masking strategy differs: BERT typically replaces a small percentage (usually around 15\%) of individual random tokens with a special `[MASK]` token, whereas we mask contiguous spans of the time series by setting their values to zero. This span-based masking encourages the model to learn the structure of longer temporal segments. Finally, the prediction tasks are distinct: BERT performs a classification task, predicting the original discrete token ID using a Cross-Entropy loss. In contrast, our model performs a regression task, predicting a continuous latent vector ($\mathbf{z}_{\text{pred}}$) and is optimised using a MSE loss against the target latent vector ($\mathbf{z}_{\text{target}}$) derived from the uncorrupted signal. During pretraining, random contiguous segments of input signals are masked (set to zero), and the model is tasked with reconstructing a latent representation of the original, uncorrupted signal from this partial input. This process encourages the model to learn temporal dependencies and feature representations. Given an input waveform $\mathbf{x} \in \mathbb{R}^{L}$, the binary mask $\mathbf{m} \in \{0,1\}^{L_{in}}$ is generated stochastically for each sample using a process modeled as like a two-state discrete-time Markov chain (see Figure \ref{fig:mask_downsampling}). The two states are $S_0$ (being inside a masked segment, where the mask value is 0) and $S_1$ (being inside an unmasked segment, where the mask value is 1). The duration of stay in each state follows a geometric distribution. The probability of transitioning out of a state at any given time step is denoted by $p$. The mean segment length is therefore $1/p$. We define the mean length for masked segments as $l_m$ and for unmasked segments as $l_u$, which in turn define the transition probabilities $p_m = 1/l_m$ and $p_u = 1/l_u$, respectively. The process is determined by two key hyperparameters: the desired mean masked length, $l_m$, and the overall masking ratio, $r$, from which the mean unmasked length is derived as $l_u = l_m \cdot (1-r)/r$. The transition matrix $\mathbf{P}$ for this Markov chain is given by
\begin{equation}
    \mathbf{P} = 
    \begin{pmatrix}
        1 - p_m & p_m \\
        p_u & 1 - p_u.
    \end{pmatrix}
\end{equation}
In practice, the implementation uses strictly alternating masked and unmasked segments, which is a close approximation to the Markov chain for the chosen parameters. For our experiments, we set a masking ratio of $r=0.15$ and a mean masked segment length of $l_m = 32$. This choice of a relatively long mean segment length is tied to our network's architecture. It serves two purposes: first, it prevents the reconstruction task from becoming trivial, as the model must learn long-range dependencies rather than simply interpolating from immediate neighbors. Second, and more critically, it ensures that the masked regions are robust to the temporal downsampling performed by the feature extractor. With a total downsampling factor of 8x in our network, a mean masked length of 32 ensures that each masked segment in the input corresponds to multiple tokens in the latent space where the reconstruction loss is computed, providing a stable and consistent learning signal.
The masked vector is then applied to the input waveform via element-wise multiplication to produce the masked input
\begin{equation}
\mathbf{x}_m = \mathbf{x} \odot \mathbf{m}, 
\end{equation}
and we also define a binary padding mask $\mathbf{p} \in \{0,1\}^{L}$, indicating valid input positions to distinguish from $0$'s in the sequence which were due to padding. 
The pretraining objective involves two parallel data processing pathways. The main pathway generates a prediction from the corrupted input, while the second generates the ground truth target from the original input. The prediction path proceeds as follows: the masked waveform is passed through a feature extractor $\mathcal{F}(\cdot)$, yielding downsampled features ($\mathbf{l}_m \in \mathbb{R}^{B,N,d_{\mathcal{F}}}$). These are subsequently transposed and linearly projected to the transformer's dimension $d_{model}$ by $f_{\text{proj}}(\cdot)$, on top of which positional encodings are added through $f_{\text{pos}}(\cdot)$. Schematically, 
\begin{equation}
\mathbf{l}_m = \mathcal{F}(\mathbf{x}_m) \in \mathbb{R}^{C \times L'} \; \; \to \; \; \mathbf{z}_m = f_{\text{proj}}(\mathbf{l}_m^\top) \in \mathbb{R}^{L' \times D} \; \; \to \; \; \mathbf{z}_{\text{pe}} = f_{\text{pos}}(\mathbf{z}_m).
\end{equation}
To prevent the self-attention mechanism from attending to padded regions, the original padding mask $\mathbf{p}$ is downsampled to the new sequence length $L'$. This downsampling must precisely mirror the reduction performed by the main feature extractor $\mathcal{F}$. We achieve this by applying a sequence of MaxPool layers (with a kernel size and stride of 2, identical to the downsamplers in $\mathcal{F}$) to the padding mask. Using Max Pooling on the binary mask correctly propagates the padding information, ensuring that a position in the resulting downsampled mask, $\mathbf{p}' \in \{0,1\}^{L'}$, is marked as padding (value = 0) only if all input positions in its receptive field were also padding. The resulting mask is then inverted to define the attention mask $\mathbf{m}_{\text{attn}} = \neg \mathbf{p}'$, which is supplied to the transformer encoder. The transformer encoder $ \mathcal{T}(\cdot)$ processes the input and the output is passed through a linear prediction head $f_{\text{head}}(\cdot)$ as
\begin{equation}
\mathbf{l}_e = \mathcal{T}(\mathbf{z}_{\text{pe}}, \mathbf{m}_{\text{attn}}) \; \;  \to  \; \; \mathbf{z}_{\text{pred}} = f_{\text{head}}(\mathbf{l}_e).
\end{equation}
\definecolor{ModernTeal}{RGB}{0,128,128}
\definecolor{MidGray}{RGB}{128,128,128}
\definecolor{DarkText}{RGB}{40,40,40}

\begin{figure}[h]
    \centering
    \caption{Illustration of the binary padding mask downsampling process. A MaxPool operation with a kernel size and stride of 2 is applied to the input mask. This ensures that a position in the downsampled mask is marked as padding (0) only if all corresponding input positions were also padding.}
    \label{fig:mask_downsampling}
    \begin{tikzpicture}[transform shape, scale=0.7,
        node distance=1.5cm,
        >=Stealth,
        cell/.style={rectangle, draw=black!60, thick, minimum size=1cm, font=\large},
        one/.style={cell, fill=ModernTeal!20},
        zero/.style={cell, fill=MidGray!30},
        arrow/.style={->, thick, color=DarkText!80},
        brace/.style={decoration={brace,amplitude=6pt}, decorate, thick, black!80}
    ]
        \matrix (input_mask) [matrix of nodes, nodes={anchor=center}, column sep=0mm]
        {
            |[one]| 1 & |[one]| 1 & |[one]| 1 & |[zero]| 0 & |[zero]| 0 & |[zero]| 0 & |[one]| 1 & |[one]| 1 \\
        };
        \node[above=0.4cm of input_mask, font=\small] {Input Padding Mask ($\mathbf{p} \in \{0,1\}^{L_{in}}$)};
        \matrix (output_mask) [matrix of nodes, below=2.5cm of input_mask, nodes={anchor=center}, column sep=0mm]
        {
            |[one]| 1 & |[one]| 1 & |[zero]| 0 & |[one]| 1 \\
        };
        \node[below=0.4cm of output_mask, font=\small] {Downsampled Mask ($\mathbf{p}' \in \{0,1\}^{L_{out}}$)};
        \coordinate (in1) at (input_mask-1-1.south);
        \coordinate (in2) at (input_mask-1-2.south);
        \coordinate (out1) at (output_mask-1-1.north);
        \draw[brace] ($(in1)+(-0.5cm,0)$) -- ($(in2)+(0.5cm,0)$) node[midway, below=8pt, font=\tiny] {max(1,1)=1};
        \draw[->, gray, thin] ($(in1)!0.5!(in2)+(0,-0.2cm)$) -- (out1);
        \coordinate (in3) at (input_mask-1-3.south);
        \coordinate (in4) at (input_mask-1-4.south);
        \coordinate (out2) at (output_mask-1-2.north);
        \draw[brace] ($(in3)+(-0.5cm,0)$) -- ($(in4)+(0.5cm,0)$) node[midway, below=8pt, font=\tiny] {max(1,0)=1};
        \draw[->, gray, thin] ($(in3)!0.5!(in4)+(0,-0.2cm)$) -- (out2);
        
        \coordinate (in5) at (input_mask-1-5.south);
        \coordinate (in6) at (input_mask-1-6.south);
        \coordinate (out3) at (output_mask-1-3.north);
        \draw[brace] ($(in5)+(-0.5cm,0)$) -- ($(in6)+(0.5cm,0)$) node[midway, below=8pt, font=\tiny] {max(0,0)=0};
        \draw[->, gray, thin] ($(in5)!0.5!(in6)+(0,-0.2cm)$) -- (out3);
        \coordinate (in7) at (input_mask-1-7.south);
        \coordinate (in8) at (input_mask-1-8.south);
        \coordinate (out4) at (output_mask-1-4.north);
        \draw[brace] ($(in7)+(-0.5cm,0)$) -- ($(in8)+(0.5cm,0)$) node[midway, below=8pt, font=\tiny] {max(1,1)=1};
        \draw[->, gray, thin] ($(in7)!0.5!(in8)+(0,-0.2cm)$) -- (out4);
        
    \end{tikzpicture}
\end{figure}
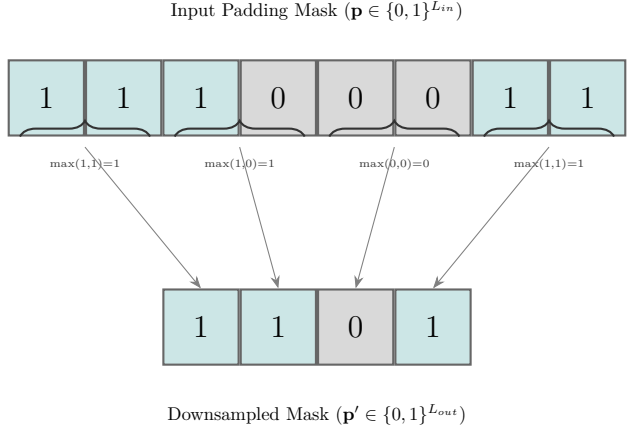
The target latent representation is computed from the original (unmasked) input using the same feature extractor and projection,
\begin{equation}
\mathbf{z}_{\text{target}} = f_{\text{proj}}(\mathcal{F}(\mathbf{x})^\top),
\end{equation}
and finally, the loss is computed only on positions that were both masked and valid (i.e., not padded). We note that the binary mask $\mathbf{m}$ is downsampled the same way as our padding mask. We define a loss mask
\begin{equation}
\mathbf{m}_{\text{loss}} = \text{downsample}(\neg \mathbf{m} \land \mathbf{p}) \in \{0,1\}^{L'}.
\end{equation}
This ensures that only valid and originally masked tokens contribute to the reconstruction loss.
Denoting with $I = \{ (b, i) \mid \mathbf{m}_{\text{loss},(b,i)} = 1 \}$ the set of batch and sequence indices where the loss should be computed, pretraining aims to optimise the mean squared error (MSE) between the predicted and target latent representations,
\begin{equation}
\mathcal{L}_{pretrain} = \frac{1}{|I|} \sum_{(b,i) \in I} \norm{ \mathbf{z}_{\text{target},(b,i,\cdot)} - \mathbf{z}_{\text{pred},(b,i,\cdot)}}_2^2.
\end{equation}

The evolution of the pretraining loss is shown in Figure~\ref{fig:pretrain_loss} in the Appendix. We select the pretrained weights at the epoch of minimum validation loss (denoted with a star in Figure~\ref{fig:pretrain_loss}). The downstream improvements reported in 
Table~\ref{tab:Performance_comparison} confirm that the representations learned during the convergence phase are effective.

Following pretraining, the model is trained/fine-tuned to predict the target physical parameters $\mathbf{y}_i$ (e.g masses, spins)
using supervised training. We minimise the MSE over batches $\mathcal{B} \subset D_{\text{train}}$,
\begin{equation} \label{eq:mse_loss_fine_tune_latex}
\mathcal{L}_{\mathrm{MSE}}(\theta; \mathcal{B}) = \frac{1}{B} \sum_{(\mathbf{x}_i, \mathbf{y}_i) \in \mathcal{B}} \| \mathbf{y}_i - f(\mathbf{x}_i; \theta) \|_2^2,
\end{equation}
where $f(\mathbf{x}_i; \theta)$ is the predicted target vector and $\theta$ the deep model's inherent parameters. Fine-tuning uses the AdamW optimiser \cite{loshchilov2019decoupledweightdecayregularization} with learning rate $5\cdot 10^{-4}$, cosine scheduler with linear 10\% warmup, batch size $B = 250$, and a maximum of $E = 120$ epochs. Gradient clipping with threshold 1.0 is applied to prevent instability. The hyperparameters of our model are  summarised in Table \ref{tab:hyperparameters}. 
We evaluate the trained model, which we denote as $f(\mathbf{x}; \theta^{**})$, on the test dataset ($D_{\text{test}}$) using two regression metrics for each predicted physical parameter. Let us denote with $y_{ij}$ and $\hat{y}_{ij}$ the true and predicted values of the $j$-th parameter for the $i$-th test sample after inverse scaling, and with $\bar{y}_j$ the mean of the true $j$-th parameter over $D_{\text{test}}$ respectively. We then define the metrics
\begin{equation}
\mathcal{L}_{\mathrm{MAE}_j} = \frac{1}{n} \sum_{i=1}^n | y_{ij} - \hat{y}_{ij} |, \; \; 
\mathcal{L}_{R^2_j} = 1 - \frac{\sum_{i=1}^n (y_{ij} - \hat{y}_{ij})^2}{\sum_{i=1}^n (y_{ij} - \bar{y}_j)^2}.
\end{equation}
The Mean Absolute Error (MAE) captures average absolute errors, while $R^2_j$ measures the proportion of variance explained by the model. Perfect predictions yield $R^2_j = 1$, while $R^2_j < 0$ indicates worse performance than predicting the mean. We note that R² is well-defined regardless of the distribution of target values. On our uniform grid, it provides a conservative assessment since the model is evaluated with equal weight across the full parameter range rather than being dominated by the most probable astrophysical configurations.


\subsection{Model uncertainty}

For uncertainty estimation we employ Monte Carlo (MC) Dropout \cite{gal2016dropoutbayesianapproximationrepresenting} applied on the positional encodings, transformer encoder and the regression head. We use it as a computationally efficient method to perform probabilistic inference and estimate the predictive uncertainty of our deep learning model. In this context, it is useful to distinguish between two primary types of uncertainty \cite{kendall2017uncertaintiesneedbayesiandeep}: epistemic uncertainty, which captures our ignorance about the model parameters and can be reduced with more data, and aleatoric uncertainty, which captures the inherent noise or randomness in the data itself.
The key idea behind MC Dropout is to retain dropout during inference, not just training. Given an input $\mathbf{x}_i$, we perform $T$ stochastic forward passes through the trained network $f(\cdot; \theta)$, applying different dropout masks in each pass. This yields a set of $T$ output samples,
\begin{equation}
\left\{ \hat{\mathbf{y}}_i^{(t)} = f(\mathbf{x}_i; \theta^{(t)}) \right\}_{t=1}^{T}.
\end{equation}
This represents a set of $T$ stochastic predictions for input $\mathbf{x}_i$,  where $\theta^{(t)}$ denotes the parameters randomly thinned by the $t$-th dropout mask. These samples approximate the posterior predictive distribution over $\mathbf{y}_i$, allowing us to estimate both the expected prediction and the associated uncertainty.
The point estimate for the target vector $\hat{\mathbf{y}}_{i,MC}$ is taken as the sample mean of the ensemble, and the uncertainty for each individual output parameter component $y_{ij}$ (the $j$-th component of $\mathbf{y}_i$) is estimated using the sample variance of the $T$ predictions for that specific component as
\begin{equation}
    \hat{\mathbf{y}}_{i, \text{MC}} = \frac{1}{T} \sum_{t=1}^T \hat{\mathbf{y}}_{i,t}, \; \; \;  \widehat{\text{Var}}(\hat{y}_{ij}) = \frac{1}{T-1} \sum_{t=1}^T (\hat{y}_{ij,t} - \hat{y}_{ij, \text{MC}})^2.
\end{equation}
The square root of this variance, $\hat{\sigma}(\hat{y}_{ij}) = \sqrt{\widehat{\text{Var}}(\hat{y}_{ij})}$ is the standard deviation for each parameter component, which serves as our primary measure of model uncertainty. The sample variance calculated via MC Dropout serves as an approximation of the total predictive uncertainty. It implicitly combines both the epistemic uncertainty, arising from the stochasticity of the dropout masks, which approximates uncertainty over the model's weights, and the aleatoric uncertainty that is propagated through the deterministic parts of the network from the input data \cite{kendall2017uncertaintiesneedbayesiandeep}. For a computation of the Bayesian posterior which captures all kinds of uncertainty, an approach such as simulation based inference would be appropriate.

\section{Results} \label{sec:results}
Here, we present our results for the parameter inference of noisy waveforms and overall performance of our model. To understand the computational requirements and training dynamics, we trained our model on a small ($\sim 20,000$), intermediate ($\sim 50,000$), and large ($\sim 80,000$) dataset. The train/validation/testing sets were created using our strategy described in the Section \ref{dataproc}, whereas the small and intermediate ones taken as subsets of the large training set. For all experiments the validation and test sets remained fixed to ensure that the performance improvements observed are attributable to the increase in training data, not variations in the evaluation sets. We note that all training and evaluation were performed on a single \texttt{NVIDIA A100 GPU}.

\subsection{Effect of pretraining} 
The results associated to the effect of pretraining at fixed detector noise (ET) are summarised in Tables ~\ref{tab:convergence_speedup1} - \ref{tab:pretraining_gain_detailed}.
Tables \ref{tab:convergence_speedup1} and \ref{tab:Performance_comparison} summarise the effect of pretraining on the convergence speed and performance metric respectively, comparing training from scratch and training/fine-tuning on top of a pretrained model. The relative performance gains are shown in Table \ref{tab:pretraining_gain_detailed}. For each combination of dataset size and signal-to-noise ratio (SNR), the “Target Loss” corresponds to the best validation loss achieved by a model trained from scratch (no pretraining). The results clearly show that pretraining substantially accelerates convergence across all SNRs and dataset sizes, with speedup factors ranging from approximately $1.4\times$ to $6.6\times$. The most pronounced benefits of pretraining are clearly observed for small and intermediate datasets, where convergence can be more than $6\times$ faster, such as for SNR = 10 in the small and intermediate dataset scenarios.  Remarkably, the speedup advantage is generally greater at lower SNRs, indicating that pretraining is especially beneficial when the signal is weaker and the learning task is more challenging.
For large datasets, the speedup effect is less pronounced, particularly at SNR = 10, though the reduction in training time remains sizable. At higher SNR with a large dataset, the speedup is still notable, suggesting that while having a large amount of data partially compensates for the absence of pretraining, the use of pretrained models continues to offer substantial gains.
Overall, pretraining in \texttt{GraviBERT} consistently reduces training times and enables more rapid model development, which is particularly valuable in computationally intensive experiments. The convergence behaviour is illustrated in Figure~\ref{fig:convergence_grid}. The pretrained model reaches the validation target in shorter time while continuing to improve, finally converging to a lower final loss in all configurations.

Pretraining enhances also the performance metrics of the model. The absolute performance metrics for the pretrained and fine-tuned models are summarised in Table~\ref{tab:Performance_comparison}, while the fractional gains compared to training from scratch in Table~\ref{tab:pretraining_gain_detailed}. The metrics are computed as the average over 100 stochastic forward passes per sample using MC Dropouts based on the test set. Across all evaluated cases, the pretrained/fine-tuned model consistently outperforms the model trained from scratch, achieving lower MAE for all parameters and, in some cases, higher $R^2$ values although the latter is typically mildly affected. At fixed dataset size, pretraining significantly improves MAE with increasing SNR, with the biggest effect in the large dataset where the percentage MAE improvement is about $3\times$ larger. For the most challenging case, specifically the case of SNR = 10 and the small dataset, the benefits of pretraining are pronounced, with a MAE reduction ranging between $6\% - 14\%$ across parameters. The best reduction is found for the case of the large dataset and SNR = 30, ranging between $11\% - 31\%$. This can be attributed to the sufficiently large amount of data in the large dataset combined with the high SNR which allow the model to adopt and train better. One notices that the intermediate dataset shows suboptimal gains compared to both smaller and larger datasets. This appears to be due to the dataset being large enough for the model to start overfitting to the training data, but not large enough to fully benefit from the pretrained representations or to learn robust patterns from scratch. Another possibility is that the pretrained features may compete with task-specific learning in a suboptimal way leading to this behavior. 
Furthermore, this could also be (partially) attributed to the use of a single set of hyperparameters across all experiments. The optimal hyperparameters for fine-tuning a pretrained model, which often benefit from a smaller learning rate, may differ from those for training a model from scratch. A more exhaustive study would involve performing a separate hyperparameter search for each condition. However, our results using a single, tuned configuration consistently demonstrate a network positive benefit from pretraining across all dataset sizes.
We notice that for all the above cases $R^2 \geq 0.90$ as shown in Table~\ref{tab:Performance_comparison}. 

Understanding how model performance scales with dataset size informs future data requirements and expected performance gains ~\cite{kaplan2020scalinglawsneurallanguage}. The test loss (measured on models trained from scratch) follows a power-law trend as a function of training dataset size $D$,
\begin{equation} \label{eq:D_powerlaw}
L(D) \propto D^{-\alpha_D}, \quad \text{with} \quad 
\alpha_D = 
\begin{cases}
0.3701, & \text{for SNR = 10} \\
0.3624, & \text{for SNR = 30},
\end{cases}
\end{equation}
suggesting \texttt{GraviBERT}'s performance continues to improve with more data, while the scaling exponent varies extremely mildly across both the SNRs studied.
\subsection{Performance on parameter inference} \label{sec:parameter-inference}
\begin{figure*}[h]
    \centering
    \caption{True vs. Predicted values for each target parameter on the test set, for the model trained from scratch on the large dataset and for waveforms at two values of SNR. The gray error bar represents the +/- 1$\sigma$ error computed through 100 MC Dropout realizations. For mass parameters, the prediction's central value represents the mean of the aggregated predictions for all masses with the same target values, while the black error bars the 1$\sigma$ spread of the mass predictions at fixed target value associated with different parameters of spin and distance. (See the discussion in Section \ref{sec:parameter-inference} for more details.) Masses are measured in $M_{\odot}$ and effective distances in Mpc.}
    \label{fig:pred_vs_true_large}

    \begin{tikzpicture}[
        column sep=0.8cm, 
        row sep=0.5cm
    ]
        \matrix (plotgrid) [matrix of nodes, nodes={anchor=center}]
        {
            \includegraphics[width=5.5cm]{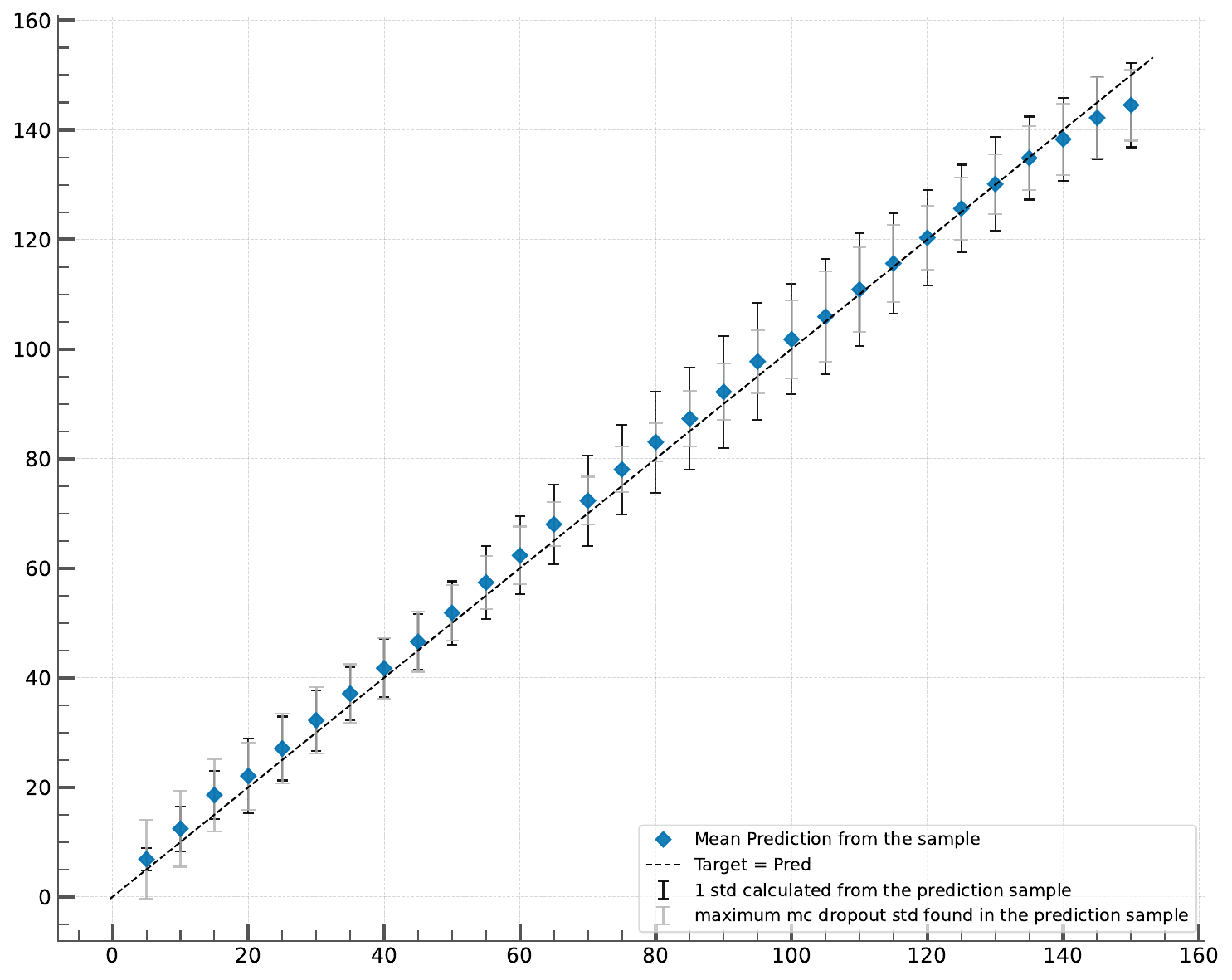} & 
            \includegraphics[width=5.5cm]{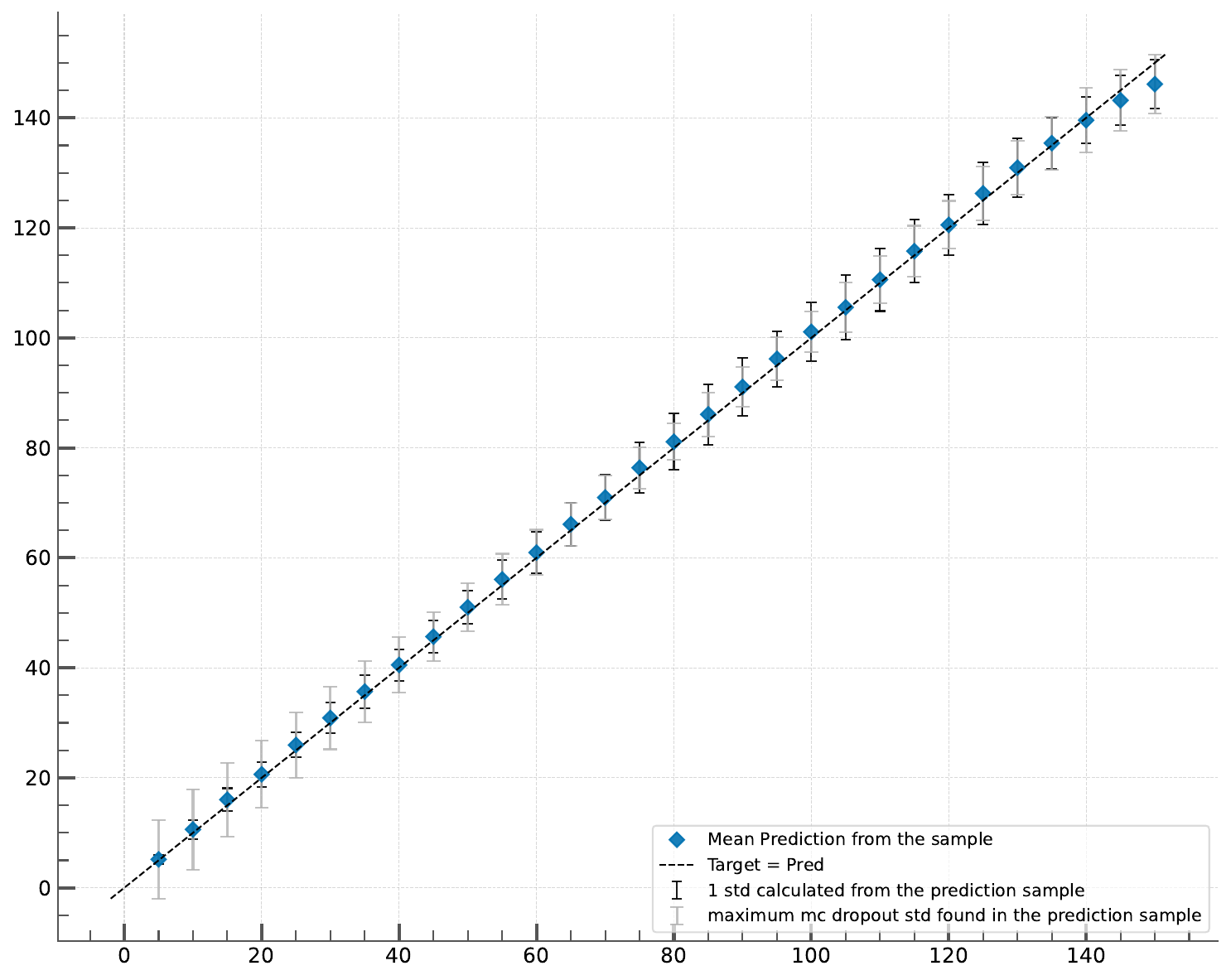} \\
            \includegraphics[width=5.5cm]{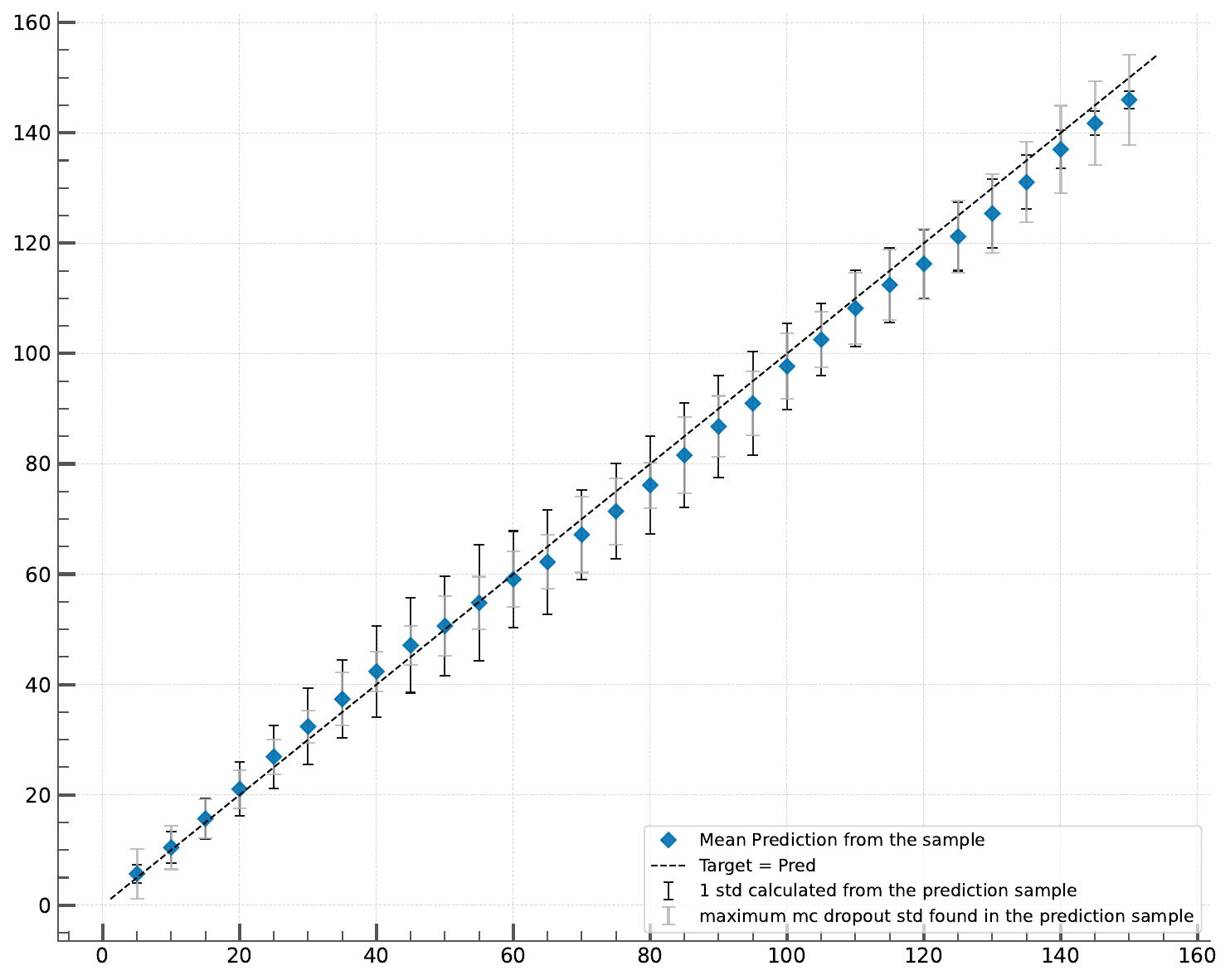} & 
            \includegraphics[width=5.5cm]{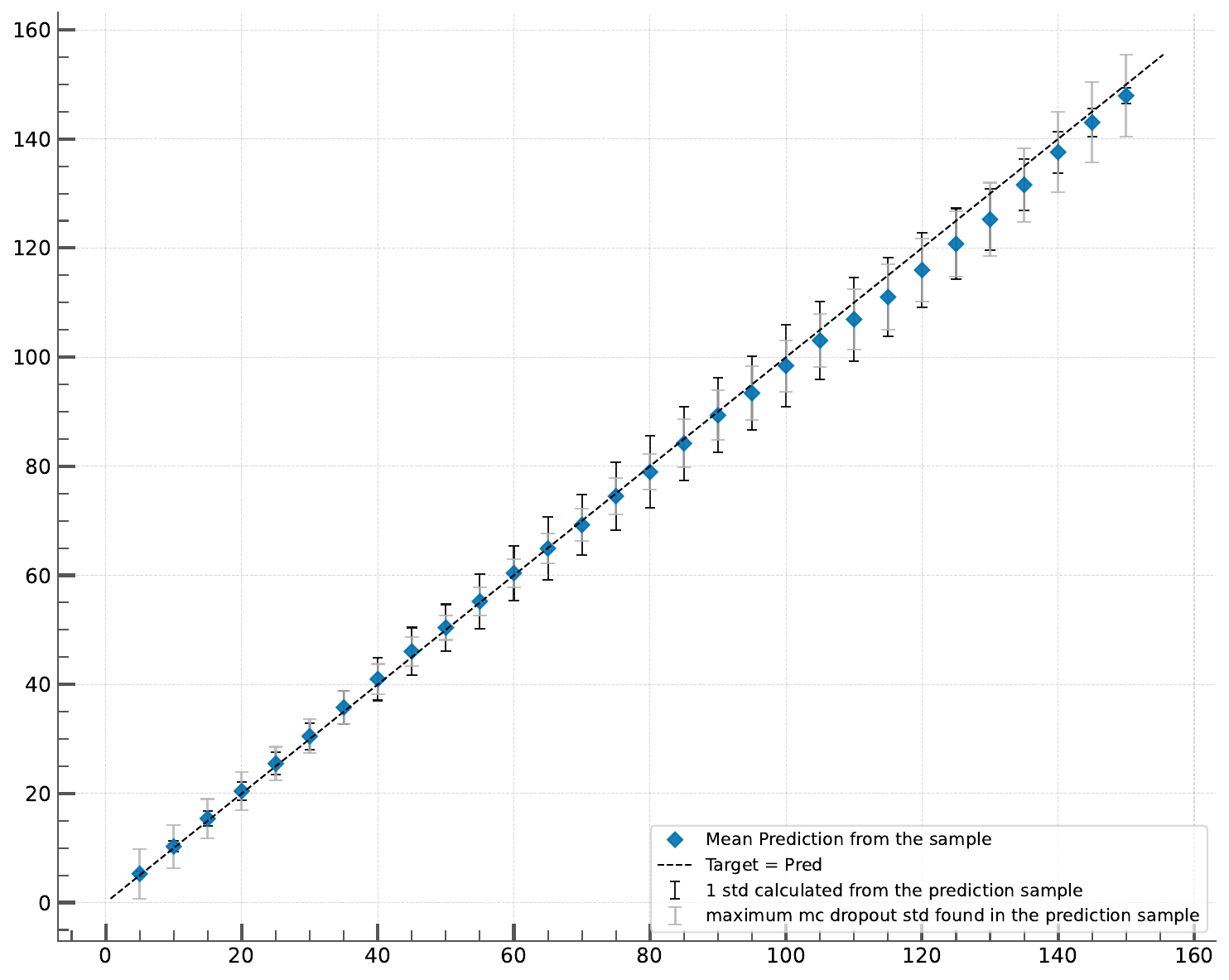} \\
            \includegraphics[width=5.5cm]{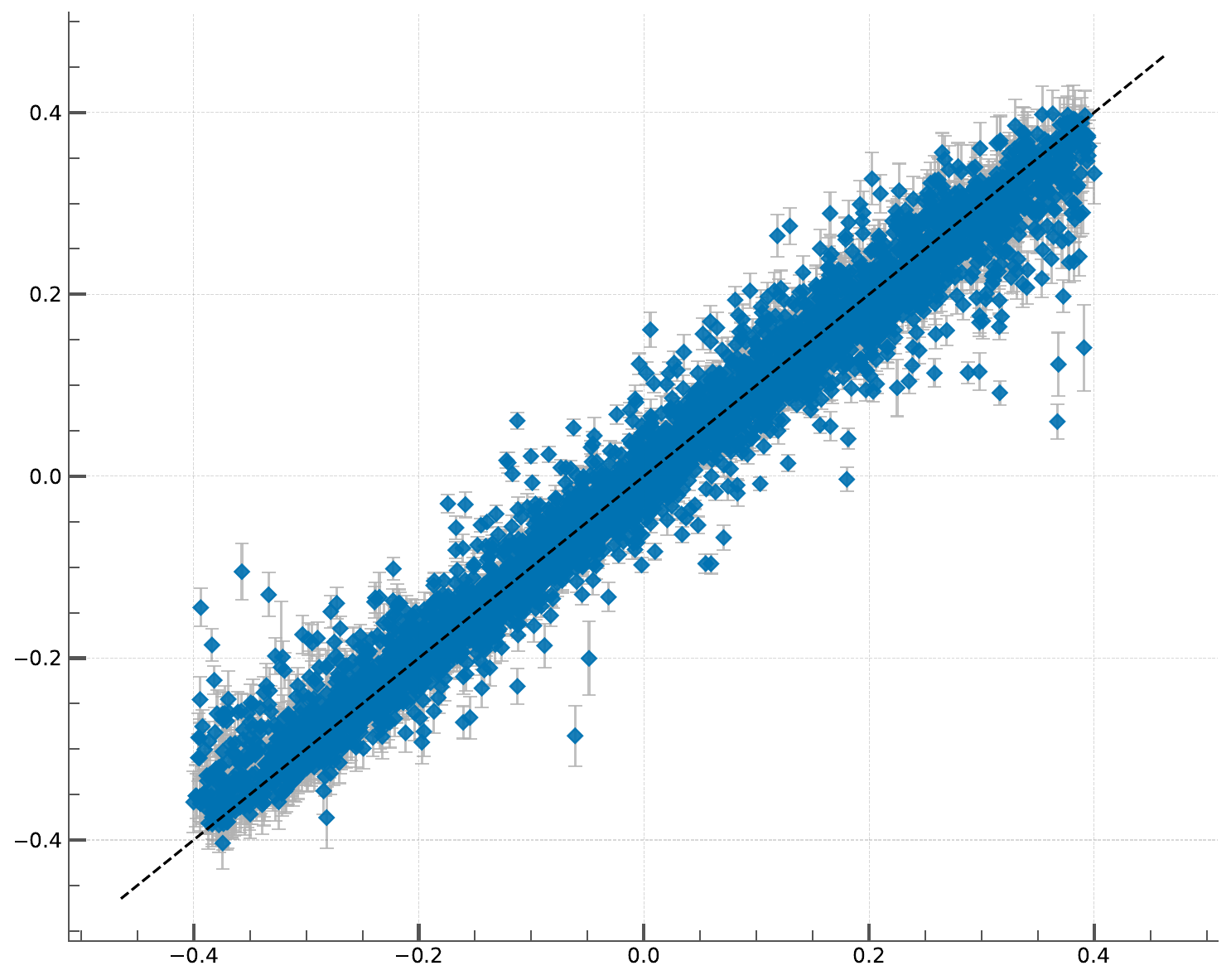} & 
            \includegraphics[width=5.5cm]{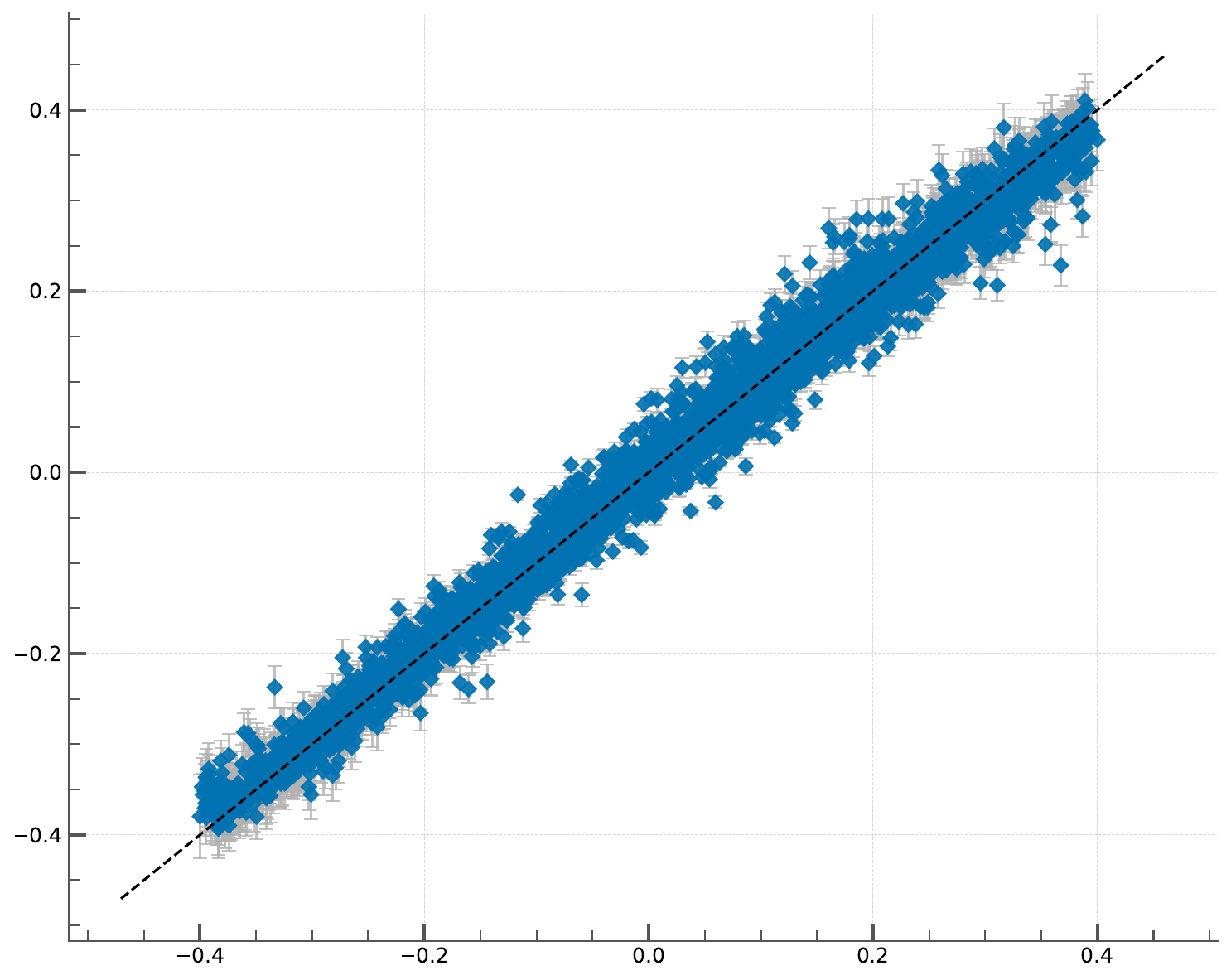} \\
            \includegraphics[width=5.5cm]{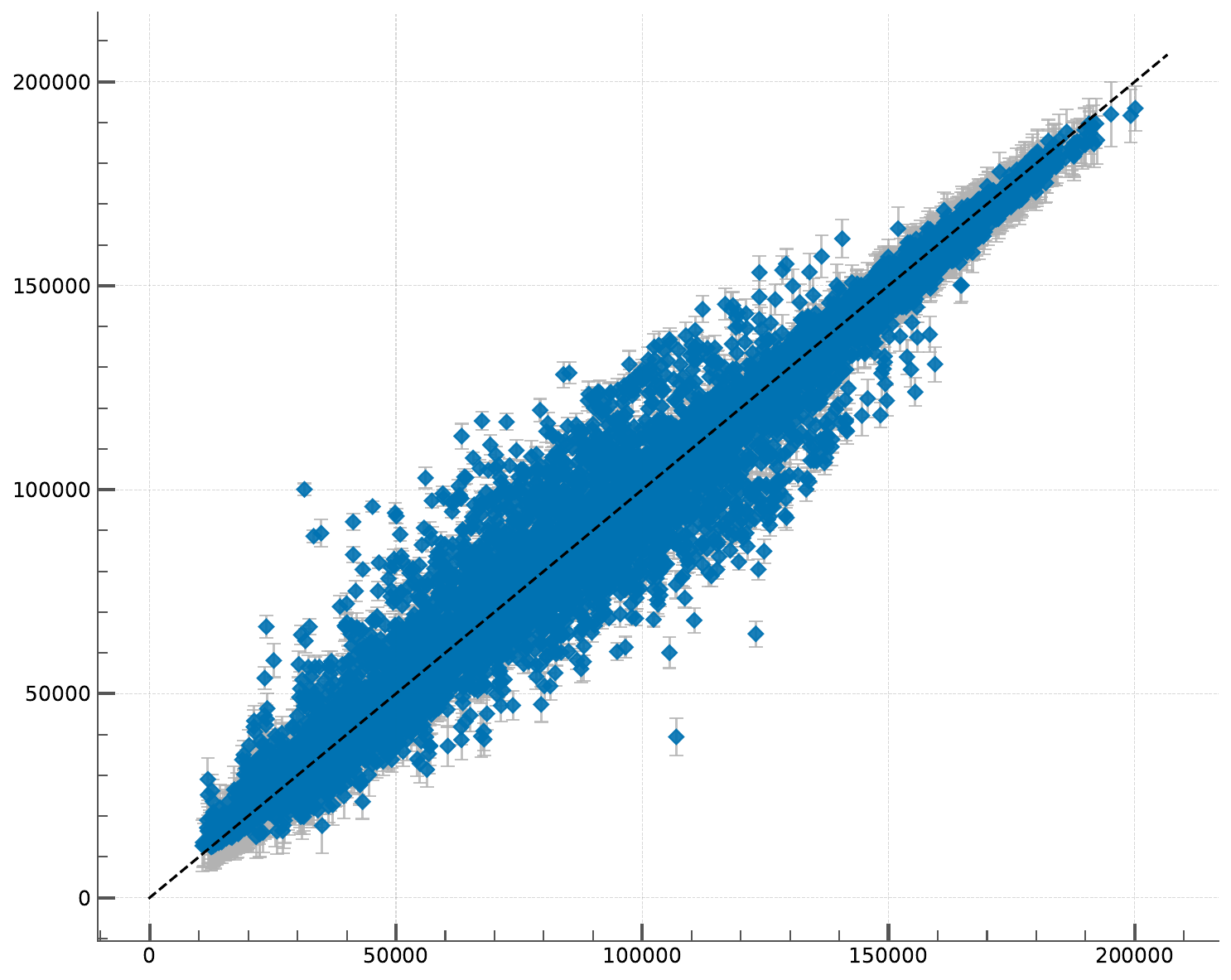} & 
            \includegraphics[width=5.5cm]{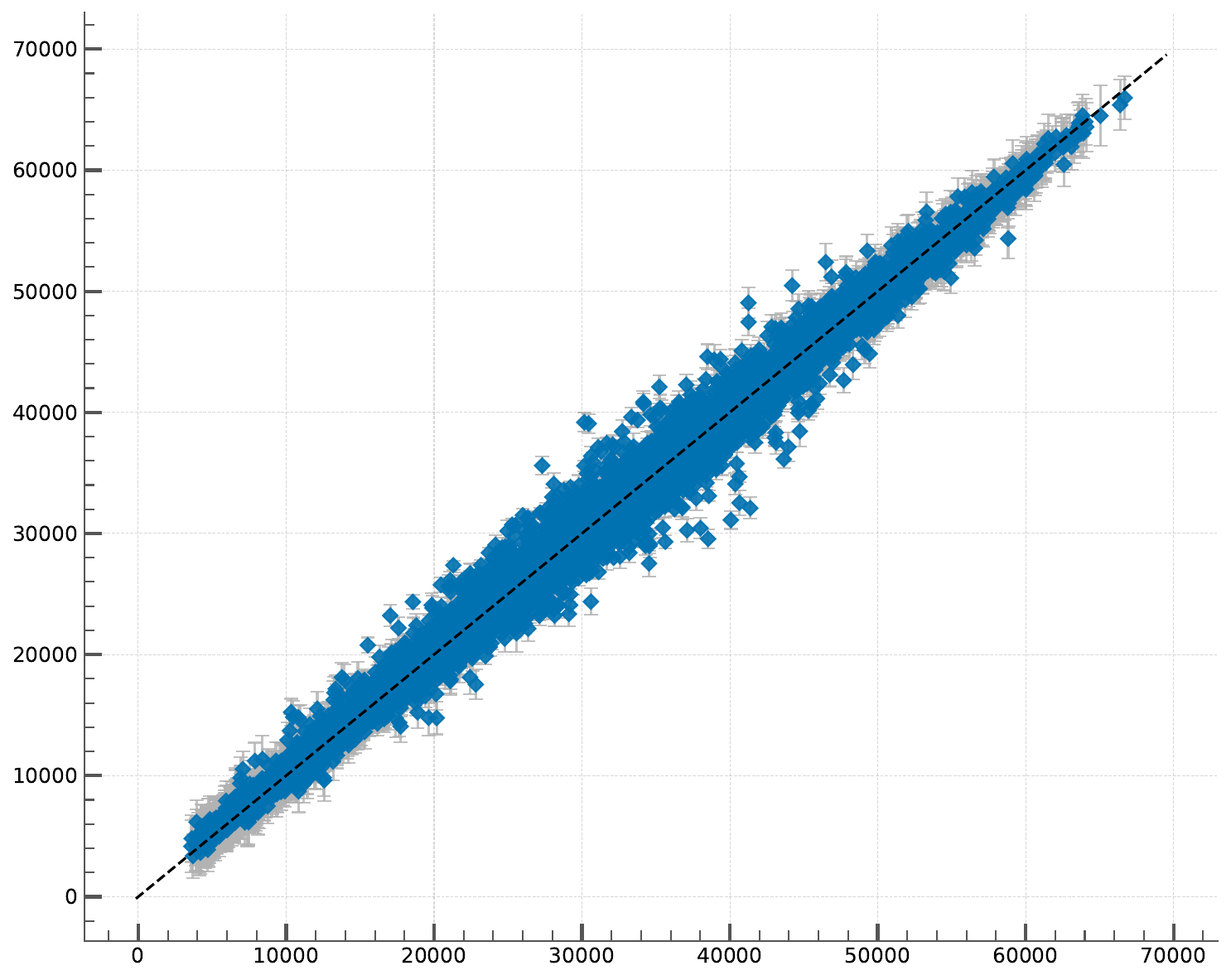} \\
        };

        \foreach \col/\snrval in {1/10, 2/30} {
            \node[font=\small, below=0.1cm of plotgrid-4-\col] {SNR \snrval};
        }
        \foreach \row/\param in {1/{m_1}, 2/{m_2}, 3/{\chi_{\rm eff}}, 4/{d}} {
            \node[font=\bfseries, left=0.6cm of plotgrid-\row-1, rotate=90, anchor=center] {Parameter $\param$};
        }

    \end{tikzpicture}
\end{figure*}

\begin{figure*}[h]
    \centering
    \caption{Same as Figure \ref{fig:pred_vs_true_large}, but for the model which was pretrained and subsequently fine-tuned on the large dataset.}
    \label{fig:pred_vs_true_large_finetuned}

    \begin{tikzpicture}[
        column sep=0.8cm, 
        row sep=1.2cm
    ]

        \matrix (plotgrid) [matrix of nodes, nodes={anchor=center}]
        {
            \includegraphics[width=5.5cm]{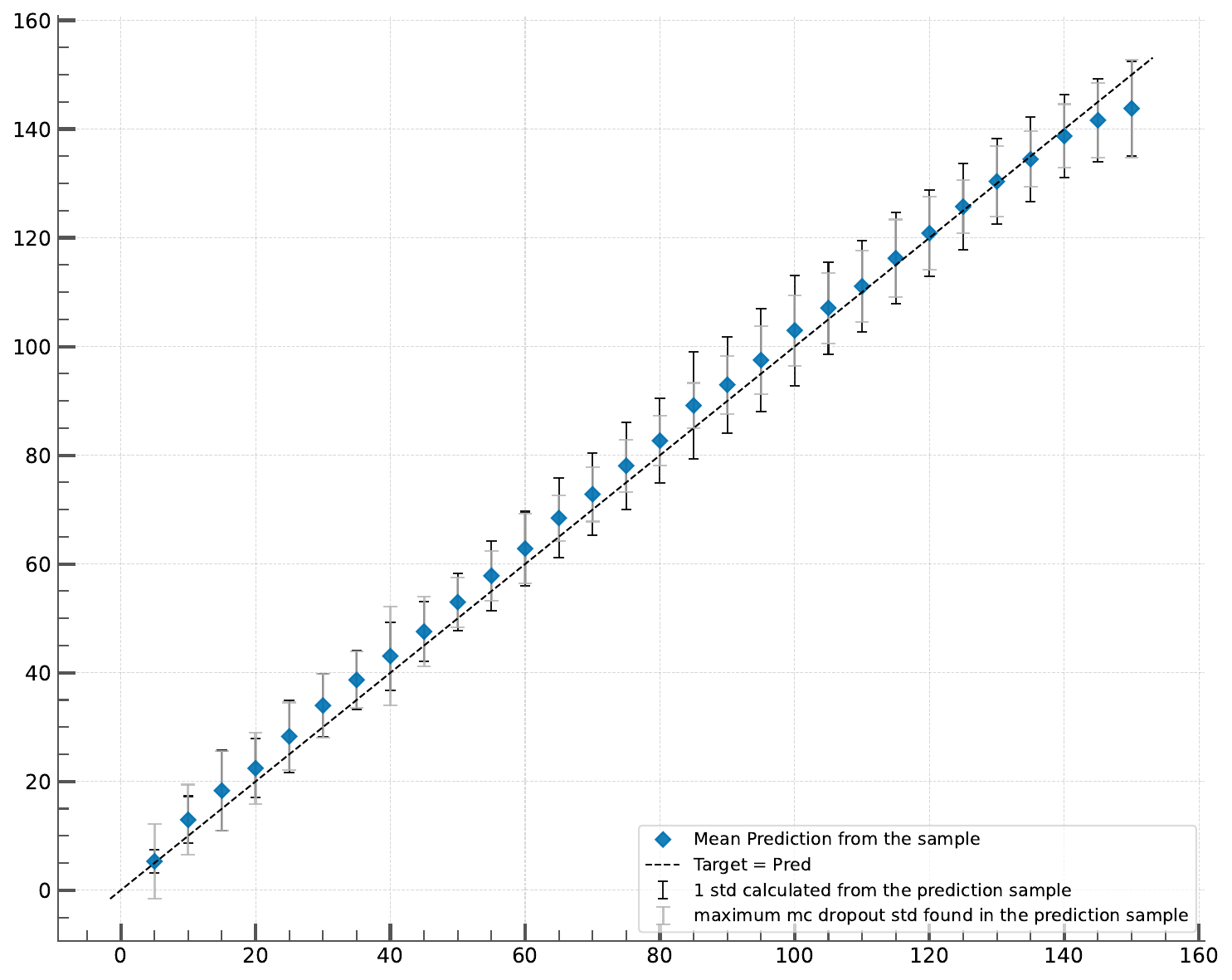} & 
            \includegraphics[width=5.5cm]{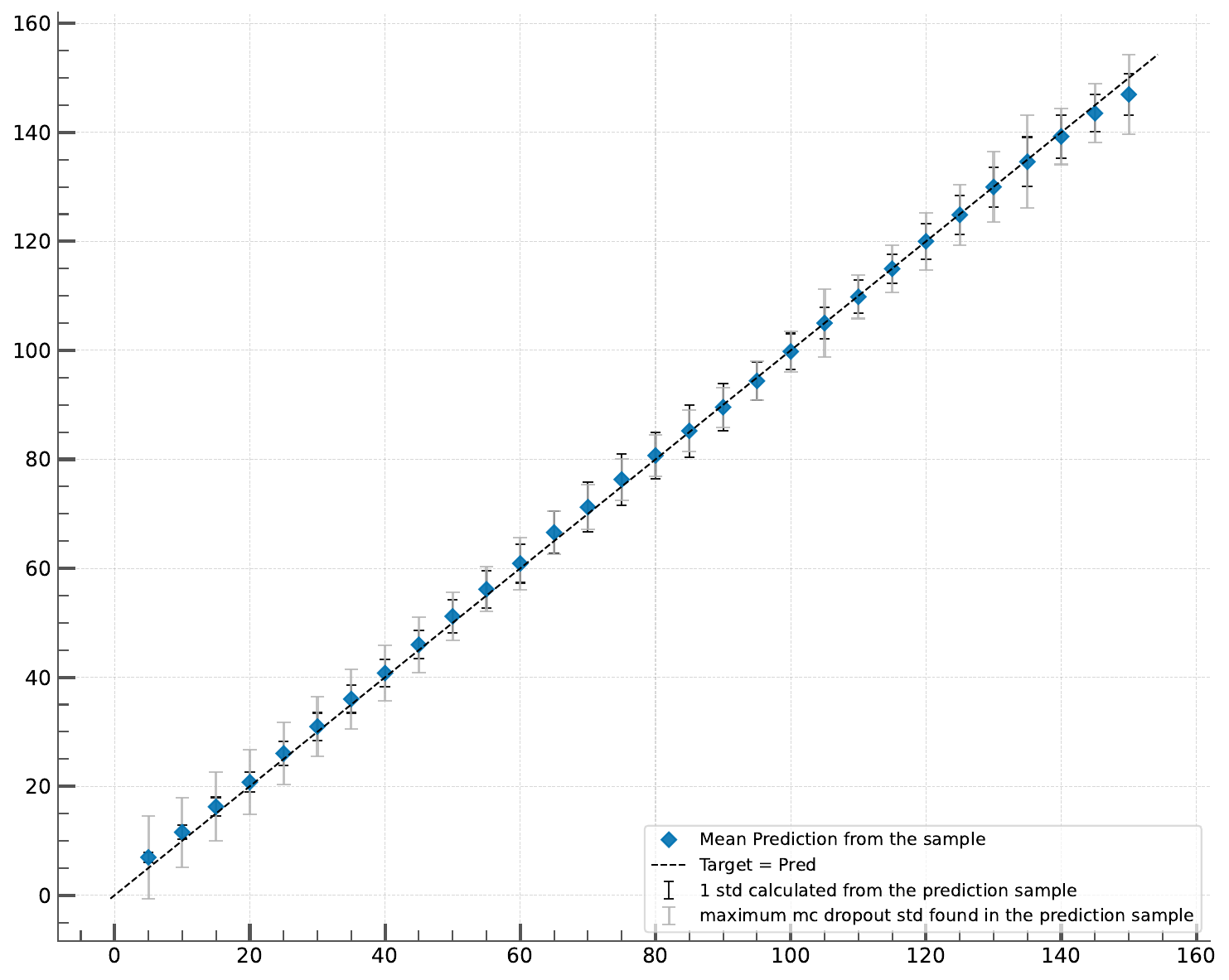} \\
            \includegraphics[width=5.5cm]{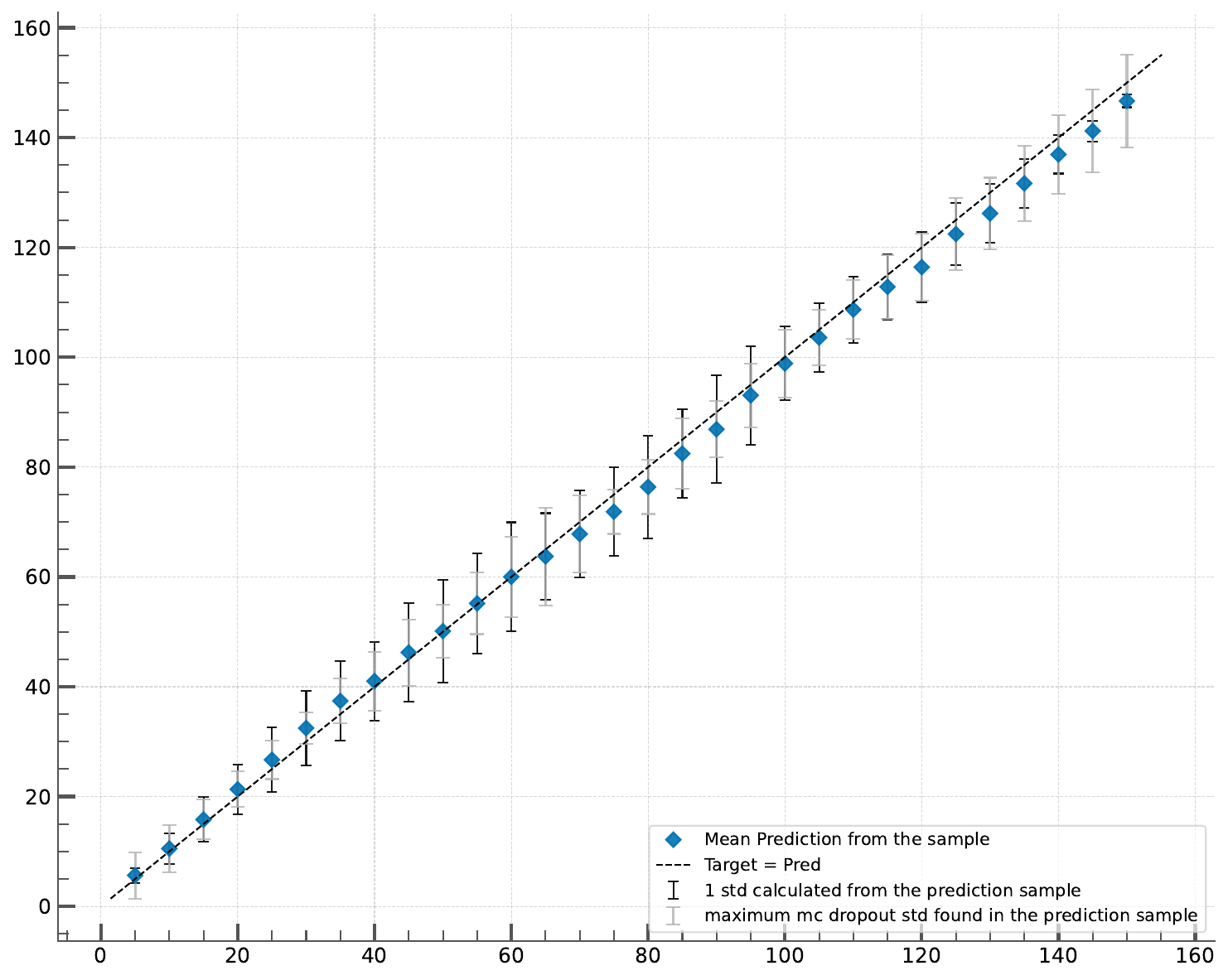} & 
            \includegraphics[width=5.5cm]{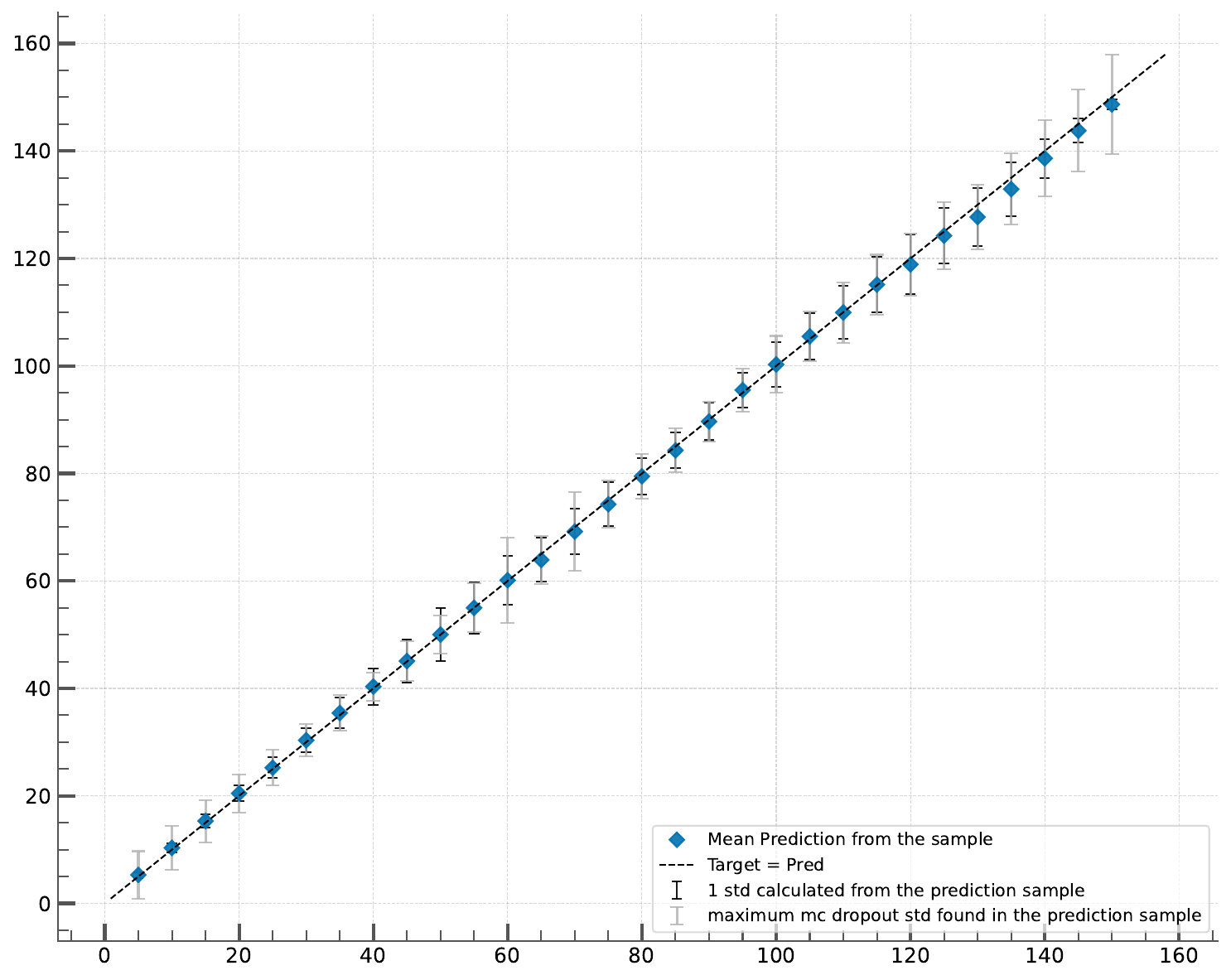} \\
            \includegraphics[width=5.5cm]{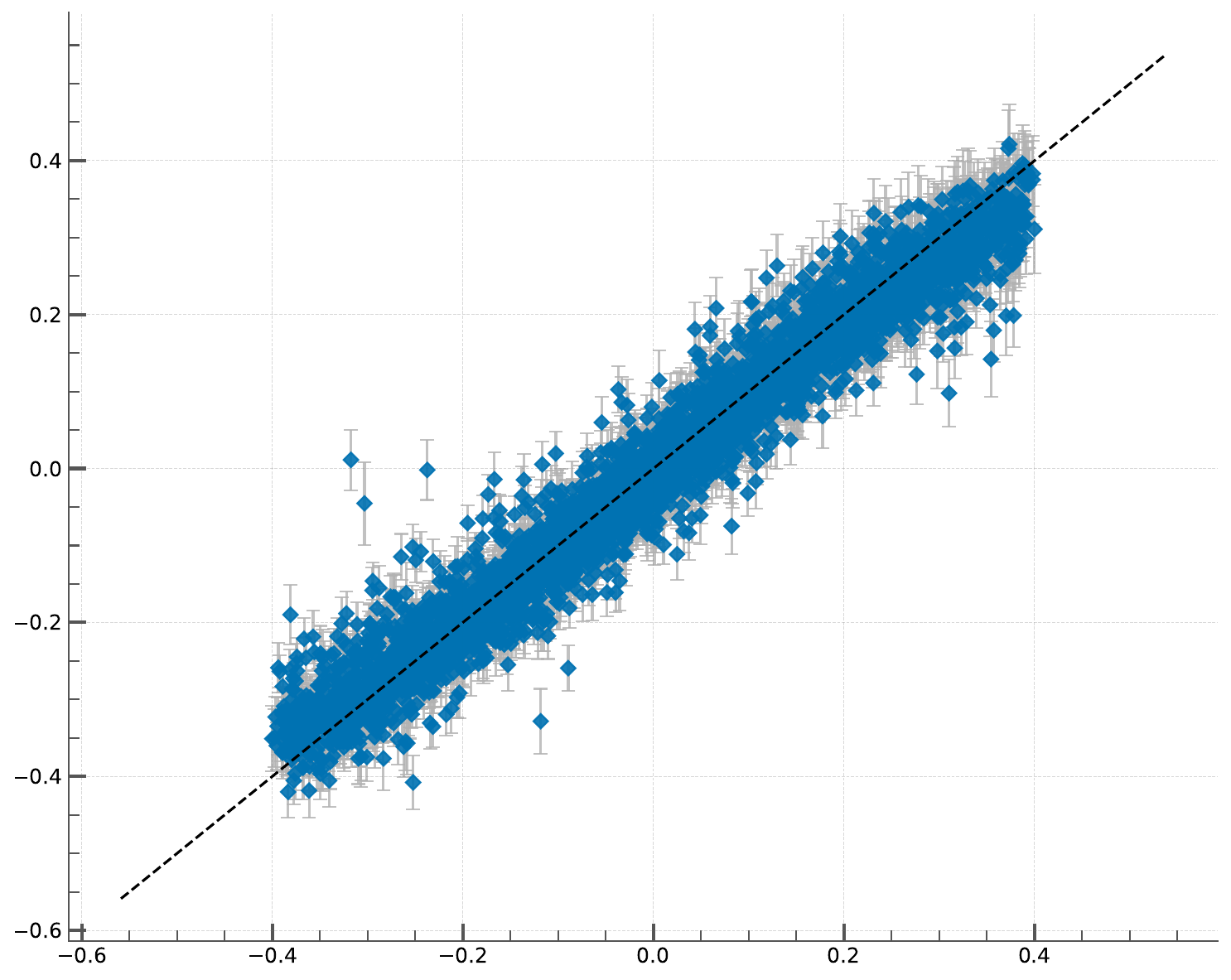} & 
            \includegraphics[width=5.5cm]{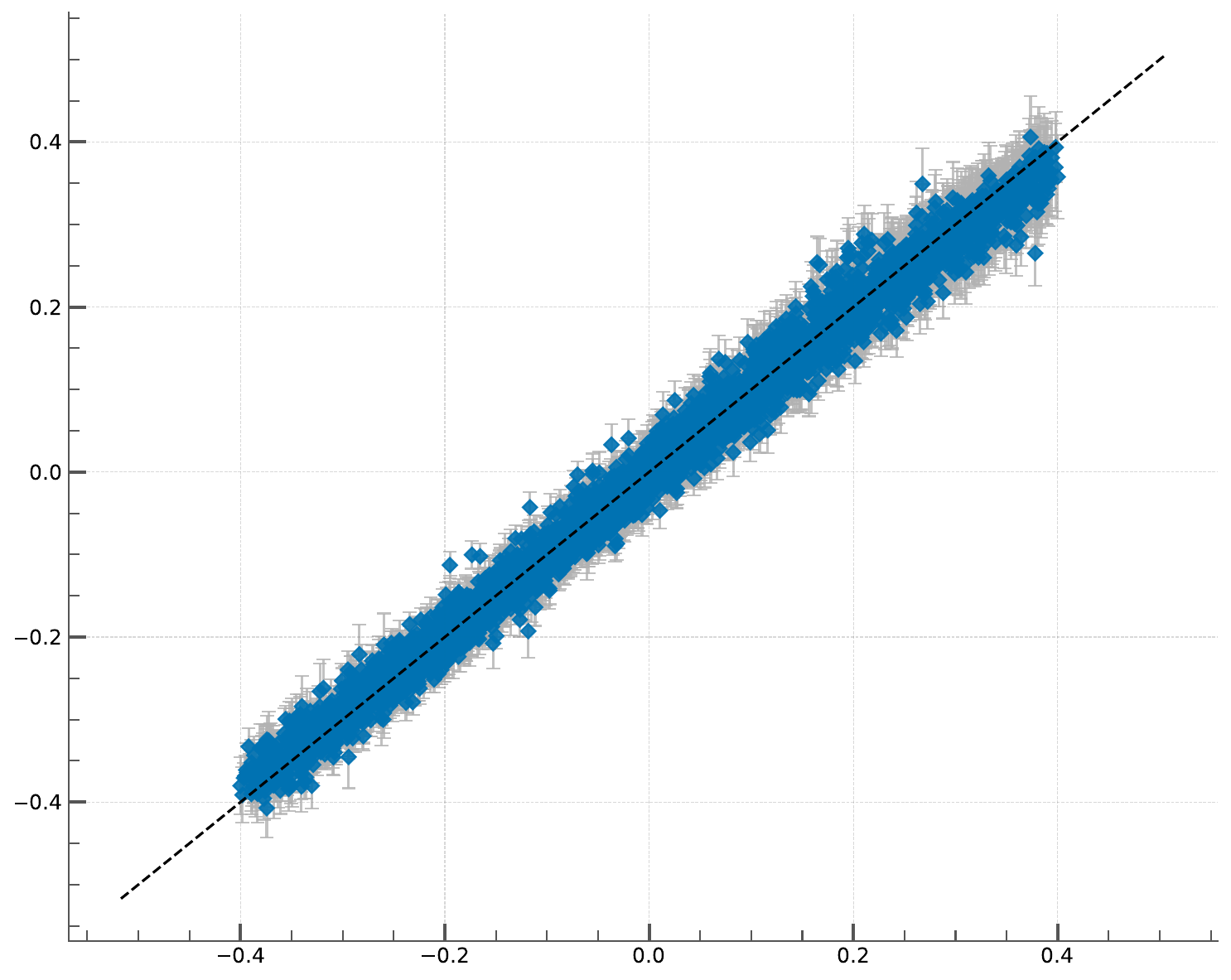} \\
            \includegraphics[width=5.5cm]{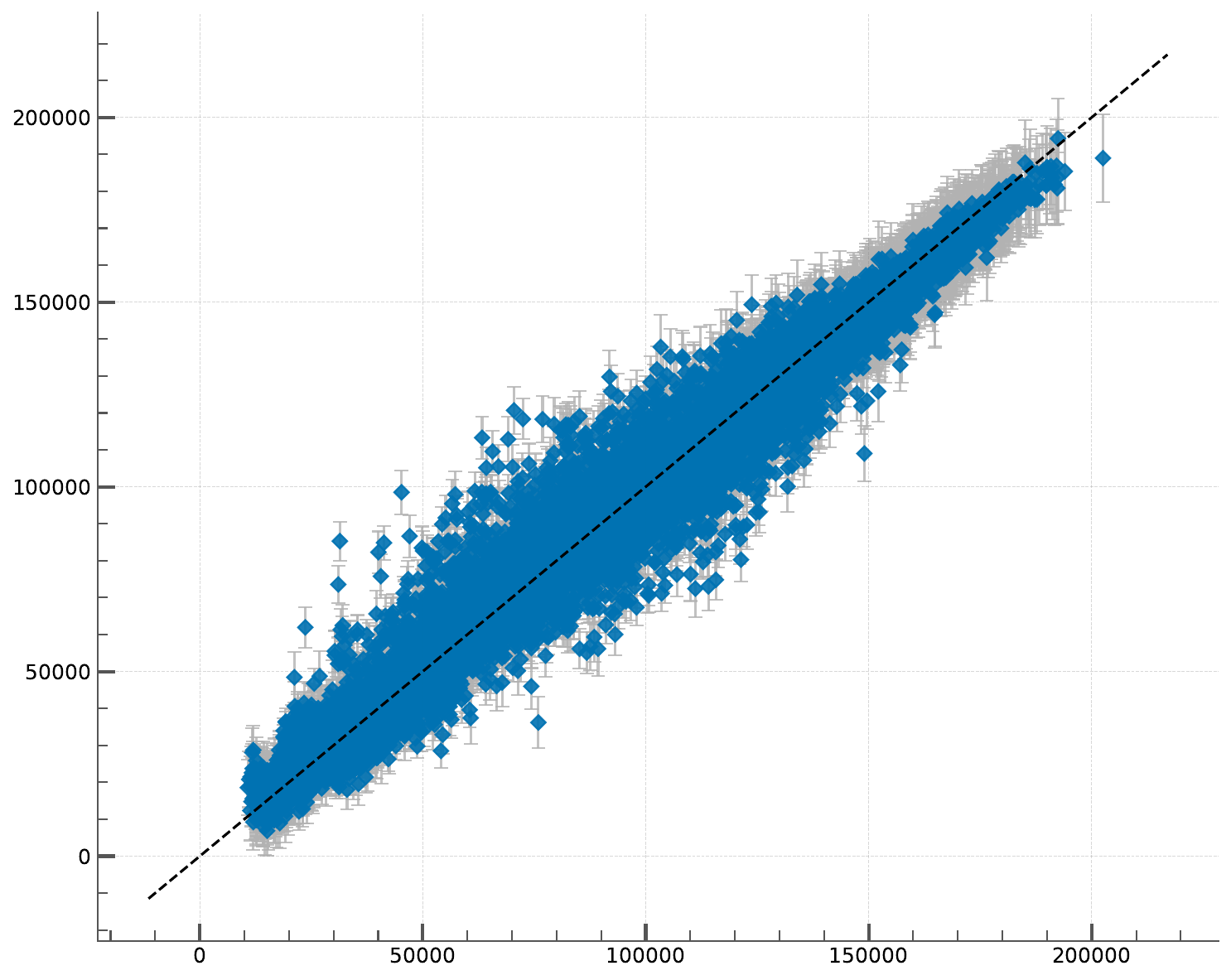} & 
            \includegraphics[width=5.5cm]{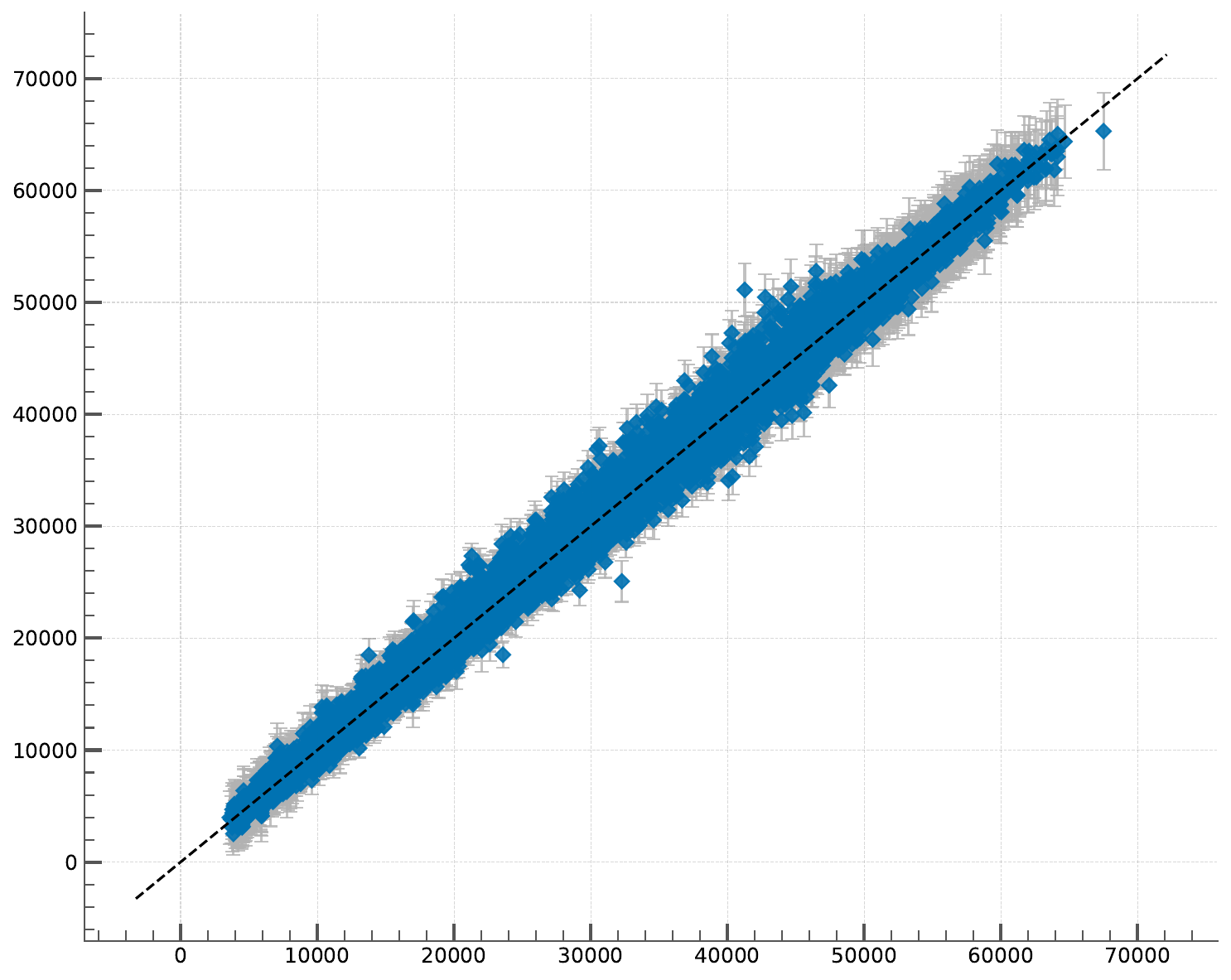} \\
        };

        \foreach \col/\snrval in {1/10, 2/30} {
            \node[font=\small, below=0.1cm of plotgrid-4-\col] {SNR \snrval};
        }
        \foreach \row/\param in {1/{m_1}, 2/{m_2}, 3/{\chi_{\rm{eff}}}, 4/{d}} {
            \node[font=\bfseries, left=0.6cm of plotgrid-\row-1, rotate=90, anchor=center] {Parameter $\param$};
        }

    \end{tikzpicture}
\end{figure*}

Figures~\ref{fig:pred_vs_true_large} and \ref{fig:pred_vs_true_large_finetuned} (see also Tables~\ref{tab:Performance_comparison}, \ref{tab:pretraining_gain_detailed}) present the true (horizontal axis) versus predicted values (vertical axis) for each of the four gravitational waveform parameters on the test set, as inferred by the model when trained from scratch and pretrained/fine-tuned respectively. Masses are shown in solar masses ($M_{\odot}$), the effective distance in megaparsecs (Mpc), and effective spin is dimensionless. For the mass plots, we visualise the model's performance by aggregating the test set results. Each point represents the mean prediction for all waveforms sharing the same true mass value. The associated error bar corresponds to the largest 1$\sigma$ uncertainty derived from MC Dropout found within that group of masses. The black error bars can be understood as the marginalised 1$\sigma$ errors for the prediction of the mass parameter at fixed target value, recalling that for a given pair of masses on our grid, there are multiple associated spin and distance parameters. Notably, the two error bars are generally of comparable size, as evident from the corresponding mass prediction plots. The same procedure is also done with the effective spin plot. We use the large dataset as our reference case.
Overall, one sees that the model fine-tuned on the large dataset is capable of accurate parameter inference for gravitational waveforms, with predicted values closely tracking the true values. The majority of predictions for all parameters cluster tightly around the $y=x$ line, and the width of the error bars decreases notably with increasing SNR.
For the masses $m_1$ and $m_2$, the typical mean value of $1\sigma$ for SNR = 10 ranges between $2-4 M_{\odot}$, with most points clustered around $2 - 3 M_{\odot}$. For SNR = 30, the error bars are similar. This suggests that the indicative precision for the masses is approximately $3 - 8\%$ of the mean mass value, assuming a typical mean value of $\sim 60 M_{\odot}$. We observe worse performance and a higher spread at the boundaries of our parameter grid. The model's uncertainty is highest for the smallest mass value of $5  M_{\odot}$ and for the largest mass value of $150 M_{\odot}$. For the spin parameter ($\chi_{\rm{eff}}$), the error bars are mostly in the range of $0.01 - 0.03$. For the effective luminosity distance ($d$), the error bars for SNR = 10 lie between $\sim 2,000 - 4,000$ Mpc, while for SNR = 30 between $\sim 1,000 - 2,000$ Mpc respectively \footnote{We emphasise that this distance is not an astrophysical luminosity distance, but an effective amplitude-scaling parameter chosen to enforce a fixed SNR.}. Given a typical value of $d \sim 50,000 - 70,000$ Mpc, corresponding to sources rescaled to fixed SNR, this translates to an indicative precision of around $3 - 7\%$ for SNR = 10, and $2 - 5\%$ for SNR = 30. In summary, for models fine-tuned on our large dataset, the typical mean precision for the model's confidence derived through MC Dropouts is at the few-percent level. 

\begin{figure}[!t]
    \centering
    \small 

    \begin{subfigure}[t]{0.47\textwidth}
        \centering
        \includegraphics[width=\linewidth]{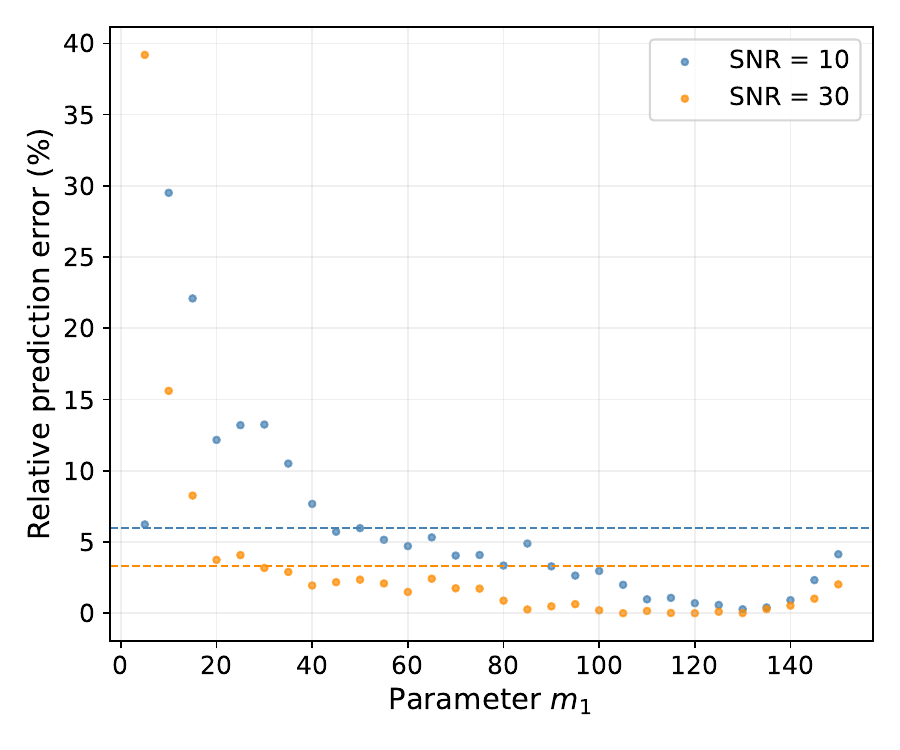}
        \caption{}
    \end{subfigure}
    \hfill
    \begin{subfigure}[t]{0.47\textwidth}
        \centering
        \includegraphics[width=\linewidth]{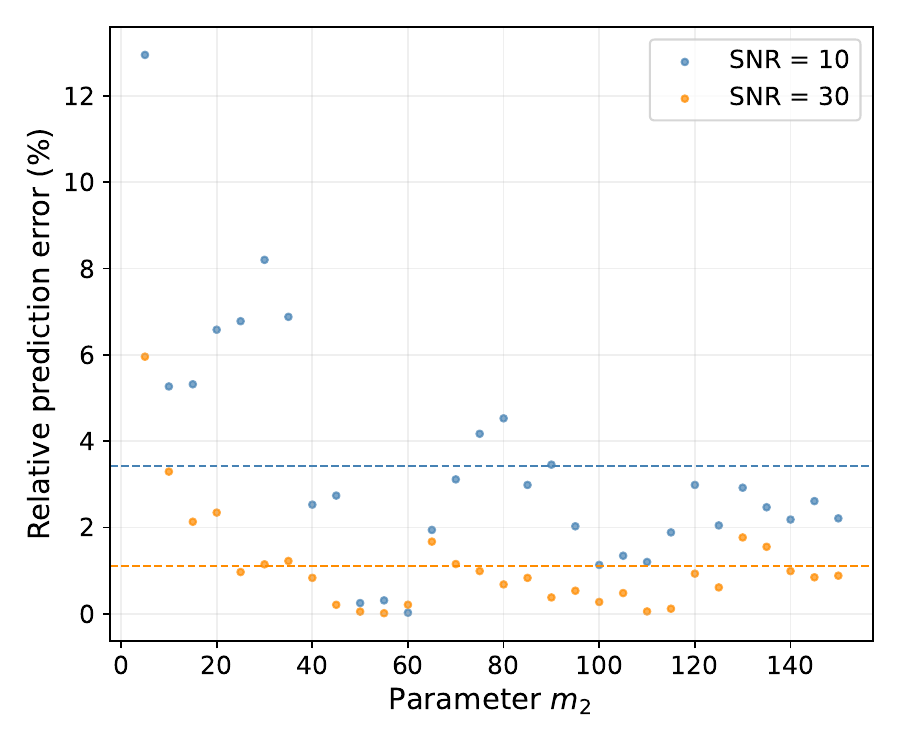}
        \caption{}
    \end{subfigure}
    \vspace{0.6em} 

    \begin{subfigure}[t]{0.47\textwidth}
        \centering
        \includegraphics[width=\linewidth]{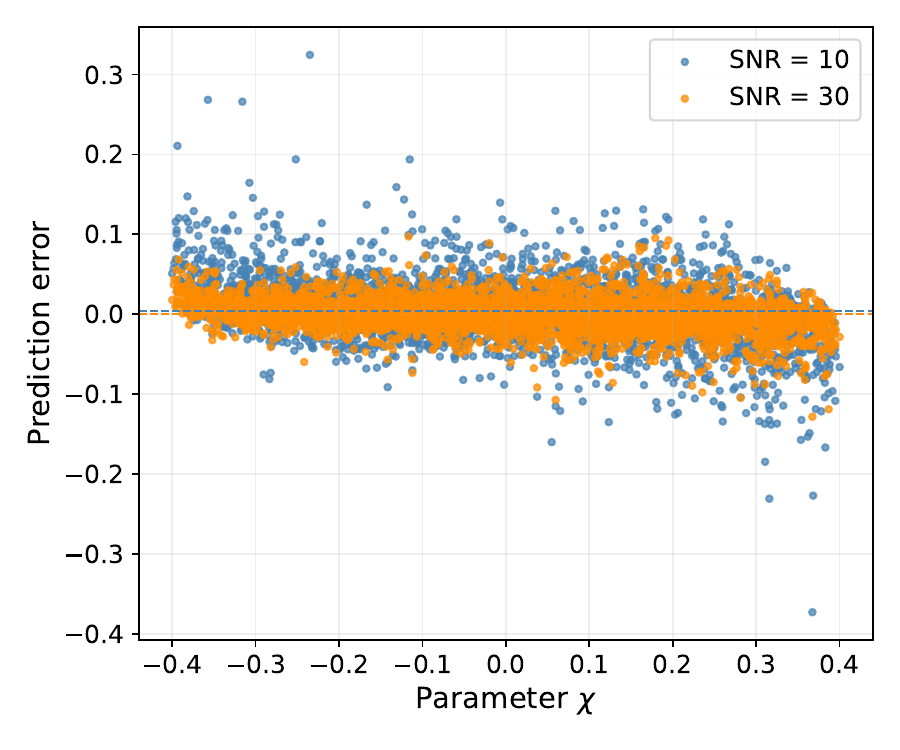}
        \caption{}
    \end{subfigure}
    \hfill
    \begin{subfigure}[t]{0.47\textwidth}
        \centering
        \includegraphics[width=\linewidth]{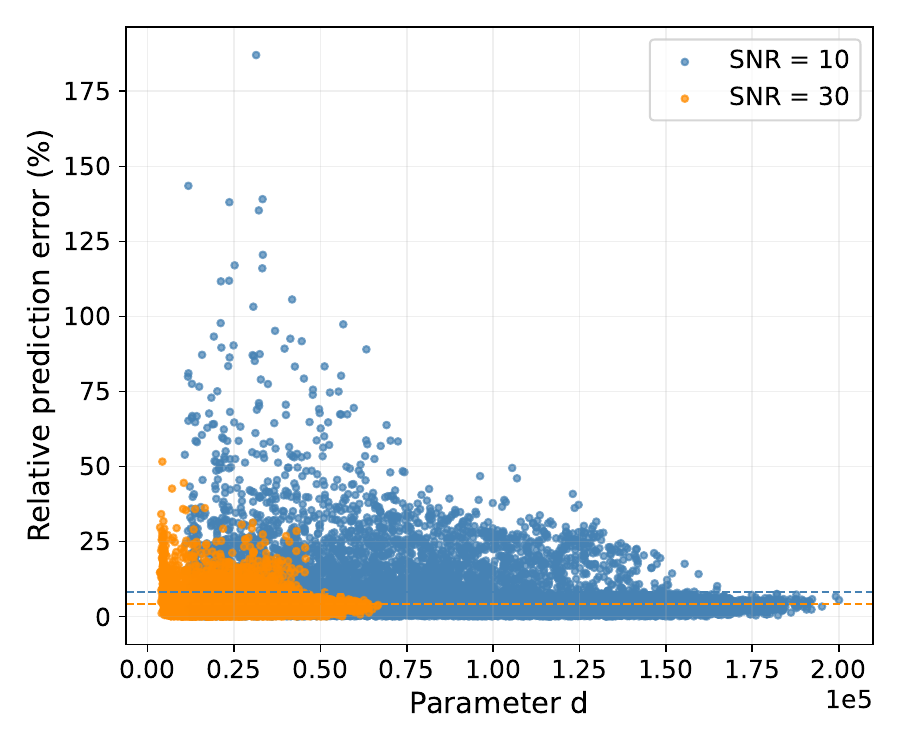}
        \caption{}
    \end{subfigure}
    \caption{\small Prediction errors across our grid sample, for the component masses, the effective spin, the distance and for both fixed-SNR cases of our work. For all parameters except the effective spin, the errors are shown as relative differences with respect to the corresponding target values, (prediction - target)/target. For the effective spin, only the actual difference is used to avoid numerical instability from dividing by small values. The dashed lines represent the mean of the respective values across the sample. We discuss an approximate comparison of these results with the typical errors of real waveform inferences at the end of Section \ref{sec:results}. }
    \label{fig:prediction_errors}
\end{figure}

Figure \ref{fig:prediction_errors} shows the precision of the model's predictions with respect to the target values across our grid. For the component masses, we evaluate the mean relative precision across our predictions, defined as $100 \times (\text{prediction} - \text{target})/\text{target}$. For the effective spin, we report the mean of the absolute difference $(\text{prediction} - \text{target})$, in order to avoid amplification due to division by values near zero. 
The mean relative precision across the sample for $m_1$ is $6\%$ at SNR = 10 and $3.3\%$ at SNR = 30. The similar values for $m_2$ are $3.4\%$ and $1.1\%$, while for distance $8.3\%$ and $4.1 \%$ respectively. For $\chi_{\rm{eff}}$, the absolute mean error value is $3.8 \times 10^{-3}$ at SNR = 10 and $6 \times 10^{-5}$ at SNR = 30. We note that in both cases the predictions are averaged within each grid point group, effectively marginalising over the remaining parameters.
%
To enable an approximate comparison with real observations, we focus on model predictions at SNR = 10, and select the event GW190828\_065509 from the third Gravitational Wave Transient Catalog (GWTC-3) \cite{Abbott2023gwtc3} of the LIGO-Virgo-Kagra (LVK) Collaboration, which has an SNR compatible with $10$ \footnote{The reported central values for masses and effective spin are $m_1 = 23.7 M_{\odot}, m_2 = 10.4 M_{\odot}, \chi_{\rm{eff}} = 0.05$.}.
To compare our model’s mean relative errors with the $90\%$ credible intervals reported by the LVK Collaboration, we convert them to equivalent $90\%$ errors under the assumption of Gaussianity, which amounts to multiplying them by $\simeq 1.645$. For $\{m_1, m_2, \chi_{\rm{eff}}, d\}$ we find for the average $90\%$ error across our sample $\{9.9\%, 5.7\%, 0.006, 13.6\% \}$, while the corresponding errors reported by the LVK Collaboration for the event GW190828\_065509 are $\{28.5\%, 28.9\%, 0.165, 44\%\}$. Notice that we have symmetrised the observation errors to facilitate comparison, i.e $\sigma_{\rm 90\%} = (\sigma_{\rm 90\% upper} + \sigma_{\rm 90\% lower})/2$.
These numbers should not be interpreted as a direct posterior-to-posterior comparison, and serves only as a back-of-the-envelope estimate. A Bayesian analysis would be needed for a proper comparison. We also emphasise that our distance refers to an effective luminosity distance rescaled to fixed SNR, not an actual astrophysical one.

\subsection{Domain adaptation} \label{sec:domain-adaptation}
Domain adaptation refers to transferring knowledge across domains and is central to foundation-style models, which learn general rather than domain-specific representations. Effective adaptation indicates that the model has captured fundamental structures in the data, with the pretraining process providing robust, transferable features which subsequently require only light fine-tuning. In our case, this invariance is assessed only within the shared ET–LVK frequency band ($\geq20$ Hz). We focus on transferring from ET to LVK noise, evaluating whether the model can adopt detector-agnostic GW features rather than memorising detector-specific noise. 

We first evaluate zero-shot inference by applying the model trained purely on ET noise directly to the LVK one, without any fine-tuning or weight updates. This results in a loss of predictive power, indicating that ET-trained features do not transfer to the LVK domain in a zero-shot setting. This is naturally explained by the model having learned a detector-conditioned latent space rather than an invariant one, and therefore, requiring domain adaptation through fine-tuning. Implementing transfer learning through fine tuning, the performance gain is significant compared to training from scratch across all inferred parameters, ranging between $9\% - 47\%$ improvement for MAE (see Tables \ref{tab:domain_adaptation_results_detailed} and \ref{tab:convergence_speedup2}). The spin parameter shows the biggest improvements ($21-47\%$), suggesting that the pretrained model learned particularly robust representations of spin-related features that transfer effectively between detectors. The mass parameters and distance also show consistent improvements of $13-27\%$, indicating that the model successfully captures detector-agnostic physical relationships. The convergence speed results show that the pretrained model achieves the same performance targets as models trained from scratch in dramatically less time, up to $15\times$ faster. This speedup is particularly pronounced at lower SNR = 10, where the pretrained model reaches target performance in just 2-3 minutes compared to 30-50 minutes for scratch training. Interestingly, the speedup is somewhat reduced SNR = 30, possibly because cleaner signals allow scratch-trained models to converge more efficiently. These results collectively demonstrate that pretraining on one detector configuration creates transferable representations that not only improve final performance, but also significantly reduce the computational cost of adapting to new detector environments. We remind that while fine-tuning on LVK noise shows successful adaptation, both detectors still share broadly similar noise structure above 20 Hz. The more demanding task of adapting LVK-trained models to ET’s full 5 Hz range remains open for future work.

To evaluate the generalisation of \texttt{GraviBERT} against waveform modelling, we performed a cross-waveform approximant validation. We first reconstructed the held-out test samples from the large dataset (originally generated with \texttt{SEOBNRv4}) using the \texttt{IMRPhenomD} approximant, adjusting the signals to match the temporal grid of the original \texttt{SEOBNRv4} training data. Inference was then performed on these new waveforms using the frozen weights of our fully trained model (pretrained and trained on ET noise and \texttt{SEOBNRv4} approximant) without any further fine-tuning, that is, at {\it zero-shot inference}. The results are shown in Table \ref{tab:approximant}. While the baseline model achieves $R^2 >0.93$ for all parameters at SNR = 10, the cross-waveform approximant inference appears to introduce a systematic bias. This effect is particularly evident for the effective spin parameter, where $R^2 = 0.65$ at SNR = 10. Spin is intimately connected with the subtle phase evolution during inspiral, which is modelled differently in \texttt{SEOBNRv4} and \texttt{IMRPhenomD}. Nevertheless, the model retains predictive power for the dominant parameters $(m_1,m_2,d)$ at zero-shot inference, with $R^2$ remaining consistently between [0.80,0.97]. This suggests that the model has extracted representations of the signal envelope and frequency chirp, which are approximant-agnostic, while the specific phase alignment remains sensitive to the training domain. 

We then moved on with cross-waveform approximant fine-tuning. The \texttt{SEOBNRv4}-pretrained model fine-tuned on \texttt{IMRPhenomD} waveforms yields consistent performance improvements across all parameters at both SNR levels (see Tables \ref{tab:approximant_results_detailed} and \ref{tab:convergence_speedup_approximant}). The performance gains range from 19--44\% reduction in MAE compared to training from scratch, with particularly strong improvements for the $\chi_{\rm{eff}}$ parameter (29--44\%). Notably, these gains are achieved with dramatically reduced training time -- the pretrained model reaches target performance 6--15$\times$ faster than scratch training. The benefits are most pronounced at lower SNR (10) and with smaller fine-tuning datasets (5,000 samples), where speedups exceed 15$\times$. However, the performance gains diminish somewhat at higher SNR (30) and with larger datasets (10,000 samples), suggesting that domain adaptation provides the greatest advantage in data-limited and low-SNR regimes where training from scratch is most challenging. The $R^2$ scores further validate the effectiveness of pretraining, with the pretrained model consistently outperforming training from scratch across all parameters and conditions. The improvements are most dramatic for the effective spin parameter $\chi_{\rm{eff}}$, where $R^2$ increases from $0.569$ to $0.813$ at SNR = 10 with $5,000$ samples. For the mass parameters $m_1$ and $m_2$, the pretrained model achieves $R^2 > 0.9$ even at SNR = 10, compared to $0.74$--$0.87$ for scratch training. The luminosity distance $d$ shows the smallest relative gains, but maintains the highest absolute $R^2$ values ($0.94$--$0.99$), indicating this parameter is well-captured by both approaches. At higher SNR (30) and with more data ($10,000$ samples), both models perform strongly ($R^2 > 0.9$ for most parameters), though the pretrained model still maintains a consistent edge, particularly for $\chi_{\rm{eff}}$ where the gap remains substantial.

\subsection{Benchmarking} \label{sec:benchmarking}

To contextualise our results, we compare \texttt{GraviBERT} with other approaches for gravitational wave parameter estimation. Since we focus on point estimates in this work, we compare with methods using similar inference targets rather than full neural posterior estimation frameworks. We should emphasise that direct comparisons with prior ML methods are complicated by differences in parameter dimensionality, waveform models, noise assumptions, signal duration, architectures, and hardware platforms. Several works focus on predicting only two intrinsic parameters (e.g., chirp mass and mass ratio), while others use neural density estimation to approximate full posterior distributions. Our model performs four-parameter regression $(m_1, m_2, \chi_{\mathrm{eff}}, d)$ with a hybrid CNN-Transformer architecture designed for representation learning and transferability. Most importantly, we have {\it not performed} a hyperparameter optimisation which could significantly improve the performance. Numerical comparisons should therefore be interpreted as contextual indicators rather than direct benchmarks.

George \& Huerta (2018) \cite{george2018deep2_inference} demonstrated CNN-based parameter estimation achieving mean relative errors below 10\% for mass parameters on real LVK data. Shen et al. (2022) \cite{shen2022deep} achieved 2 ms inference time using Bayesian neural networks trained on over 10 million waveforms for 4-parameter estimation, with relative errors of 7--12\% for masses and 5--13\% for remnant properties at SNR $\geq$ 15.

Without a hyperparameter optimisation, \texttt{GraviBERT} achieves a single-event inference time of $10.6 \pm 0.5$ ms at batch size 1, corresponding to a throughput of approximately 94 events per second (see Table \ref{tab:benchmark}). More importantly, at higher batch sizes the model reaches a maximum throughput of 1895 events per second (0.48 ms per waveform at batch size 250, including MC Dropout for uncertainty quantification), demonstrating efficient parallel processing capabilities suitable for large-scale population studies and systematic analyses. This throughput substantially exceeds the effective rates of single-event methods like Shen et al. (2 ms per event, approximately 500 events per second assuming no batching overhead). With a throughput of 1895 events per second, \texttt{GraviBERT} can process GW detections in real-time, exceeding projected ET peak detection rates ($\sim 50$ events/s) by approximately 40$\times$. The single-event latency of 10.6 ms is well below the sub-second requirements for rapid electromagnetic follow-up alerts. 

We remind that we did not perform hyperparameter optimisation, as our aim was to introduce the architecture and demonstrate that (i) BERT-style masked pretraining greatly speeds up convergence and improves accuracy, (ii) pretrained models enable strong transfer learning, and (iii) hybrid CNN-Transformer models capture the essential features of the GW signals studied in this work. These strategies support both point estimation and Bayesian inference.

\section{Conclusions and Outlook} \label{sec:conclusions}
We have introduced \texttt{GraviBERT}, a novel deep learning model for GW signal analysis that fuses an Inception-inspired multi-scale convolutional frontend with a transformer encoder. The key novelty of \texttt{GraviBERT} lies in its self-supervised pretraining strategy, which enables learning of transferable representations of the underlying physics, followed by fine-tuning on labeled data. This is facilitated by the feature extractor, which learns local temporal features, and the transformer's self-attention mechanism, which captures global dependencies. In the current work, we demonstrate its capabilities using a regression head that outputs point estimates for source parameters backed by MC dropouts.

Our two-stage training strategy---leveraging BERT-style pretraining followed by targeted fine-tuning---yields substantial improvements in both training efficiency and inference accuracy, while demonstrating promising adaptation to new detector configurations and waveform approximants. When evaluated on in-domain data (ET), pretraining reduces the MAE by up to $31\%$ and accelerates training convergence by up to $6\times$. Furthermore, the mean relative precision of the inferred masses and distances reaches the few-percent level, while the MAE in effective spin is as low as $\sim 10^{-3}$ for the challenging case of SNR = 10\footnote{These results were obtained without exhaustive hyperparameter optimisation.}. We note that the reported distance corresponds to an effective luminosity distance rescaled to fixed SNR, rather than the actual astrophysical luminosity distance.

The advantages are even more pronounced in domain adaptation from ET to LVK data. By fine-tuning on a small target dataset on LVK data, the pretrained model on ET noise converges up to $15\times$ faster and achieves an error reduction of up to $47\%$ compared to a model trained from scratch on LVK. This enhanced predictive precision is particularly notable in the low SNR regime, while the flexibility of the model’s architecture paves the way for broad applicability across multi-messenger astronomy. We note that while our estimation of inherent model uncertainty relied on the probabilistic approach of MC Dropouts, our parameter inference itself was not Bayesian. A Bayesian treatment could be pursued by replacing the regression head with normalising flows.

Cross-waveform approximant adaptation from \texttt{SEOBNRv4} to \texttt{IMRPhenomD} (at fixed detector noise) reveals that pretrained features transfer effectively for dominant parameters ($m_1$, $m_2$, $d$) even at zero-shot, while effective spin benefits significantly from fine-tuning. This demonstrates that GraviBERT learns physically meaningful representations that generalise across waveform modelling choices, achieving comparable gains ($19$--$44\%$ MAE reduction, $6$--$15\times$ speedup) to detector noise adaptation.

The motivation for developing foundation-style modelling lies in the ability to generalise across a wide range of tasks and datasets, serving as adaptable tools for diverse downstream applications. Foundation models have transformed fields such as natural language processing by enabling rapid fine-tuning and transfer learning. For GW astrophysics, a similar approach could accelerate scientific discovery by enabling fast, accurate, and robust inference as new detectors, noise environments, and physical scenarios emerge. In this regard, \texttt{GraviBERT} represents a first step forward towards establishing a pretrained baseline framework for GW astronomy, which can be subsequently fine-tuned efficiently.

Natural extensions of the present work include applying GraviBERT to full Bayesian parameter estimation, inference of the complete set of source parameters for binary mergers, and comprehensive hyperparameter optimisation — all of which follow directly from the current framework. Additionally, our modelling could be expanded to cover neutron star mergers, alternative waveform approximants, and multi-channel detector configurations such as the full triangular ET design. More broadly, the architecture and pretraining approach presented here could in principle be extended to account for multi-messenger data, and to probing deviations from GR through their imprint on the waveform amplitude and phase evolution \cite{Berti2015, Saltas:2018fwy,LISA:2022kgy, Barausse:2020rsu}. On the architectural side, enhancements such as frequency-domain inputs, variable sampling rate, and comprehensive hyperparameter optimization could further improve performance and generality. We leave these directions for future work.

\section*{Acknowledgements \& code}
We thank François Charton, Nikos Karnesis, Fillipos Kokkinos, Nick Stergioulas and David Trestini for discussions and feedback. The authors were funded by the Czech Grant Agency (GAČR) under the grant number 21-16583M. The bulk of our computations were carried out at the \texttt{MetaCentrum} cluster in Czechia. The code and pretrained models are publicly available at this \href{https://zenodo.org/records/18671949}{link}.

\clearpage
\begin{table}[h]
\centering
\begin{threeparttable}
\caption{Comparison of convergence speed between models trained from scratch and pretrained models at fixed detector configuration (ET). For each condition, the ``Target Loss" is the best validation loss achieved by the corresponding model trained from scratch. The speedup factor quantifies how faster the pretrained model reaches this same performance target. Results are discussed in Section \ref{sec:results}.}
\label{tab:convergence_speedup1}
\footnotesize
\sisetup{round-mode=places, round-precision=2}
\begin{tabular}{l S[table-format=1.4] S[table-format=3.1] S[table-format=2.1] S[table-format=1.1, table-space-text-post=x]}
\toprule
\textbf{SNR} & {\textbf{Target Loss}} & {\textbf{Time (Scratch)}} & {\textbf{Time (Pretrained)}} & {\textbf{Speedup}} \\
& {\textbf{(Best Val Loss)}} & {\textbf{(minutes)}} & {\textbf{(minutes)}} & \\
\midrule
\multicolumn{5}{c}{\textit{\textbf{Dataset Size: Small}}} \\
\addlinespace
SNR 10 & {0.005474} & {79.64} & {13.05} & { 6.10}x \\
SNR 30 & {0.001822} & {98.61} & {20.86} & {4.73}x \\
\addlinespace[1em]
\multicolumn{5}{c}{\textit{\textbf{Dataset Size: Intermediate}}} \\
\addlinespace
SNR 10 & {0.004029} & {196.07} & {29.75} & {6.59}x \\
SNR 30 & {0.001299} & {222.02} & {51.51} & {4.31}x \\
\addlinespace[1em]
\multicolumn{5}{c}{\textit{\textbf{Dataset Size: Large}}} \\
\addlinespace
SNR 10 & {0.003172} & {242.82} & {169.63} & {1.43}x \\
SNR 30 & {0.000952} & {348.69} & {87.92} & {3.97}x \\
\bottomrule
\end{tabular}
\begin{tablenotes}
    \item[*] \footnotesize Time is measured as the cumulative training time until the validation loss first drops below the corresponding target loss for that specific SNR and dataset size.
\end{tablenotes}
\end{threeparttable}
\end{table}

\begin{table}[H]
\centering
\caption{Per-parameter metrics across different dataset sizes and SNR values for the model trained from scratch (not pretrained) and the pretrained and fine-tuned model, evaluated on the held-out test set with size of $10,000$ samples. The same detector configuration is kept throughout (ET). The reported metrics are measured on the mean values obtained from an ensemble of $100$ stochastic forward passes using MC Dropout for each test sample. We remind that $d$ corresponds to an effective luminosity distance rescaled to fixed SNR. For each metric, the best value across the two training configurations is shown in \textbf{bold}. The results are rounded to two decimal points.}
\label{tab:Performance_comparison}
\begin{minipage}[t]{0.48\textwidth}
\centering
\begin{threeparttable}
\caption*{\textbf{(a) Trained from Scratch}}
\footnotesize
\begin{tabular}{l cc cc}
\toprule
& \multicolumn{2}{c}{\textbf{MAE}} & \multicolumn{2}{c}{\textbf{R²}} \\
\cmidrule(lr){2-3} \cmidrule(lr){4-5}
\textbf{Parameter} & \textbf{SNR 10} & \textbf{SNR 30}  & \textbf{SNR 10} & \textbf{SNR 30} \\
\midrule
\multicolumn{5}{c}{\textit{\textbf{Dataset Size: Small}}} \\
$m_1$ & 8.51 & 4.79 & 0.90 & 0.97 \\
$m_2$ & 6.97 & 4.24 & 0.93 & 0.97 \\
$\chi_{\rm{eff}}$ & 0.05 & 0.03 & 0.86 & 0.96 \\
$d$ & 7401.28 & 1238.39 & 0.96 & 0.99 \\
\addlinespace
\multicolumn{5}{c}{\textit{\textbf{Intermediate}}} \\
$m_1$ & 7.21 & 4.19 & 0.92 & 0.97 \\
$m_2$ & 6.18 & 3.80 & 0.94 & 0.98 \\
$\chi_{\rm{eff}}$ & 0.04 & 0.02 & 0.91 & 0.98 \\
$d$ & 6232.91 & 1025.78 & 0.96 & 0.99 \\
\addlinespace
\multicolumn{5}{c}{\textit{\textbf{Large}}} \\
$m_1$ & 6.02 & 3.49 & 0.94 & 0.98 \\
$m_2$ & 5.10 & 3.24 & 0.95 & 0.98 \\
$\chi_{\rm{eff}}$ & 0.03 & 0.02 & 0.93 & 0.98 \\
$d$ & 5414.76 & 867.37 & 0.96 & 0.99 \\
\bottomrule
\end{tabular}
\end{threeparttable}
\end{minipage}
\hfill
\begin{minipage}[t]{0.48\textwidth}
\centering
\begin{threeparttable}
\caption*{\textbf{(b) Pretrained + Fine-tuned}}
\footnotesize
\begin{tabular}{l cc cc}
\toprule
& \multicolumn{2}{c}{\textbf{MAE}} & \multicolumn{2}{c}{\textbf{R²}} \\
\cmidrule(lr){2-3} \cmidrule(lr){4-5}
\textbf{Parameter} & \textbf{SNR 10} & \textbf{SNR 30}  & \textbf{SNR 10} & \textbf{SNR 30} \\
\midrule
\multicolumn{5}{c}{\textit{\textbf{Dataset Size: Small}}} \\
$m_1$ & \textbf{7.80} & \textbf{4.25} & \textbf{0.91} & 0.97 \\
$m_2$ & \textbf{6.58} & \textbf{3.82} & 0.93 & \textbf{0.98} \\
$\chi_{\rm{eff}}$ & \textbf{0.04} & \textbf{0.02} & \textbf{0.90} & \textbf{0.97} \\
$d$ & \textbf{6807.13} & \textbf{1144.08} & \textbf{0.95} & 0.99 \\
\addlinespace
\multicolumn{5}{c}{\textit{\textbf{Intermediate}}} \\
$m_1$ & \textbf{7.06} & \textbf{4.01} & 0.92 & 0.97 \\
$m_2$ & \textbf{5.98} & \textbf{3.69} & 0.94 & 0.98 \\
$\chi_{\rm{eff}}$ & 0.04 & 0.02 & \textbf{0.92} & 0.98 \\
$d$ & \textbf{6154.66} & \textbf{986.82} & 0.96 & 0.99 \\
\addlinespace
\multicolumn{5}{c}{\textit{\textbf{Large}}} \\
$m_1$ & \textbf{5.44} & \textbf{2.44} & 0.94 & \textbf{0.99} \\
$m_2$ & \textbf{4.51} & \textbf{2.23} & \textbf{0.96} & \textbf{0.99} \\
$\chi_{\rm{eff}}$ & 0.03 & 0.02 & 0.93 & 0.98 \\
$d$ & \textbf{5043.25} & \textbf{639.72} & 0.96 & 0.99 \\
\bottomrule
\end{tabular}
\end{threeparttable}
\end{minipage}
\end{table}

\begin{table}[H]
\centering
\begin{threeparttable}  
\caption{Relative MAE improvement from pretraining followed by fine-tuning, compared to training from scratch, according to the results of Table \ref{tab:Performance_comparison}. Positive values indicate the percentage reduction in error achieved through pretraining. The gains are calculated with original unrounded numbers.}
\label{tab:pretraining_gain_detailed}
\footnotesize 
\begin{tabular}{l cc}
\toprule
& \multicolumn{2}{c}{\textbf{Performance Gain on MAE (\%)}} \\
\cmidrule(lr){2-3}
\textbf{Parameter} & \textbf{SNR 10} & \textbf{SNR 30} \\
\midrule
\multicolumn{3}{c}{\textit{\textbf{Dataset Size: Small}}} \\
\addlinespace
$m_1$     & $\approx 8\%$ & $\approx 11\%$ \\
$m_2$     & $\approx 6\%$ & $\approx 10\%$ \\
$\chi_{\rm{eff}}$    & $\approx 14\%$ & $\approx 5\%$ \\
$d$       & $\approx 8\%$ & $\approx 8\%$ \\
\addlinespace[1em] 
\multicolumn{3}{c}{\textit{\textbf{Dataset Size: Intermediate}}} \\ 
\addlinespace
$m_1$     & $\approx 2\%$ & $\approx 4\%$ \\
$m_2$     & $\approx 3\%$  & $\approx 3\%$ \\
$\chi_{\rm{eff}}$    & $\approx 4\%$ & $\approx 7\%$ \\
$d$       & $\approx 1\%$  & $\approx 4\%$ \\
\addlinespace[1em]
\multicolumn{3}{c}{\textit{\textbf{Dataset Size: Large}}} \\ 
\addlinespace
$m_1$     & $\approx 10\%$ & $\approx 30\%$ \\
$m_2$     & $\approx 11\%$  & $\approx 31\%$ \\
$\chi_{\rm{eff}}$    & $\approx 3\%$ & $\approx 11\%$ \\
$d$       & $\approx 7\%$  & $\approx 26\%$ \\
\bottomrule
\end{tabular}
\end{threeparttable}
\end{table} 

\begin{table}[h]
\centering
\begin{threeparttable}
\caption{Domain adaptation performance of the pretrained model evaluated with MAE. The comparison is between the model pretrained on ET noise and fine-tuned on LVK noise waveforms, and a model trained on LVK noise waveforms from scratch. The clean waveforms used cross both datasets were generated using the same procedure, with only the detector noise differing between domains. The results are rounded to two decimal points. The gains are calculated with original unrounded numbers.}
\label{tab:domain_adaptation_results_detailed}
\scriptsize 
\sisetup{round-mode=places, round-precision=2} 
\begin{tabular}{l l S[table-format=2.2] S[table-format=2.2] S[table-format=2.1, table-space-text-post={\,\%}] c S[table-format=2.2] S[table-format=2.2] S[table-format=2.1, table-space-text-post={\,\%}]}
\toprule
& & \multicolumn{3}{c}{\textbf{SNR 10}} & \phantom{abc} & \multicolumn{3}{c}{\textbf{SNR 30}} \\
\cmidrule(lr){3-5} \cmidrule(lr){7-9}
\textbf{Fine-tuning} & \textbf{Parameter} & {\textbf{MAE}} & {\textbf{MAE}} & {\textbf{Perf.}} && {\textbf{MAE}} & {\textbf{MAE}} & {\textbf{Perf.}} \\
\textbf{Dataset Size} & & {\textbf{(Scratch)}} & {\textbf{(Fine-tuned)}} & {\textbf{Gain}} && {\textbf{(Scratch)}} & {\textbf{(Fine-tuned)}} & {\textbf{Gain}} \\
\midrule
\textit{5,000 Samples} & $m_1$ & {12.54} & {9.22} & {$\approx 26\%$} && {6.39} & {4.91} & {$\approx 23\%$} \\
                      & $m_2$ & {9.42} & {7.75} & {$\approx 18\%$} && {5.64} & {4.41} & {$\approx 22\%$} \\
                      & $\chi_{\rm{eff}}$ & {0.10} & {0.05} & {$\approx 43\%$} && {0.05} & {0.03} & {$\approx 47\%$} \\
                      & $d$ & {543.63} & {406.15} & {$\approx 25\%$} && {87.85} & {63.82} & {$\approx 27\%$} \\
\addlinespace[1em]
\textit{10,000 Samples} & $m_1$ & {10.43} & {8.64} & {$\approx 17\%$} && {5.80} & {4.79} & {$\approx 18\%$} \\
                       & $m_2$ & {8.18} & {7.12} & {$\approx 13\%$} && {5.02} & {4.36} & {$\approx 13\%$} \\
                       & $\chi_{\rm{eff}}$ & {0.07} & {0.05} & {$\approx 21\%$} && {0.04} & {0.03} & {$\approx 32\%$} \\
                       & $d$ & {448.28} & {406.89} & {$\approx 9\%$} && {72.77} & {58.81} & {$\approx 19\%$} \\
\bottomrule
\end{tabular}
\end{threeparttable}
\end{table}

\vspace{2cm}

\begin{table}[h]
\centering
\begin{threeparttable}
\caption{Same as Table \ref{tab:domain_adaptation_results_detailed}, but comparing the convergence speed between the model pretrained on ET noise and fine-tuned on LVK noise waveforms, and a model trained on LVK noise waveforms from scratch. As before, the ``Target Loss" is the best validation loss achieved by the corresponding model trained from scratch and the speedup factor quantifies how many times faster the pretrained model reaches this same performance target.}
\label{tab:convergence_speedup2}
\footnotesize
\sisetup{round-mode=places, round-precision=2}
\begin{tabular}{l S[table-format=1.4] S[table-format=3.1] S[table-format=2.1] S[table-format=1.1, table-space-text-post=x]}
\toprule
\textbf{SNR} & {\textbf{Target Loss}} & {\textbf{Time (Scratch)}} & {\textbf{Time (Pretrained)}} & {\textbf{Speedup}} \\
& {\textbf{(Best Val Loss)}} & {\textbf{(minutes)}} & {\textbf{(minutes)}} & \\
\midrule
\multicolumn{5}{c}{\textit{\textbf{Dataset Size: 5,000 Samples}}} \\
\addlinespace
SNR 10 & {0.012263} & {31.97} & {2.12} & {15.05}x \\
SNR 30 & {0.004212} & {29.65} & {5.13} & {5.78}x \\
\addlinespace[1em]
\multicolumn{5}{c}{\textit{\textbf{Dataset Size: 10,000 Samples}}} \\
\addlinespace
SNR 10 & {0.008057} & {52.17} & {3.42} & {15.24}x \\
SNR 30 & {0.002486} & {46.97} & {7.79} & {6.03}x \\
\bottomrule
\end{tabular}
\begin{tablenotes}
    \item[*] \footnotesize Time is measured as the cumulative training time until the validation loss first drops below the corresponding target loss for that specific SNR and dataset size.
\end{tablenotes}
\end{threeparttable}
\end{table}

\begin{table}[h]
\centering
\caption{Zero-shot performance of the model on a new waveform approximant at fixed detector noise (ET), quantified using MAE and \(R^2\). The model was pretrained and subsequently trained on \texttt{SEOBNRv4}. It can be seen that zero-shot inference (without fine-tuning) already yields encouraging performance in the cross-waveform approximant setting (see also Section \ref{sec:domain-adaptation}). The results are rounded to two decimal points.}
\label{tab:approximant}
\footnotesize
\begin{tabular}{l cc cc}
\toprule
& \multicolumn{2}{c}{\textbf{MAE}} & \multicolumn{2}{c}{\textbf{R²}} \\
\cmidrule(lr){2-3} \cmidrule(lr){4-5}
\textbf{Parameter} & \textbf{SNR 10} & \textbf{SNR 30} & \textbf{SNR 10} & \textbf{SNR 30} \\
\midrule
$m_1$    & 11.16 & 6.44 & 0.80 & 0.93 \\
$m_2$    & 9.27 & 6.67 & 0.86 & 0.92 \\
$\chi_{\rm{eff}}$   & 0.08 & 0.07 & 0.65 & 0.70 \\
$d$      & 10320.96 & 1840.13 & 0.87 & 0.97 \\
\bottomrule
\end{tabular}
\end{table}

\begin{table}[h]
\centering
\begin{threeparttable}
\caption{Domain adaptation performance of the pretrained model on a new waveform approximant at fixed detector noise (ET), evaluated with MAE. The comparison is between the model pretrained on \texttt{SEOBNRv4} waveforms and fine-tuned on \texttt{IMRPhenomD} waveforms, and a model trained on \texttt{IMRPhenomD} waveforms from scratch. The results are rounded to two decimal points. The gains are calculated with original unrounded numbers.}
\label{tab:approximant_results_detailed}
\scriptsize 
\sisetup{round-mode=places, round-precision=2} 
\begin{tabular}{l l S[table-format=2.2] S[table-format=2.2] S[table-format=2.1, table-space-text-post={\,\%}] c S[table-format=2.2] S[table-format=2.2] S[table-format=2.1, table-space-text-post={\,\%}]}
\toprule
& & \multicolumn{3}{c}{\textbf{SNR 10}} & \phantom{abc} & \multicolumn{3}{c}{\textbf{SNR 30}} \\
\cmidrule(lr){3-5} \cmidrule(lr){7-9}
\textbf{Fine-tuning} & \textbf{Parameter} & {\textbf{MAE}} & {\textbf{MAE}} & {\textbf{Perf.}} && {\textbf{MAE}} & {\textbf{MAE}} & {\textbf{Perf.}} \\
\textbf{Dataset Size} & & {\textbf{(Scratch)}} & {\textbf{(Fine-tuned)}} & {\textbf{Gain}} && {\textbf{(Scratch)}} & {\textbf{(Fine-tuned)}} & {\textbf{Gain}} \\
\midrule
\textit{5,000 Samples} & $m_1$ & {13.85} & {8.80} & {$\approx 36\%$} && {6.36} & {4.64} & {$\approx 27\%$} \\
                      & $m_2$ & {9.51} & {6.91} & {$\approx 27\%$} && {5.42} & {4.10} & {$\approx 24\%$} \\
                      & $\chi_{\rm{eff}}$ & {0.09} & {0.06} & {$\approx 39\%$} && {0.06} & {0.03} & {$\approx 44\%$} \\
                      & $d$ & {11420.42} & {7471.91} & {$\approx 35\%$} && {1839.72} & {1291.50} & {$\approx 30\%$} \\
\addlinespace[1em]
\textit{10,000 Samples} & $m_1$ & {11.44} & {8.19} & {$\approx 28\%$} && {5.46} & {4.29} & {$\approx 21\%$} \\
                       & $m_2$ & {8.14} & {6.48} & {$\approx 20\%$} && {4.56} & {3.70} & {$\approx 19\%$} \\
                       & $\chi_{\rm{eff}}$ & {0.07} & {0.05} & {$\approx 28\%$} && {0.04} & {0.03} & {$\approx 29\%$} \\
                       & $d$ & {9371.46} & {6793.89} & {$\approx 27\%$} && {1454.00} & {1095.89} & {$\approx 25\%$} \\
\bottomrule
\end{tabular}
\end{threeparttable}
\end{table}

\begin{table}[h]
\centering
\begin{threeparttable}
\caption{Same as Table \ref{tab:approximant_results_detailed}, but comparing the speedup gain. The ``Target Loss" is the best validation loss achieved by the corresponding model trained from scratch and the speedup factor quantifies how faster the pretrained model reaches the same performance target.}
\label{tab:convergence_speedup_approximant}
\footnotesize
\sisetup{round-mode=places, round-precision=2}
\begin{tabular}{l S[table-format=1.4] S[table-format=3.1] S[table-format=2.1] S[table-format=1.1, table-space-text-post=x]}
\toprule
\textbf{SNR} & {\textbf{Target Loss}} & {\textbf{Time (Scratch)}} & {\textbf{Time (Pretrained)}} & {\textbf{Speedup}} \\
& {\textbf{(Best Val Loss)}} & {\textbf{(minutes)}} & {\textbf{(minutes)}} & \\
\midrule
\multicolumn{5}{c}{\textit{\textbf{Dataset Size: 5,000 Samples}}} \\
\addlinespace
SNR 10 & {0.014031} & {34.12} & {2.24} & {15.21}x \\
SNR 30 & {0.004849} & {30.37} & {3.52} & {8.63}x \\
\addlinespace[1em]
\multicolumn{5}{c}{\textit{\textbf{Dataset Size: 10,000 Samples}}} \\
\addlinespace
SNR 10 & {0.009436} & {52.46} & {3.65} & {14.36}x \\
SNR 30 & {0.002799} & {53.36} & {8.29} & {6.44}x \\
\bottomrule
\end{tabular}
\begin{tablenotes}
    \item[*] \footnotesize Time is measured as the cumulative training time until the validation loss first drops below the corresponding target loss for that specific SNR and dataset size.
\end{tablenotes}
\end{threeparttable}
\end{table}

\begin{table}[h]
\centering
\caption{Inference latency comparison for different batch sizes for the current version of \texttt{GraviBERT}, without hyperparameter optimisation. We consider a standard forward pass (no MC Dropout activated), and one with MC Dropout turned on. The results are discussed in Section \ref{sec:benchmarking}.}
\label{tab:benchmark}
\begin{tabular}{lcccc}
\hline
Batch Size & Standard (ms) & MC Dropout (ms) & Overhead (\%) & Throughput (samples/s) \\
\hline
1 & $10.6 \pm 0.5$ & $10.4 \pm 0.5$ & -2.2 & 96.3 \\
8 & $11.0 \pm 0.5$ & $10.8 \pm 0.3$ & -2.0 & 743.9 \\
32 & $18.8 \pm 0.2$ & $20.1 \pm 0.2$ & 7.1 & 1590.2 \\
64 & $33.8 \pm 0.3$ & $36.8 \pm 0.2$ & 8.8 & 1741.1 \\
128 & $62.9 \pm 0.3$ & $69.5 \pm 0.2$ & 10.4 & 1843.0 \\
250 & $119.4 \pm 0.3$ & $132.0 \pm 0.3$ & 10.5 & 1894.5 \\
\hline
\end{tabular}
\end{table}

\clearpage

\FloatBarrier
\clearpage
\section{Appendix}
\setcounter{figure}{0}
\renewcommand{\thefigure}{A\arabic{figure}}
\setcounter{table}{0}
\renewcommand{\thetable}{A\arabic{table}}

In this Appendix we present training loss curves and a table overviewing the hyperparameter choice for our analysis. 
\begin{figure}[!htbp]
    \centering

    \begin{subfigure}{0.6\linewidth}
        \centering
        \includegraphics[width=\linewidth]{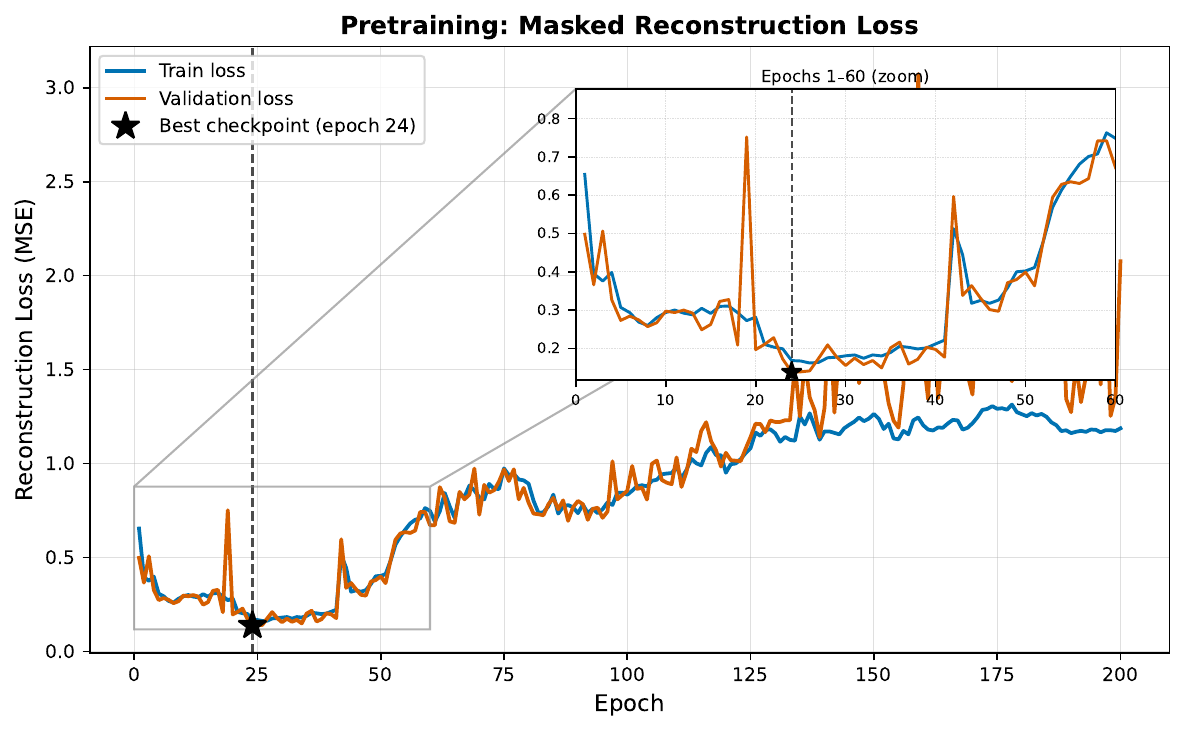}
        \caption{Pretraining loss curve as a function of epochs.}
        \label{fig:pretrain_loss}
    \end{subfigure}

    \vspace{0.8em}

    \begin{subfigure}{\linewidth}
        \centering
        \includegraphics[width=\linewidth]{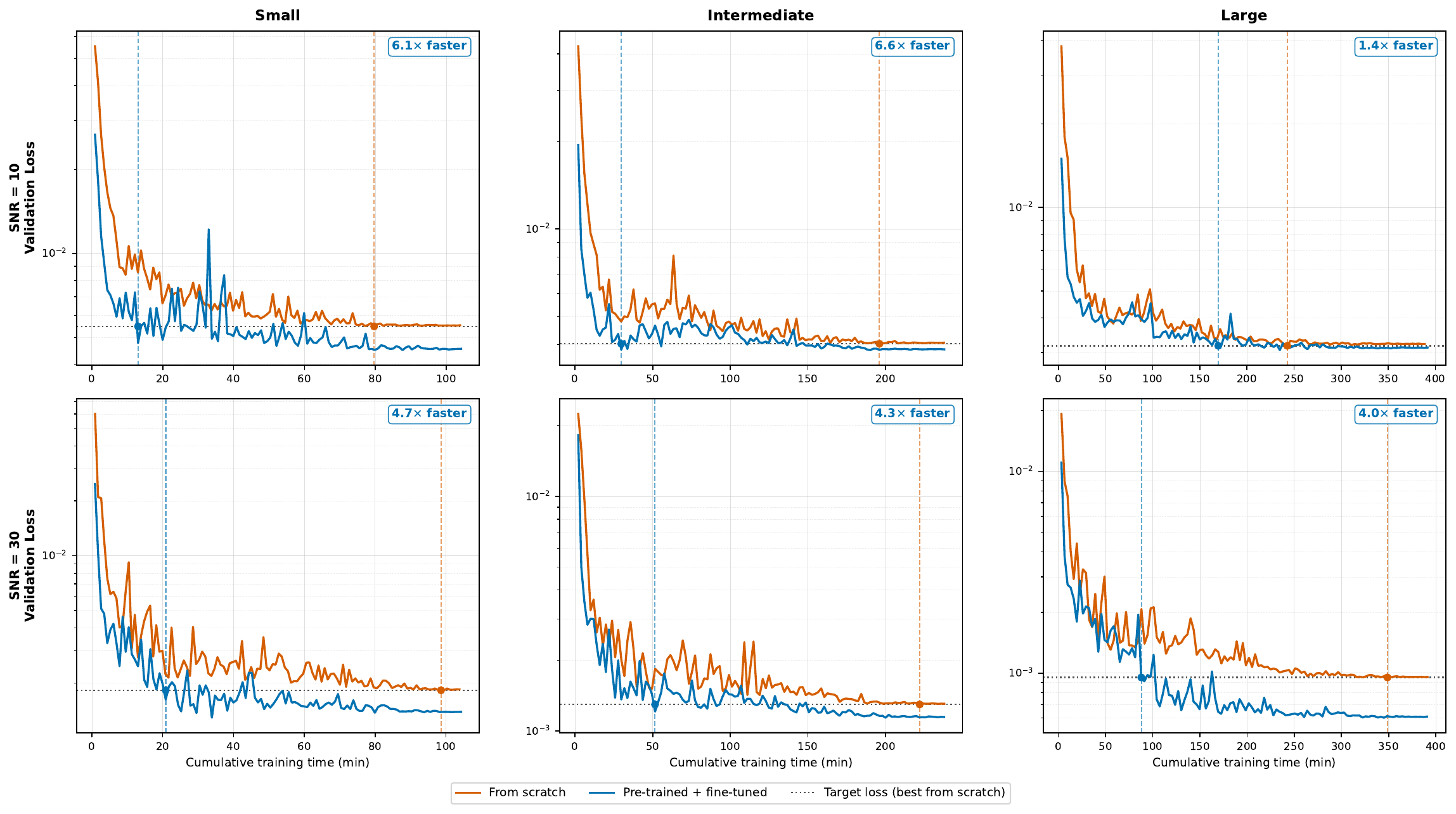}
        \caption{Val loss as a function of cumulative training time for models trained from scratch (orange) and fine-tuned (blue), across all configurations. The dotted horizontal lines (orange, blue) and the associated bullet points indicate the target loss. 
        We refer to Table~\ref{tab:convergence_speedup1} for detailed numerical values. }
        \label{fig:convergence_grid}
    \end{subfigure}

    \caption{Training dynamics during pretraining, and fine-tuning. During pretraining, training and validation losses decrease rapidly during the early stages of training, followed by a gradual increase at later epochs. This behaviour is consistent with the non-stationary nature of the self-targeting pretraining objective, where the reconstruction target is generated by the same evolving network.}
    \label{fig:training_dynamics}
\end{figure}

\begin{table*}[h]
\centering
\caption{Hyperparameter configuration for the pretraining and training stages of the model.}
\label{tab:hyperparameters}
\begin{threeparttable}
\begin{tabular}{llc}
\toprule
\textbf{Stage / Component} & \textbf{Parameter} & \textbf{Value} \\
\midrule
\multicolumn{3}{l}{\textit{\textbf{Pretraining}}} \\
\addlinespace
& Optimiser & AdamW \\
& Learning Rate & $1 \times 10^{-4}$ \\
& LR Scheduler & Cosine Annealing with Warmup \\
& Warmup Ratio & 0.1 \\
& Batch Size & 250 \\
& Max Epochs & 200 (with saving the best Val checkpoint)\\
& Masking Ratio ($r$) & 0.15 \\
& Mean Masked Length ($l_m$) & 32 \\
\addlinespace[1em]
\multicolumn{3}{l}{\textit{\textbf{Fine-tuning and Training from scratch}}}\\
\addlinespace
\textit{Training} & Optimiser & AdamW \\
& Learning Rate & $5 \times 10^{-4}$ \\
& Weight Decay & $1 \times 10^{-2}$ \\
& LR Scheduler & Cosine Annealing with Warmup \\
& Warmup Ratio & 0.1 \\
& Batch Size & 250 \\
& Max Epochs & 120 (with saving the best Val checkpoint) \\
\addlinespace
\textit{Feature Extractor} & Initial Stem Channels & 64 \\
& Num. Main Blocks ($N_B$) & 3 \\
& Modules per Block ($M$) & 3 \\
& Bottleneck Channels & 32 \\
& Kernel Sizes & [10, 20, 40] \\
& Filters per Branch ($d_{branch}$) & [32, 48, 64] \\
\addlinespace
\textit{Transformer Encoder} & Model Dimension ($d_{model}$) & 256 \\
& Num. Layers & 4 \\
& Num. Heads ($n_{heads}$) & 4 \\
& Feed-forward Dim. ($d_{ff}$) & 512 \\
& Dropout Rate & 0.10 \\
\addlinespace
\textit{MLP Head} & Num. Hidden Layers & 2 \\
& Hidden Dims & [256, 64] \\
& Output layer Dim & 4\\
& Dropout Rate & 0.10 \\
\bottomrule
\end{tabular}
\end{threeparttable}
\end{table*}
\clearpage

\bibliographystyle{unsrt}
\bibliography{ref}

@article{Dax:2024mcn,
    author = {Dax, Maximilian and Green, Stephen R. and Gair, Jonathan and Gupte, Nihar and P{\"u}rrer, Michael and Raymond, Vivien and Wildberger, Jonas and Macke, Jakob H. and Buonanno, Alessandra and Sch{\"o}lkopf, Bernhard},
    title = "{Real-time inference for binary neutron star mergers using machine learning}",
    eprint = "2407.09602",
    archivePrefix = "arXiv",
    primaryClass = "gr-qc",
    reportNumber = "LIGO-P2400294",
    doi = "10.1038/s41586-025-08593-z",
    journal = "Nature",
    volume = "639",
    number = "8053",
    pages = "49--53",
    year = "2025"
}

@misc{kofler2025,
      title={Flexible Gravitational-Wave Parameter Estimation with Transformers}, 
      author={Annalena Kofler and Maximilian Dax and Stephen R. Green and Jonas Wildberger and Nihar Gupte and Jakob H. Macke and Jonathan Gair and Alessandra Buonanno and Bernhard Schölkopf},
      year={2025},
      eprint={2512.02968},
      archivePrefix={arXiv},
      primaryClass={gr-qc},
      url={https://arxiv.org/abs/2512.02968}, 
}

@article{Wang:2025aqk,
  title = {Anatomy of parameter-estimation biases in overlapping gravitational-wave signals: Detector network},
  author = {Wang, Ziming and Liang, Dicong and Shao, Lijing},
  journal = {Phys. Rev. D},
  volume = {113},
  issue = {2},
  pages = {023044},
  numpages = {19},
  year = {2026},
  publisher = {American Physical Society},
  doi = {10.1103/3f5p-q6tq},
  url = {https://link.aps.org/doi/10.1103/3f5p-q6tq}
}

@article{Chan:2025pdf,
    author = "Chan, Juno C. L. and Maga{\~n}a Zertuche, Lorena and Ezquiaga, Jose Mar{\'\i}a and Lo, Rico K. L. and Vujeva, Luka and Bowman, Joey",
    title = "{Identification and characterization of distorted gravitational waves by lensing using deep learning}",
    eprint = "2511.07186",
    archivePrefix = "arXiv",
    primaryClass = "gr-qc",
    month = "11",
    year = "2025",
    note = {arXiv:2511.07186}
}

@article{Inguglia:2025cig,
    author = "Inguglia, Gianluca and Haigh, Huw and Vitulova, Kristyna and Dupletsa, Ulyana",
    title = "{Towards an anomaly detection pipeline for gravitational waves at the Einstein Telescope}",
    eprint = "2511.13154",
    archivePrefix = "arXiv",
    primaryClass = "gr-qc",
    month = "11",
    year = "2025",
    note = {arXiv:2511.13154}
}

@article{shen2022deep,
  title={Statistically-informed deep learning for gravitational wave parameter estimation},
  author={Shen, Hongyu and Huerta, EA and O'Shea, Eamonn and Kumar, Prayush and Zhao, Zhizhen},
  journal={Machine Learning: Science and Technology},
  volume={3},
  number={1},
  pages={015007},
  year={2022},
  publisher={IOP Publishing},
  note={}
}

@article{papalini2025transformer,
  title={Can Transformers help us perform parameter estimation of overlapping signals in gravitational wave detectors?},
  author={Papalini, Lucia and De Santi, Federico and Razzano, Massimiliano and Heng, Ik Siong and Cuoco, Elena},
  journal={arXiv:2505.02773},
  year={2025},
  note={}
}

@article{Abbott2023gwtc3,
  title={GWTC-3: Compact Binary Coalescences Observed by LIGO and Virgo during the Second Part of the Third Observing Run},
  author={Abbott, R and Abbott, TD and Acernese, F and Ackley, K and Adams, C and Adhikari, N and Adhikari, RX and Adya, VB and Affeldt, C and Agarwal, D and others},
  collaboration={LIGO Scientific Collaboration and Virgo Collaboration and KAGRA Collaboration},
  journal={Physical Review X},
  volume={13},
  number={4},
  pages={041039},
  year={2023},
  publisher={APS},
  doi={10.1103/PhysRevX.13.041039}
}

@article{Abbott2016,
  title={Observation of gravitational waves from a binary black hole merger},
  author={Abbott, Benjamin P and Abbott, Richard and Abbott, TD and Abernathy, MR and Acernese, Fausto and Ackley, K and Adams, C and Adams, T and Addesso, P and Adhikari, RX and others},
  collaboration={LIGO Scientific Collaboration and Virgo Collaboration},
  journal={Physical Review Letters},
  volume={116},
  number={6},
  pages={061102},
  year={2016},
  publisher={APS},
  doi={10.1103/PhysRevLett.116.061102}
}

@article{Abbott2021,
  title={GWTC-2: Compact Binary Coalescences Observed by LIGO and Virgo during the First Half of the Third Observing Run},
  author={Abbott, R and Abbott, TD and Abraham, S and Acernese, F and Ackley, K and Adams, A and Adams, C and Adhikari, RX and Adya, VB and Affeldt, C and others},
  collaboration={LIGO Scientific Collaboration and Virgo Collaboration},
  journal={Physical Review X},
  volume={11},
  number={2},
  pages={021053},
  year={2021},
  publisher={APS},
  doi={10.1103/PhysRevX.11.021053}
}

@article{TheLIGOScientificCollaboration2021,
  title={GWTC-2.1: Deep extended catalog of compact binary coalescences observed by LIGO and Virgo during the first half of the third observing run},
  author={Abbott, R and Abbott, TD and Acernese, F and Ackley, K and Adams, C and Adhikari, N and others},
  collaboration={LIGO Scientific Collaboration and Virgo Collaboration},
  journal={Physical Review D},
  volume={109},
  number={2},
  pages={022001},
  year={2024},
  publisher={APS},
  doi={10.1103/PhysRevD.109.022001}
}

@article{Berti2015,
  title={Testing general relativity with present and future astrophysical observations},
  author={Berti, Emanuele and Barausse, Enrico and Cardoso, Vitor and Gualtieri, Leonardo and Pani, Paolo and Sperhake, Ulrich and Stein, Leo C and Wex, Norbert and Yagi, Kent and Baker, Tessa and others},
  journal={Classical and Quantum Gravity},
  volume={32},
  number={24},
  pages={243001},
  year={2015},
  publisher={IOP Publishing},
  doi={10.1088/0264-9381/32/24/243001}
}

@article{Saltas:2018fwy,
  title={Anisotropic stress as a signature of nonstandard propagation of gravitational waves},
  author={Saltas, Ippocratis D and Sawicki, Ignacy and Amendola, Luca and Kunz, Martin},
  journal={Physical Review Letters},
  volume={113},
  number={19},
  pages={191101},
  year={2014},
  publisher={APS},
  doi={10.1103/PhysRevLett.113.191101}
}

@article{LISA:2022kgy,
  title={Laser Interferometer Space Antenna},
  author={Amaro-Seoane, Pau and Audley, Heather and Babak, Stanislav and Baker, John and Barausse, Enrico and Bender, Peter and others},
  journal={arXiv:1702.00786},
  year={2017},
  eprint={1702.00786},
  archivePrefix={arXiv},
  primaryClass={astro-ph.IM}
}

@article{Barausse:2020rsu,
  title={Prospects for fundamental physics with LISA},
  author={Barausse, Enrico and Berti, Emanuele and Hertog, Thomas and Hughes, Scott A and Jetzer, Philippe and Pani, Paolo and Sotiriou, Thomas P and Tamanini, Nicola and Witek, Helvi and Yagi, Kent and others},
  journal={General Relativity and Gravitation},
  volume={52},
  pages={81},
  year={2020},
  publisher={Springer},
  doi={10.1007/s10714-020-02691-1}
}

@article{Blanchet2014,
  title={Gravitational radiation from post-Newtonian sources and inspiralling compact binaries},
  author={Blanchet, Luc},
  journal={Living Reviews in Relativity},
  volume={17},
  pages={2},
  year={2014},
  publisher={Springer},
  doi={10.12942/lrr-2014-2}
}

@article{Maggiore2000,
  title={Gravitational wave experiments and early universe cosmology},
  author={Maggiore, Michele},
  journal={Physics Reports},
  volume={331},
  number={6},
  pages={283--367},
  year={2000},
  publisher={Elsevier},
  doi={10.1016/S0370-1573(99)00102-7}
}

@article{Buonanno99,
  title={Effective-one-body approach to general relativistic two-body dynamics},
  author={Buonanno, Alessandra and Damour, Thibault},
  journal={Physical Review D},
  volume={59},
  number={8},
  pages={084006},
  year={1999},
  publisher={APS},
  doi={10.1103/PhysRevD.59.084006}
}

@article{Bohe2017,
  title={Improved effective-one-body model of spinning, nonprecessing binary black holes for the era of gravitational-wave astrophysics with advanced detectors},
  author={Boh{\'e}, Alejandro and Shao, Lijing and Taracchini, Andrea and Buonanno, Alessandra and Babak, Stanislav and Harry, Ian W and Hinder, Ian and Ossokine, Serguei and P{\"u}rrer, Michael and Raymond, Vivien and others},
  journal={Physical Review D},
  volume={95},
  number={4},
  pages={044028},
  year={2017},
  publisher={APS},
  doi={10.1103/PhysRevD.95.044028}
}

@article{Hild2011,
  title={Sensitivity studies for third-generation gravitational wave observatories},
  author={Hild, Stefan and Abernathy, M and Acernese, F and Amaro-Seoane, P and Andersson, N and Arun, K and Barone, F and Barr, B and Barsuglia, M and Beker, M and others},
  journal={Classical and Quantum Gravity},
  volume={28},
  number={9},
  pages={094013},
  year={2011},
  publisher={IOP Publishing},
  doi={10.1088/0264-9381/28/9/094013}
}

@article{Cutler1994,
  title={Gravitational waves from merging compact binaries: How accurately can one extract the binary's parameters from the inspiral waveform?},
  author={Cutler, Curt and Flanagan, {\'E}anna E},
  journal={Physical Review D},
  volume={49},
  number={6},
  pages={2658},
  year={1994},
  publisher={APS},
  doi={10.1103/PhysRevD.49.2658}
}

@article{Vallisneri2008,
  title={Use and abuse of the Fisher information matrix in the assessment of gravitational-wave parameter-estimation prospects},
  author={Vallisneri, Michele},
  journal={Physical Review D},
  volume={77},
  number={4},
  pages={042001},
  year={2008},
  publisher={APS},
  doi={10.1103/PhysRevD.77.042001}
}

@article{allen2005chi2,
  title={$\chi^2$ time-frequency discriminator for gravitational wave detection},
  author={Allen, Bruce},
  journal={Physical Review D},
  volume={71},
  number={6},
  pages={062001},
  year={2005},
  publisher={APS},
  doi={10.1103/PhysRevD.71.062001}
}

@article{skilling2006nested,
  title={Nested sampling for general Bayesian computation},
  author={Skilling, John},
  journal={Bayesian Analysis},
  volume={1},
  number={4},
  pages={833--859},
  year={2006},
  publisher={International Society for Bayesian Analysis},
  doi={10.1214/06-BA127}
}

@article{veitch2010bayesian,
  title={Bayesian coherent analysis of in-spiral gravitational wave signals with a detector network},
  author={Veitch, John and Vecchio, Alberto},
  journal={Physical Review D},
  volume={81},
  number={6},
  pages={062003},
  year={2010},
  publisher={APS},
  doi={10.1103/PhysRevD.81.062003}
}

@article{vandersluys2008parameter,
  title={Parameter estimation of spinning binary inspirals using Markov chain Monte Carlo},
  author={van der Sluys, Marc and Raymond, Vivien and Mandel, Ilya and R{\"o}ver, Christian and Christensen, Nelson and Kalogera, Vicky and Meyer, Renate and Vecchio, Alberto},
  journal={Classical and Quantum Gravity},
  volume={25},
  number={18},
  pages={184011},
  year={2008},
  publisher={IOP Publishing},
  doi={10.1088/0264-9381/25/18/184011}
}

@article{cornish2015bayeswave,
  title={BayesWave: Bayesian inference for gravitational wave bursts and instrument glitches},
  author={Cornish, Neil J and Littenberg, Tyson B},
  journal={Classical and Quantum Gravity},
  volume={32},
  number={13},
  pages={135012},
  year={2015},
  publisher={IOP Publishing},
  doi={10.1088/0264-9381/32/13/135012}
}

@article{veitch2015parameter,
  title={Parameter estimation for compact binaries with ground-based gravitational-wave observations using the LALInference software library},
  author={Veitch, John and Raymond, Vivien and Farr, Ben and Farr, Will and Graff, Philip and Vitale, Salvatore and Aylott, Ben and Blackburn, Kent and Christensen, Nelson and Coughlin, Michael and others},
  journal={Physical Review D},
  volume={91},
  number={4},
  pages={042003},
  year={2015},
  publisher={APS},
  doi={10.1103/PhysRevD.91.042003}
}

@article{ashton2019bilby,
  title={BILBY: A user-friendly Bayesian inference library for gravitational-wave astronomy},
  author={Ashton, Gregory and H{\"u}bner, Moritz and Lasky, Paul D and Talbot, Colm and Ackley, Kendall and Biscoveanu, Sylvia and Chu, Qi and Divakarla, Atul and Easter, Paul J and Goncharov, Boris and others},
  journal={The Astrophysical Journal Supplement Series},
  volume={241},
  number={2},
  pages={27},
  year={2019},
  publisher={IOP Publishing},
  doi={10.3847/1538-4365/ab06fc}
}

@article{GWsequential,
  title = {Sequential simulation-based inference for gravitational wave signals},
  author = {Bhardwaj, Uddipta and Alvey, James and Miller, Benjamin Kurt and Nissanke, Samaya and Weniger, Christoph},
  journal = {Phys. Rev. D},
  volume = {108},
  issue = {4},
  pages = {042004},
  numpages = {21},
  year = {2023},
  month = {Aug},
  publisher = {American Physical Society},
  doi = {10.1103/PhysRevD.108.042004},
  url = {https://link.aps.org/doi/10.1103/PhysRevD.108.042004}
}

@article{thrane2019introduction,
  title={An introduction to Bayesian inference in gravitational-wave astronomy: Parameter estimation, model selection, and hierarchical models},
  author={Thrane, Eric and Talbot, Colm},
  journal={Publications of the Astronomical Society of Australia},
  volume={36},
  pages={e010},
  year={2019},
  publisher={Cambridge University Press},
  doi={10.1017/pasa.2019.2}
}

@article{george2018deep2_inference,
  title={Deep neural networks to enable real-time multimessenger astrophysics},
  author={George, Daniel and Huerta, EA},
  journal={Physical Review D},
  volume={97},
  number={4},
  pages={044039},
  year={2018},
  publisher={APS},
  doi={10.1103/PhysRevD.97.044039}
}

@misc{chatterjee2024transformer,
      title={Pre-trained Audio Transformer as a Foundational AI Tool for Gravitational Waves}, 
      author={Chayan Chatterjee and Abigail Petulante and Karan Jani and Jesse Spencer-Smith and Yang Hu and Roy Lau and Haowei Fu and Trang Hoang and Stephen Chong Zhao and Suyash Deshmukh},
      year={2025},
      eprint={2412.20789},
      archivePrefix={arXiv},
      primaryClass={gr-qc},
      url={https://arxiv.org/abs/2412.20789}, 
}

@article{williams2021nested,
  title={Nested sampling with normalizing flows for gravitational-wave inference},
  author={Williams, Michael J and Veitch, John and Messenger, Chris},
  journal={Physical Review D},
  volume={103},
  number={10},
  pages={103006},
  year={2021},
  publisher={APS},
  doi={10.1103/PhysRevD.103.103006}
}

@article{williams2023importance,
  title={Importance nested sampling with normalising flows},
  author={Williams, Michael J and Veitch, John and Messenger, Chris},
  journal={Machine Learning: Science and Technology},
  volume={4},
  number={3},
  pages={035011},
  year={2023},
  publisher={IOP Publishing},
  doi={10.1088/2632-2153/acd5aa}
}

@article{green2020gravitational,
  title={Gravitational-wave parameter estimation with autoregressive neural network flows},
  author={Green, Stephen R and Simpson, Christine and Gair, Jonathan},
  journal={Physical Review D},
  volume={102},
  number={10},
  pages={104057},
  year={2020},
  publisher={APS},
  doi={10.1103/PhysRevD.102.104057}
}

@article{dax2021real,
  title={Real-time gravitational wave science with neural posterior estimation},
  author={Dax, Maximilian and Green, Stephen R and Gair, Jonathan and Macke, Jakob H and Buonanno, Alessandra and Sch{\"o}lkopf, Bernhard},
  journal={Physical Review Letters},
  volume={127},
  number={24},
  pages={241103},
  year={2021},
  publisher={APS},
  doi={10.1103/PhysRevLett.127.241103}
}

@article{dax2023neural,
  author    = {Maximilian Dax and Stephen R. Green and Jonathan Gair and Michael P{\"u}rrer and Jonas Wildberger and Jakob H. Macke and Alessandra Buonanno and Bernhard Sch{\"o}lkopf},
  title     = {Neural Importance Sampling for Rapid and Reliable Gravitational-Wave Inference},
  journal   = {Physical Review Letters},
  volume    = {130},
  number    = {17},
  pages     = {171403},
  year      = {2023},
  doi       = {10.1103/PhysRevLett.130.171403},
  eprint    = {2210.05686},
  archivePrefix = {arXiv}
}

@article{gabbard2022bayesian,
  title={Bayesian parameter estimation using conditional variational autoencoders for gravitational-wave astronomy},
  author={Gabbard, Hunter and Messenger, Chris and Heng, Ik Siong and Tonolini, Francesco and Murray-Smith, Roderick},
  journal={Nature Physics},
  volume={18},
  number={1},
  pages={112--117},
  year={2022},
  publisher={Nature Publishing Group},
  doi={10.1038/s41567-021-01425-7}
}

@article{chua2019reducing,
  title = {Learning Bayesian Posteriors with Neural Networks for Gravitational-Wave Inference},
  author = {Chua, Alvin J. K. and Vallisneri, Michele},
  journal = {Phys. Rev. Lett.},
  volume = {124},
  issue = {4},
  pages = {041102},
  numpages = {6},
  year = {2020},
  month = {Jan},
  publisher = {American Physical Society},
  doi = {10.1103/PhysRevLett.124.041102},
  url = {https://link.aps.org/doi/10.1103/PhysRevLett.124.041102}
}

@article{cuoco2021enhancing,
  title={Enhancing gravitational-wave science with machine learning},
  author={Cuoco, Elena and Powell, Jade and Cavagli{\`a}, Marco and Ackley, Kendall and Bejger, Micha{\l} and Chatterjee, Chayan and Coughlin, Michael and Coughlin, Scott and Easter, Paul and Essick, Reed and others},
  journal={Machine Learning: Science and Technology},
  volume={2},
  number={1},
  pages={011002},
  year={2021},
  publisher={IOP Publishing},
  doi={10.1088/2632-2153/abb93a}
}

@incollection{stergioulas2024machine,
  title={Machine Learning Applications in Gravitational Wave Astronomy},
  author={Stergioulas, Nikolaos},
  booktitle={Compact Objects in the Universe},
  editor={Papantonopoulos, Eleftherios and Mavromatos, Nikolaos},
  pages={329--356},
  year={2024},
  publisher={Springer},
  address={Cham},
  doi={10.1007/978-3-031-55098-0_12},
  eprint={2401.07406},
  archivePrefix={arXiv},
  primaryClass={astro-ph.IM},
  url={https://arxiv.org/abs/2401.07406}
}

@article{vaswani2017attention,
  title={Attention is all you need},
  author={Vaswani, Ashish and Shazeer, Noam and Parmar, Niki and Uszkoreit, Jakob and Jones, Llion and Gomez, Aidan N and Kaiser, {\L}ukasz and Polosukhin, Illia},
  journal={Advances in Neural Information Processing Systems},
  volume={30},
  pages={5998--6008},
  year={2017},
  publisher={Curran Associates, Inc.}
}

@misc{Devlin2018,
  title={{BERT}: Pre-training of Deep Bidirectional Transformers for Language Understanding},
  author={Devlin, Jacob and Chang, Ming-Wei and Lee, Kenton and Toutanova, Kristina},
  year={2018},
  eprint={1810.04805},
  archivePrefix={arXiv},
  primaryClass={cs.CL},
  url={https://arxiv.org/abs/1810.04805},
  note = {arXiv:1810.04805}
}

@misc{Radford2018,
  title={Improving Language Understanding by Generative Pre-Training},
  author={Radford, Alec and Narasimhan, Karthik and Salimans, Tim and Sutskever, Ilya},
  year={2018},
  url={https://cdn.openai.com/research-covers/language-unsupervised/language_understanding_paper.pdf}
}

@misc{Radford2019,
  title={Language Models are Unsupervised Multitask Learners},
  author={Radford, Alec and Wu, Jeffrey and Child, Rewon and Luan, David and Amodei, Dario and Sutskever, Ilya},
  year={2019},
  url={https://cdn.openai.com/better-language-models/language_models_are_unsupervised_multitask_learners.pdf}
}

@article{inceptiontime,
  title={InceptionTime: Finding AlexNet for Time Series Classification},
  author={Fawaz, Hassan Ismail and Lucas, Benjamin and Forestier, Germain and Pelletier, Charlotte and Schmidt, Daniel F and Weber, Jonathan and Webb, Geoffrey I and Idoumghar, Lhassane and Muller, Pierre-Alain and Petitjean, Fran{\c{c}}ois},
  journal={Data Mining and Knowledge Discovery},
  volume={34},
  number={6},
  pages={1936--1962},
  year={2020},
  publisher={Springer},
  doi={10.1007/s10618-020-00710-y},
  eprint={1909.04939},
  archivePrefix={arXiv},
  primaryClass={cs.LG},
  url={http://arxiv.org/abs/1909.04939}
}

@article{Hochreiter1997,
  title={Long short-term memory},
  author={Hochreiter, Sepp and Schmidhuber, J{\"u}rgen},
  journal={Neural Computation},
  volume={9},
  number={8},
  pages={1735--1780},
  year={1997},
  publisher={MIT Press},
  doi={10.1162/neco.1997.9.8.1735}
}

@inproceedings{loshchilov2019decoupledweightdecayregularization,
  title={Decoupled Weight Decay Regularization},
  author={Loshchilov, Ilya and Hutter, Frank},
  booktitle={International Conference on Learning Representations},
  year={2019},
  url={https://openreview.net/forum?id=Bkg6RiCqY7},
  eprint={1711.05101},
  archivePrefix={arXiv},
  primaryClass={cs.LG}
}

@inproceedings{zerveas2020transformerbasedframeworkmultivariatetime,
  title={A Transformer-based Framework for Multivariate Time Series Representation Learning},
  author={Zerveas, George and Jayaraman, Srideepika and Patel, Dhaval and Bhamidipaty, Anuradha and Eickhoff, Carsten},
  booktitle={Proceedings of the 27th ACM SIGKDD Conference on Knowledge Discovery \& Data Mining},
  pages={2114--2124},
  year={2021},
  publisher={ACM},
  doi={10.1145/3447548.3467401},
  eprint={2010.02803},
  archivePrefix={arXiv},
  primaryClass={cs.LG},
  url={https://arxiv.org/abs/2010.02803}
}

@article{tay2022efficienttransformerssurvey,
  title={Efficient Transformers: A Survey},
  author={Tay, Yi and Dehghani, Mostafa and Bahri, Dara and Metzler, Donald},
  journal={ACM Computing Surveys},
  volume={55},
  number={6},
  pages={1--28},
  year={2022},
  publisher={ACM},
  doi={10.1145/3530811},
  eprint={2009.06732},
  archivePrefix={arXiv},
  primaryClass={cs.LG},
  url={https://arxiv.org/abs/2009.06732}
}

@inproceedings{gal2016dropoutbayesianapproximationrepresenting,
  title={Dropout as a Bayesian Approximation: Representing Model Uncertainty in Deep Learning},
  author={Gal, Yarin and Ghahramani, Zoubin},
  booktitle={Proceedings of the 33rd International Conference on Machine Learning},
  pages={1050--1059},
  year={2016},
  publisher={PMLR},
  eprint={1506.02142},
  archivePrefix={arXiv},
  primaryClass={stat.ML},
  url={https://arxiv.org/abs/1506.02142}
}

@inproceedings{kendall2017uncertaintiesneedbayesiandeep,
  title={What Uncertainties Do We Need in Bayesian Deep Learning for Computer Vision?},
  author={Kendall, Alex and Gal, Yarin},
  booktitle={Advances in Neural Information Processing Systems},
  volume={30},
  pages={5574--5584},
  year={2017},
  publisher={Curran Associates, Inc.},
  eprint={1703.04977},
  archivePrefix={arXiv},
  primaryClass={cs.CV},
  url={https://arxiv.org/abs/1703.04977}
}

@misc{kaplan2020scalinglawsneurallanguage,
  title={Scaling Laws for Neural Language Models},
  author={Kaplan, Jared and McCandlish, Sam and Henighan, Tom and Brown, Tom B and Chess, Benjamin and Child, Rewon and Gray, Scott and Radford, Alec and Wu, Jeffrey and Amodei, Dario},
  year={2020},
  eprint={2001.08361},
  archivePrefix={arXiv},
  primaryClass={cs.LG},
  url={https://arxiv.org/abs/2001.08361},
  note = {arXiv:2001.08361}
}

\end{document}